\documentclass[sn-mathphys]{sn-jnl}

\usepackage{lipsum}
\usepackage{isotope}
\usepackage{comment}

\jyear{2022}

\begin{document}

\title[Article Title]{Perspectives on few-body cluster structures in exotic nuclei}

\author[1,2]{\fnm{Daniel} \sur{Bazin}}\email{bazin@frib.msu.edu}
\author[3]{\fnm{Kevin} \sur{Becker}}\email{kbeck13@lsu.edu}
\author[4]{\fnm{Francesca} \sur{Bonaiti}}\email{fbonaiti@uni-mainz.de}
\author[5] {{\fnm{Charlotte} \sur{Elster}}} \email{}
\author*[6,1,7]{\fnm{Kévin} \sur{Fossez}}\email{kfossez@fsu.edu}
\author[8]{\fnm{Tobias} \sur{Frederico}}\email{tobias@ita.br}
\author[9]{\fnm{Alex} \sur{Gnech}}\email{agnech@ectstar.eu}
\author*[1,10]{\fnm{Chloë} \sur{Hebborn}}\email{hebborn@frib.msu.edu}
\author[{}]{\fnm{Michael} \sur{Higgins$^{\text{11}}$}}\email{higgin45@purdue.edu}
\author[10] {\fnm{Linda} \sur{Hlophe}} \email{hlophe1@llnl.gov}
\author[7] {\fnm{Benjamin} \sur{Kay}} \email{kay@anl.gov}
\author*[12]{\fnm{Sebastian} \sur{König}}\email{skoenig@ncsu.edu}
\author[10] {{\fnm{Konstantinos} \sur{Kravvaris}}} \email{kravvaris1@llnl.gov}
\author[13] {\fnm{Jesus} \sur{Lubian}}\email{jlubian@id.uff.br}
\author[14] {{\fnm{Augusto} \sur{Macchiavelli}}} \email{}
\author[1,2] {\fnm{Filomena} \sur{Nunes}} \email{nunes@frib.msu.edu}
\author*[14,15]{\fnm{Lucas}
\sur{Platter}}
\email{lplatter@utk.edu}
\author[10] {\fnm{Gregory} \sur{Potel}} \email{potelaguilar1@llnl.gov}
\author[1] {\fnm{Xilin} \sur{Zhang}} \email{zhangx@frib.msu.edu}

\affil[1]{\orgname{Facility for Rare Isotope Beams}, \city{East Lansing}, \postcode{48824}, \state{MI}, \country{USA}}

\affil[2]{\orgname{Department of Physics and Astronomy}, \orgaddress{Michigan State University, \city{East Lansing}, \postcode{48824}, \state{MI}, \country{USA}}}

\affil[3]{\orgname{Department of Physics and Astronomy}, \orgaddress{Louisiana State University, \city{Baton Rouge}, \postcode{70803}, \state{LA}, \country{USA}}}

\affil[4]{Institut f\"ur Kernphysik and PRISMA$^+$ Cluster of Excellence, Johannes Gutenberg-Universit\"at, 55128
  Mainz, Germany}

  \affil[5]{Institute of Nuclear and Particle Physics, and Department of Physics and Astronomy, Ohio University, Athens, OH 45701, USA}

  \affil*[6]{\orgname{Florida State University}, \orgaddress{\city{Tallahassee}, \postcode{32306}, \state{FL}, \country{USA}}}

\affil[7]{\orgdiv{Physics Division}, \orgname{Argonne National Laboratory}, \city{Lemont}, \postcode{60439}, \state{IL}, \country{USA}}

\affil[8]{\orgname{Instituto Tecnol\'ogico de Aeron\'autica, DCTA}, 
\postcode{12228-900},
\city{S\~ao Jos\'e dos Campos},  
\country{Brazil}}

\affil[9]{\orgname{Theory Center, Jefferson Lab}, \city{Newport News}, \postcode{23606}, \state{VA}, \country{USA}}
\affil[10]{\orgname{Lawrence Livermore National Laboratory}, \orgaddress{P.O. Box 808, L-414, \city{Livermore}, \postcode{94551}, \state{CA}, \country{USA}}}

\affil[11]{\orgname{Department of Physics and Astronomy}, \orgaddress{Purdue University, \city{West Lafayette}, \postcode{47906}, \state{IN}, \country{USA}}}

\affil[12]{\orgname{Department of Physics, North Carolina State University}, \orgaddress{\city{Raleigh}, \postcode{27695}, \state{NC}, \country{USA}}}

\affil[13]{\orgname{Department of Physics},  \orgaddress{Federal Fluminense University}, \city{Niterói}, \postcode{24210-340}, \state{R.J.}, \country{Brazil}}

\affil[14]{\orgname{Physics Division},
\orgaddress{Oak Ridge National Laboratory},
\city{Oak Ridge},
\postcode{37831}
\state{TN},
\country{USA}}

\affil[15]{\orgname{Department of Physics and Astronomy},
\orgaddress{University of Tennessee, Knoxville},
\city{Knoxville},
\postcode{37996},\state{TN}, \country{USA}}

\abstract{
It is a fascinating phenomenon in nuclear physics that states with a pronounced few-body structure can emerge from the complex dynamics of many nucleons.  Such halo or cluster states often appear near the boundaries of nuclear stability.  As such, they are an important part of the experimental program beginning at the Facility for Rare Isotope Beams (FRIB).  A concerted effort of theory and experiment is necessary both to analyze experiments involving effective few-body states, as well as to constrain and refine theories of the nuclear force in light of new data from these experiments.  As a contribution to exactly this effort, this paper compiles a collection of ``perspectives'' that emerged out of the Topical Program ``Few-body cluster structures in exotic nuclei and their role in FRIB experiments'' that was held at FRIB in August 2022 and brought together theorists and experimentalists working on this topic.
}


\maketitle

\newpage
\tableofcontents
\newpage

\newpage
\section*{
Preface}
\addcontentsline{toc}{section}{Preface}

\label{sec0}
\textbf{L.~Platter, K.~Fossez, C.~Hebborn, and S.~König}

Nuclear physics is rich with complexity, emerging ultimately out of the Standard Model of particle physics and in particular out of Quantum Chromodynamics (QCD), the fundamental theory of the strong interaction.  Protons and neutrons that bind together to form atomic nuclei are the initial effective degrees of freedom in nuclear physics and \textit{ab initio} nuclear structure theory uses these to compute observables for a wide range of nuclei.  However, interesting emergent phenomena appear near the boundaries of nuclear stability, the so-called driplines, where many nuclei display a transition to new effective degrees of freedom by hosting phenomena which indicate a dominant few-body cluster or halo structure.  This fascinating feature of nuclear systems leads frequently to dynamics that experimentally are similar to those of few-body systems in atomic and particle physics and are considered to be signatures of low-energy universality.  The signatures of cluster states are also particularly pronounced in nuclear reactions that, in most cases, cannot be accessed directly in \textit{ab initio} approaches at present.

The impact of clustering in nuclei close to the driplines affects ongoing and proposed experiments at the Facility for Rare Isotope Beams (FRIB) and other radioactive beam facilities around the world.  Halo nuclei, i.e., systems that can be described as a tightly bound core nucleus with weakly bound valence nucleons, are an important example of such clustered systems that have been studied extensively both experimentally and theoretically.  In these systems, a clear separation of scales exists facilitating the construction of a model-independent low-energy effective field theory (EFT), a systematic expansion with quantified uncertainty estimates.  However, clustering plays a role for a much larger group of nuclei and for those it is at present not clearly understood how to construct such EFTs.

An important task in this exciting time with new facilities starting to operate is therefore to develop new models and EFTs that treat cluster effects explicitly, consistently, and efficiently.  These approaches should be constrained by experimental data and be able to predict observables with reliable uncertainty estimates.  In addition, it is important to understand the mechanisms underlying the transition from many-nucleon systems to new effective degrees of freedom, and how the nuclear force gives rise to these phenomena.  However, the interplay between few-body effects and nuclear shell structure is not yet well understood and requires a combined effort from theory and experiment to advance our understanding.

Clearly the low-energy nuclear physics community needs to build bridges between different groups, involving both theorists and experimentalists, to join together in order to design, analyze, and explain current and future experiments and to refine and extend nuclear theories.  Experimental data can guide theorists in constructing better models and approaches that improve our understanding and interpretation of nuclear observables.  In turn, a strong theory program can assist experimental groups in identifying observables that can provide new insights into the underlying mechanisms of emergent phenomena in nuclear physics.

In this collection of ``perspectives,'' we give an overview of thoughts on the requirements for effective progress in this critical time that will provide a significant amount of new data.  The articles compiled here are the result of the FRIB Topical Program ``Few-Body Clusters in Exotic Nuclei and Their Role in FRIB Experiments'' held at FRIB/Michigan State University in August 2022. This program supported by the FRIB Theory Alliance brought together theorists and experimentalists working on this topic.  Each perspective was written by a small subgroup of participants as indicated for each piece.  It is our hope that this collection will be useful to the broader community and serve as a continuation of the stimulating discussions that we had during the program.

\clearpage
\section{Building the nuclear theory to experiment pipeline}
\label{sec1}
\textbf{K. Fossez, C. Hebborn, D. Bazin, and A. Macchiavelli}
\smallbreak

\begin{figure}[h]
\centering
\includegraphics[width=1\textwidth]{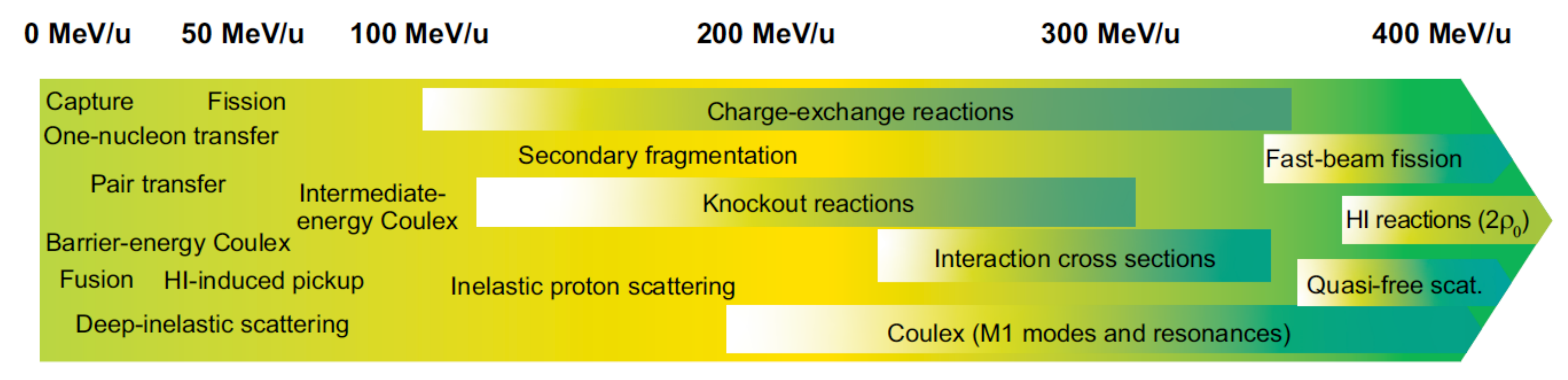}
\caption{Schematic representation of the possible reaction mechanisms that can be used to probe exotic nuclei at FRIB as a function of energy. Reactions beyond 200 MeV/u will require the FRIB400 upgrade. Figure adapted from Ref.~\cite{FRIBupgrade}}
\label{fribta1}
\end{figure}

The DOE-NP flagship FRIB has started operations at MSU and will push forward the exploration of the nuclear landscape by giving access to more than a thousand new exotic nuclei, mostly using nuclear reactions. 
Depending on the reaction considered and the nuclear degrees of freedom to be probed, the required beam energies can vary from keV/u to GeV/u. 
The different types of reactions that can be used as a function of the energy are shown schematically in Fig.~\ref{fribta1}. 
Reactions at energies up to about 200 MeV/u are already accessible using the current nominal FRIB beam energy, but those up to 400 MeV/u will require the so-called high-energy upgrade~\cite{FRIBupgrade}. 
In addition, in anticipation of the coming online of FRIB, critical detector development programs were started in the last decade~\cite{Equ}. 
These programs led to new state-of-the-art detector systems having unparalleled resolving power, and which will play a central role in addressing essentially all of the challenges put forth in the National Academies Decadal Study~\cite{NAS} and the 2015 NSAC Long Range Plan~\cite{LRP1}, \textit{i.e.} by answering fundamental questions in both nuclear structure and nuclear astrophysics such as ``How does subatomic matter organize itself and what phenomena emerge?" and ``How did visible matter come into being and how does it evolve"?

\begin{figure}
\centering
\includegraphics[width=0.9\textwidth]{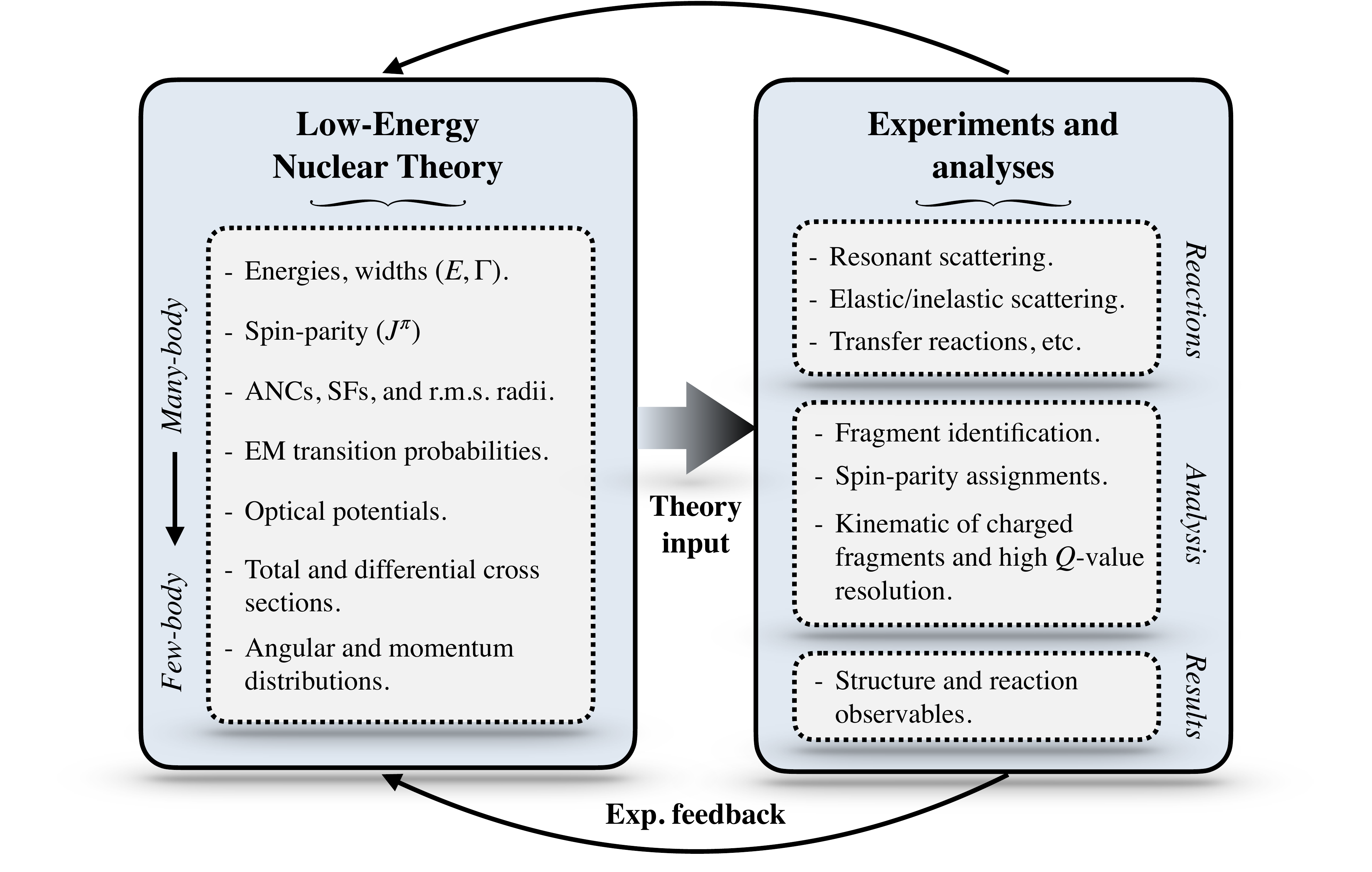}
\caption{Schematic representation of (some of) the theory inputs required to fully exploit experimental capabilities at FRIB.}
\label{fig_pipeline}
\end{figure}

However, the ultimate success of the FRIB scientific program will require a concerted effort between experiment and theory. 
Indeed, to infer accurate and precise nuclear properties of exotic nuclei from experimental data, reliable nuclear structure and reaction models will be needed. 
The schematic pipeline shown in Fig.~\ref{fig_pipeline} depicts some of required theory inputs necessary to guide experiments or even analyse data in some cases. 
Of course, data can then be used to provide feedback on theory models.

For example, the Active Target Time Projection Chamber (AT-TPC)~\cite{AT-TPC} is ideally suited to probe the cluster structure of low-energy resonances since it allows to perform resonant scattering reactions by scanning different beam energies in a single experiment. 
In order to determine complementary information such a the spins, parities, or widths of the resonances observed, it might be necessary to directly compare with structure models, or to look at transfer angular distributions which require the use of reaction models to be analysed. 
Moreover, the absolute value of transfer cross sections is directly related to the cluster content of the populated states, usually quantified in terms of spectroscopic factors. 

On the theory side, confrontation with experimental data should ideally be done with a theory using nucleons as degrees of freedom, and the nuclear forces between nucleons should be derived from effective field theories (EFTs) of the underlying theory of quantum chromodynamics (QCD)  and electroweak theory \cite{hergert20_2412,HammerReview2020}.
However, solving this canonical \textit{ab initio} nuclear many-body problem exactly for the description of both nuclear structure and reactions in a unified picture is a particularly daunting task, which so far has only be achieved in selected light nuclei with $A < 12$ \cite{Navratil_2016,quaglioni08_755,quaglioni09_756,lee09_949,nollett12_835,flores22}.
To make matters worse, issues remain in the construction of EFTs of nuclear forces due to the lack of clear separation of scales, effectively reducing the predictive power of any theory of the atomic nucleus beyond light nuclei \cite{stroberg21_2483}. 
This situation leads to the conundrum of low-energy nuclear theory: making accurate predictions by solving a difficult problem using an uncertain input. 
In that regard, the use of simplified effective approaches such as few-body reaction models or the widely successful nuclear shell model \cite{caurier05_424} to go beyond light nuclei comes as no surprise. 

In the present context of few-body cluster structures in exotic nuclei, one important line of development is the move towards a unification of nuclear structure and reactions, in which  1-, 2-, ..., $A$-nucleons decay channels are included in the description of the $A$-body problem~\cite{rotter91_448,michel09_2,volya05_470,baroni13_385}.  
It has been recognized early on that, far from the valley of stability, weak binding and couplings to continuum states play an important role in shaping the properties of exotic nuclei \cite{matsuo10_1238,michel10_4}. 
As a consequence, closer to the drip lines, \textit{i.e.} the regions of greatest interest for FRIB, nuclear models have shortcomings due to, for instance, the influence of couplings to continuum states \cite{okolowicz03_21,michel10_4}, the increased number of particles involved, or the presence of new emergent phenomena \cite{otsuka20_2383} (or interplays of them) having collective degrees of freedom and presenting a complex few-body dynamics \cite{fossez16_1335,fossez16_1793,fossez18_2171,kravvaris17_1960,wang18_2144,fossez22_2540} (see also Sec.~\ref{secEmergence}). 
Similar comments can also be made for unbound excited states in well bound nuclei. 
As a result, it is not rare to find halo states \cite{jensen04_233,frederico12_372,tanihata13_549}, clustered states \cite{freer18_2138} often near decay thresholds \cite{ikeda68_2329,oertzen06_1017,freer07_1018,okolowicz13_241,okolowicz12_998}, or the decay of two or more nucleons \cite{guillemaud90_1761,blank08_449,pfutzner12_1169,spyrou12_1216,thoennessen13_1776} in exotic nuclei. 
The need to unify nuclear structure and reactions has been formally recognized during a previous FRIB Theory Alliance Topical Program \cite{johnson20_2389}.

From the structure side of the problem, the inclusion of continuum states within  many-body calculations has essentially been achieved in, notably, continuum shell model approaches \cite{rotter91_448,okolowicz03_21,volya06_94} and many-body methods generalized in the complex-energy plane using the Berggren basis \cite{michel09_2,rotureau06_15,hagen07_976} or the uniform complex-scaling technique \cite{carbonell14_1240}. 
However, describing the emergence of collective degrees of freedom in exotic nuclei within such many-body approaches is still a challenge (see more details in Sec.~\ref{SecEmergenceFewBody}), and extracting reaction observables requires additional treatment as explained hereafter. 

On the reaction side, several strategies have been employed to include structure information explicitly, for example by combining the resonating group method with structure methods \cite{Navratil_2016,quaglioni08_755,quaglioni09_756,jaganathen14_988,fossez15_1119,kravvaris17_1960}  to treat the relative motion of a target and a projectile directly within a many-body framework. 
Such approaches can thus provide reaction cross sections in a  unified picture of nuclear structure and reactions, but so far their \textit{ab initio} formulation limits them to light nuclei and at best includes only selected two- or three-body reaction channels. 
Other \textit{ab initio} approaches that can provide scattering cross sections in light nuclei include, for instance, the lattice-EFT (L-EFT) \cite{lee09_949} and the quantum Monte Carlo (QMC) \cite{nollett12_835,flores22} methods. 
Both techniques are formulated in radial space and thus offer an alternative to traditional cluster approaches based on nucleons as degrees of freedom \cite{ono92_2334,enyo01_2335,feldmeier00_2337} to study the impact of clustering in exotic nuclei.
In the near future, symmetry adapted \textit{ab initio} methods \cite{launey16_2403} will hopefully provide a way to extend \textit{ab initio} reactions to medium-mass nuclei while capturing nuclear shapes and collective modes naturally. For all these methods, additional developments will be needed to include more continuum couplings within the target and projectile, e.g. through the inclusion of many-body reaction channels.

Such first-principles methods can provide and have provided valuable insight about the emergence of few-body dynamics in nuclei, but few of these methods can be used in a practical way to give reliable spin and parity assignments in medium-mass nuclei due to their prohibitive computational cost.

\begin{figure}
    \centering
    \includegraphics[width=0.8\linewidth]{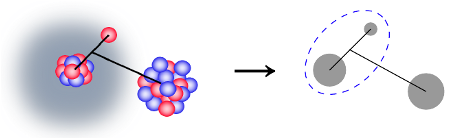}
    \vspace{0.2cm}
    \caption{Reduction of the many-body problem into a few-body one.}
    \label{Fig3}
\end{figure}

An alternative to solving the full $A$-body problem for structure and reactions consists in using effective degrees of freedom, as is illustrated in Fig.~\ref{Fig3} showing a schematic reduction of the many-body problem into a few-body one, composed of tightly bound clusters assumed structureless. 
This strategy has the obvious advantage of dramatically reducing the computational cost of calculations, and thus to allow for studies beyond light nuclei.  

In this few-body picture, the structures of both the projectile and the target are described using effective potentials reproducing their low-energy single-particle spectra, and the cluster-cluster interactions are simulated using optical potentials~\cite{feshbach1958unified,TN09,BC12}. 
Additionally, only pairwise cluster interactions are usually considered and other irreducible interactions beyond this level are neglected, even though there is some evidence from deuteron-induced reactions that this approximation leads to sizeable uncertainties~\cite{Hlophe:2022fdn,PhysRevC.74.044304,PhysRevC.104.034614,PhysRevC.104.024612}. 
For that reason, efforts are underway to quantify the effect of irreducible three-body cluster-cluster interactions (see Sec. \ref{SecUniversalCorr}).
In general, additional investigations on the reduction of the many-body problem into a few-body one should be supported, as they will provide paths  to efficiently inform few-body reaction models by many-body predictions.

For most reactions, the few-body problem is solved using approximate schemes such as  continuum-discretized coupled-channel methods~\cite{AUSTERN1987125,Yetal86,Kawai86,PhysRevC.68.064609}, the eikonal approximation~\cite{G59,Hansen03,PhysRevLett.95.082502}, the distorted wave Born approximation, or the adiabatic distorted wave approximation~\cite{JOHNSON197456,TIMOFEYUK2020103738}. 
These models are usually applied to the three-body problem, but some have recently been extended up to the five-body problem (see for example Ref.~\cite{PhysRevC.101.064611}). 
Moreover, due to the high cost of treating the few-body problem in the continuum, most of these models rely on additional assumptions such as neglecting excitations of the clusters during the reaction, considering only one-step processes, or they neglect excitations of the projectile and the target during the reaction. 
These approximations limit the accuracy of the information inferred from experiments (see for example Refs~\cite{PhysRevC.78.054602,CooperPairPotel,PhysRevLett.109.232502}) and it is thus crucial to develop more complex reaction models treating dynamical excitations explicitly (see more detailed discussion in Secs.~\ref{SecReactionTheory} and~\ref{SecQuenching}).

Besides the treatment of the reaction dynamics, another important element is the quality of the cluster-cluster effective potentials. 
For bound states, these interactions are typically adjusted to reproduce properties of the low-energy spectra of the relevant nuclei, i.e., the energy of the states, their spins, parities, partial widths, and electromagnetic transition probabilities, as well as information about the overlap functions, their asymptotic normalization constants (ANCs), root mean square (rms) radii, and spectroscopic factors (SFs) (see Fig.~\ref{fig_pipeline}).
The optical potentials between the projectile's and target's clusters have been historically fitted on elastic-scattering data on the same system~\cite{koning03,becchetti69}. 
However, most of the scattering data used to constrain these phenomenological interactions were from stable targets, and therefore their quality for unstable systems is uncertain. 
With the start of FRIB, it is thus urgent to develop optical potentials that can reliably be extrapolated away from stability to the driplines.

As a consequence, recent efforts have been undertaken to enhance our predictive power for reactions involving exotic nuclei by strengthening the connection between many- and few-body approaches. 
One promising avenue rests on the EFT framework, which provides a rigorous way to construct theories. 
By definition, EFTs are as good as the underlying theory within a limit range defined by the effective separation of scales exploited, and they can be improved systematically with quantified systematic errors at each order. 
Examples of successful EFTs related to few-body physics are the so-called halo EFTs~\cite{bertulani02_869,ryberg14_997,ji14_1101,Hammer2017,Capel:2018kss} (see Sec.~\ref{SecUnderstandingFewBody}) and the EFT formulation of the particle-plus rotor model \cite{papenbrock11_1310,papenbrock14_1312,papenbrock15_1311}. 
What makes EFTs interesting in the reduction of the many-body problem is that when they are applied to the few-body problem, their low-energy constants can be fixed using results from experiments or many-body approaches. 
While this limits their predictive power, in principle, it allows for the rigorous construction of cluster-cluster interactions for few-body reaction models. 
For that reason, we propose to prioritize two research programs to address this problem: i) to reformulate phenomenological models into EFTs to ensure a proper connection with the underlying theory, and ii) to match all the EFTs with each other when appropriate, and with the underlying theory.
Such programs will bring few-body approaches to a new level of predictive power and ensure consistency across the description of all few-body effects in exotic nuclei.

\begin{figure}
\centering
\includegraphics[width=0.9\textwidth]{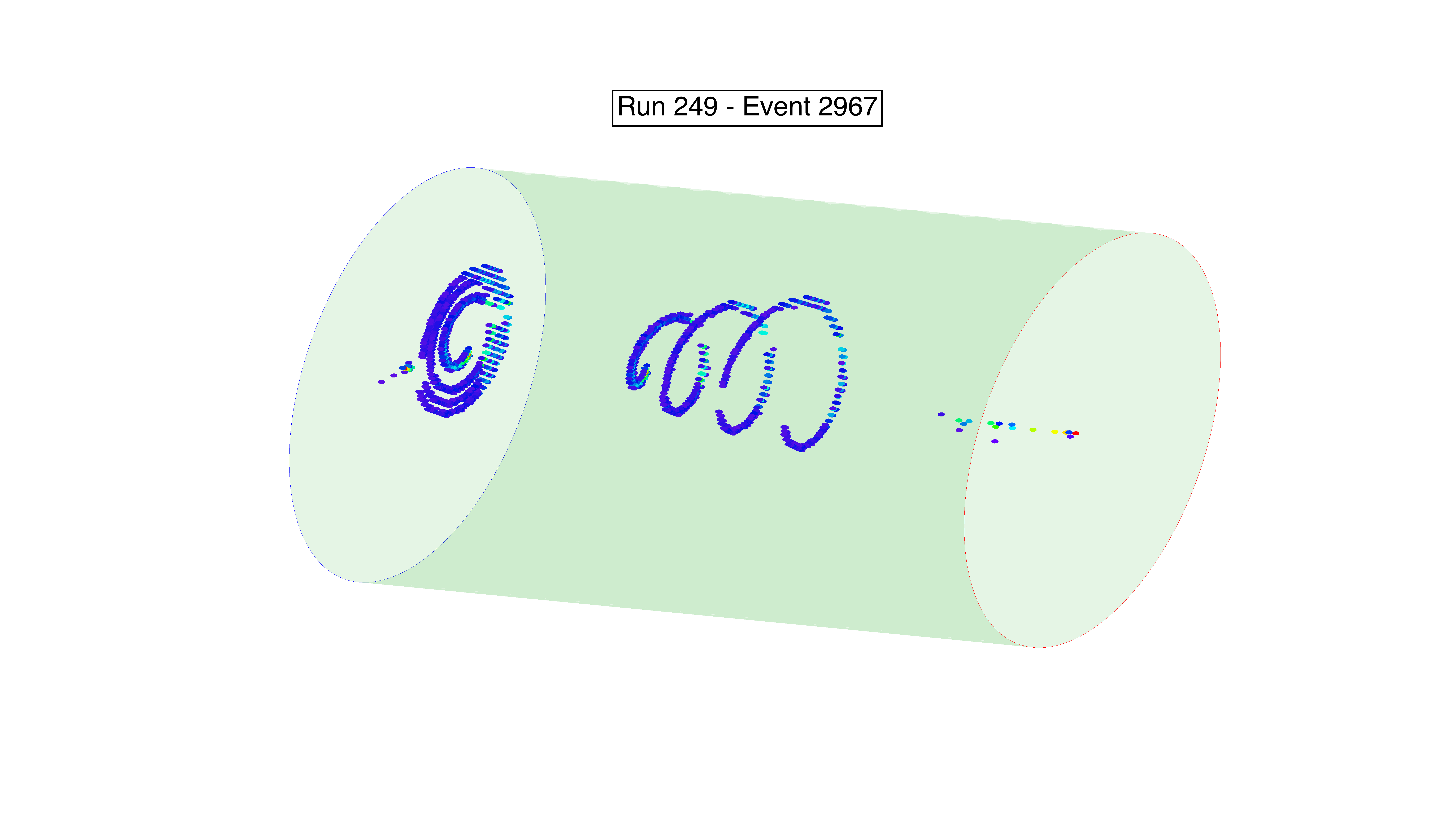}
\caption{Example of a proton track recorded in the AT-TPC from a $(d,p)$ reaction between a $^{10}$Be beam and a deuteron from the deuterium gas filling the active volume of the detector.}
\label{fig_attpc}
\end{figure}

In situations where EFTs cannot \textit{a priori} be developed, i.e., when they are no clear separation of scales that can be exploited, cluster-cluster interactions for few-body reaction models can be built from many-body theory using Watson multiple scattering theory, the Feshbach projection operator formalism, Green's function theory, or by starting from nuclear matter predictions (see recent white paper~\cite{WhitePaperOptPot22} and references therein). 
Optical potentials obtained from many-body approaches have the advantage of carrying the correct isospin dependence and are therefore more likely to extrapolate properly in the regions of greatest interest for FRIB away from stability. 
In addition, when an optical potential obeys dispersion relations, bound and scattering states are described on the same footing, providing that a consistent description of the reaction (see discussion in Sec.~\ref{SecQuenching}). 
Nevertheless, a common shortcoming of these microscopic potentials is their lack of absorption strength, which can be improved by informing them  elastic and inelastic experimental data from, for example, transfer, charge-exchange and breakup reactions.  
This experimental feedback will become particularly important for experiments at FRIB using, for instance, the previously mentioned AT-TPC detector, in which plenty of high-quality data for different reaction channels and at different energies will be obtained (Fig.~\ref{fig_attpc} shows an example of trajectory in the AT-TPC chamber).

To conclude, in light of the experimental needs at FRIB and the state of theoretical developments, the success of the FRIB scientific mission will require a concerted effort between theory and experiment, and in particular important developments to pursue the unification of nuclear structure and reactions in exotic nuclei to provide reliable predictions during the exploration of the drip lines. 
We believe that close collaborations between experimentalists and theorists will benefit the planning of new experiments and maximize our discovery power of new exotic phenomena in nuclei.

\clearpage

\section{
Emergence of few-body effects in exotic nuclei}
\label{secEmergence}
\textbf{K. Kravvaris, F. Bonaiti, A. Gnech, K. Fossez, and K. Becker} \label{SecEmergenceFewBody}
\smallbreak

\textit{How do few-body effects emerge in atomic nuclei?} This remains one of the most intriguing open questions of low-energy nuclear physics \cite{LRP1,NAS}. Collective phenomena appear when a transition from (many) single-particle degrees of freedom to (few) collective ones occurs, leading to the emergence of few-body dynamics, which can have a direct impact on nuclear observables and astrophysical processes. It is thus critical to develop a proper theoretical understanding of collective phenomena and their associated few-body effects.
However, a comprehensive description of collectivity in nuclei is challenging because it appears at various nearby energy scales comparable to single-particle energies, and leads to a rich variety of phenomena such as deformation, clustering, rotational and vibrational motions, or pairing dynamics. 
In exotic nuclei near the drip lines, which the Facility for Rare Isotope Beams (FRIB) will be exploring in the coming decades, couplings to the continuum of scattering states and decay channels \cite{matsuo10_1238,michel10_4,johnson20_2389} can dramatically impact collective phenomena, and lead to new ones such as halo structures, the superradiance effect~\cite{volya03_858}, or exotic decay modes like two-neutron or two-proton decay~\cite{guillemaud90_1761,blank08_449,pfutzner12_1169,spyrou12_1216,thoennessen13_1776}.
Moreover, the delicate interplay between various collective phenomena can fundamentally shape the properties of a system, as seen, for example, in the case of a halo structure above a deformed core, or when a rotational band extends into the continuum forming resonant rotational states \cite{garrido13_1171,fossez16_1335}. 
In this perspective piece, we will mention some recent works related to the emergence of few-body effects in exotic nuclei to illustrate more general ideas which, we will argue, should be investigated to ensure the success of the FRIB scientific program. Essentially, these ideas fall into two main topics which are i) the proper theoretical description of collectivity in exotic nuclei and the associated few-body dynamics, and ii) the connection between nuclear forces and emergent properties in exotic nuclei.

\subsection*{Deformation} 

Nuclear deformation is one of the most common emergent phenomena in atomic nuclei. It has been shown to naturally arise from the underlying strong nuclear force in large-scale ab initio calculations~\cite{DytrychLDRWRBB20}, and yet its fundamental origin and link to the elementary particle dynamics is still not fully understood. While enhanced deformation is evident in both heavy nuclei and those away from closed shells, as recognized early on~\cite{Mottelson_NP,Elliott58}, deformed configurations are present even in a nucleus such as the doubly closed-shell  $^{16}$O, which is commonly treated as spherical in its ground state, and whose first excited state is the head of a strongly deformed rotational band. This shape coexistence of states of widely differing deformation in many nuclei is now well established~\cite{HeydeW11} and it has been argued that it occurs in nearly all nuclei~\cite{Wood16}.

Moreover, while vibrational spectra are often associated with spherical nuclei, the fact that a nucleus has a zero quadrupole moment in its $0^+$ ground state does not imply that it is spherical in its intrinsic frame; for quantum mechanical reasons, it appears spherical in the lab frame. With an expanding body of experimental evidence, it is becoming clear that non-zero deformation is far more widespread than zero deformation~\cite{RoweW2010book}. 
Deformation is also found in the so-called ``islands of inversion", the regions of the nuclear chart where shell evolution brings two major shell blocks, such as $sd$ and $fp$ shells, together (for further discussion, see, e.g., \cite{physics4030048}). The preference of the nuclear force for spatially symmetric and low-spin configurations, which are energetically favored, leads to an enhancement of quadrupole couplings and a spontaneous symmetry breaking in the intrinsic frame, as observed in Elliott's model (similarly in the Jahn-Teller effect). 
The emergence of deformation and the subsequent dominance of collective rotational degrees of freedom in the nuclear spectrum reflects the presence of an effective separation of scales, which can be exploited to simplify the description of deformed nuclei. 

Interestingly, unambiguously claiming the presence of deformation in an atomic nucleus based on energy spectra can be challenging. For instance, phenomena like shape coexistence can affect energy levels, and disentangling vibrational and deformation effects can be difficult.

On the theory side, one of the best ways to detect deformation is to study the neutron and proton quadrupole (or higher) moments. 
When such quantities can be extracted from, for example, \textit{ab initio} calculations, results can then be matched with simpler collective models to extract additional information. 
Such matching can be made rigorous using the effective field theory framework \cite{weinberg1990,ordonez1992,HammerReview2020}, as was done, for instance, in Ref.~\cite{papenbrock11_1310,papenbrock2016} for the particle-plus-rotor model. 
However, in exotic nuclei, and in particular in unbound states, measuring quadrupole moments might not be an option, and one has to rely on the observation of rotational bands extending in the continuum. The study of how continuum couplings affect deformation \cite{luo21_2394,fossez22_2540} and rotational motion \cite{fossez16_1335,fossez16_1793} will be critical for the interpretation of experimental data at FRIB. 

Apart from quadrupole moments, but directly related to them, certain electromagnetic transitions--such as the electric quadrupole (E2)--between states at low excitation energy are also typically considered signatures of collective behaviour~\cite{eisenberg1976}. 
However, such calculations remain challenging in various first-principle approaches scaling polynomially with model space sizes, as can be seen in, for instance, valence-space in-medium similarity renormalization group~\cite{stroberg2017} and coupled-cluster theory~\cite{hagen2014} calculations, which systematically underestimate experimental data~\cite{stroberg2022}. 
One of the culprits behind this discrepancy is the difficulty to capture both "static" and "dynamical" correlations \cite{cordoba2016} and hence, collectivity. Dynamical correlations involve the mixing of a dominant configuration with small (but energetically significant) contributions coming from other configurations. This is what \textit{ab initio} many-body methods as the ones mentioned above are designed to do. In this framework, in fact, one starts from a reference state (usually a Hartree-Fock Slater determinant), and includes dynamical correlations adding particle-hole excitations on top of the reference. Static correlations, instead, are harder to capture: they emerge when there are two or more equally relevant configurations, as, for instance, the degenerate states produced by rotating a symmetry-breaking reference.   
Some progress in this direction has already been made. As an example, static correlations can be included in the reference state, as was done in the in-medium generator coordinate method~\cite{yao2018}.  Restoration of the broken rotational symmetry, e.g. via angular momentum projection is another strategy that can be pursued to capture such correlations. For instance, recent coupled-cluster computations, starting from axially-symmetric reference states and including angular momentum projection, reproduced the ground-state rotational band of $^{34}$Mg~\cite{hagen2022}. This result is particularly relevant considering the ongoing experimental investigations of magnesium isotopes, e.g. $^{40}$Mg, at FRIB.


\begin{figure}[t]
    \centering
    \includegraphics[width=0.8\textwidth]{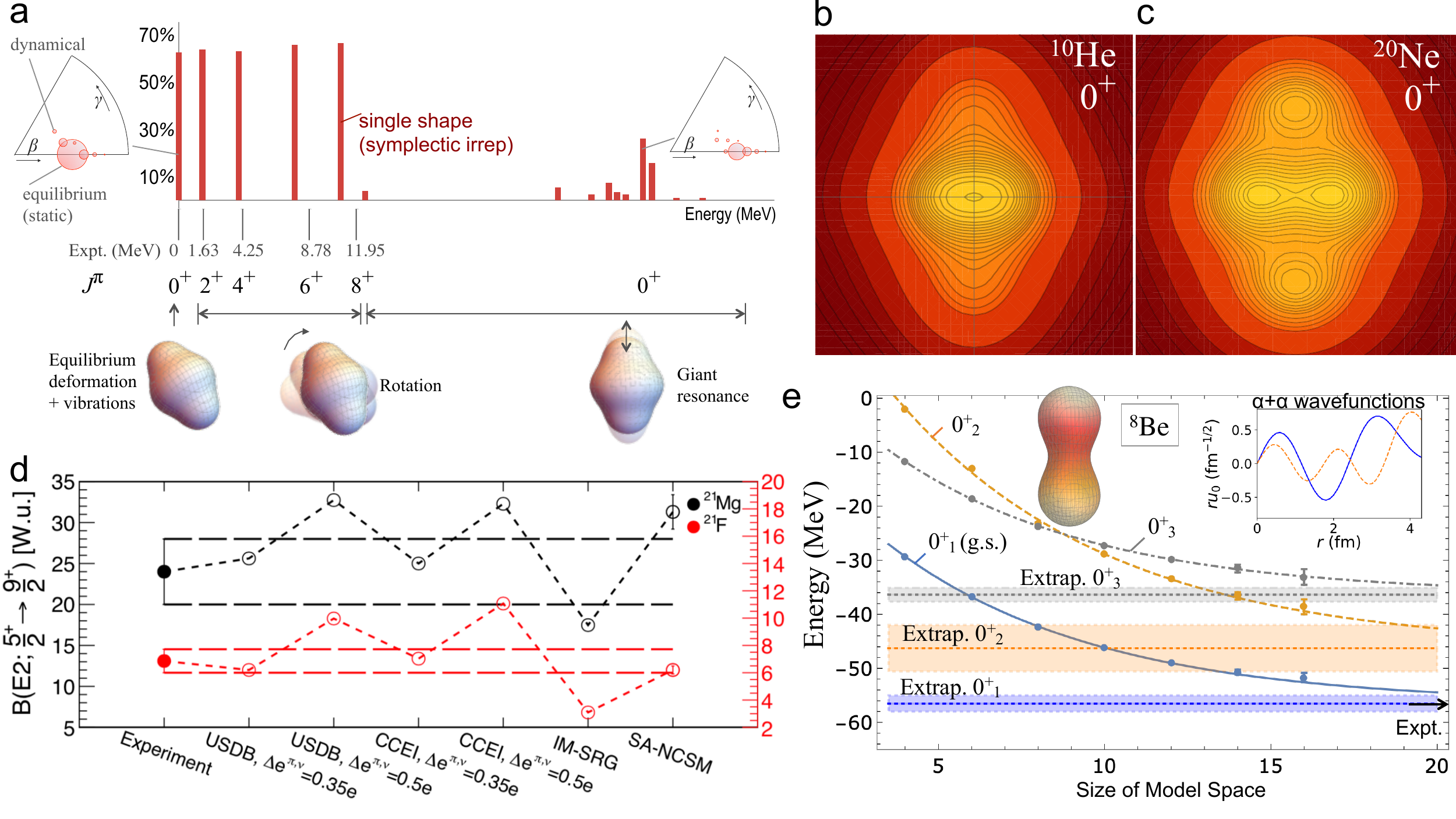}
    \caption{\textbf{a.} Emergence of almost perfect symplectic symmetry in nuclei from first principles \cite{launey16_2403,LauneyMD_ARNPS21} (figure adapted from Ref.~\cite{DytrychLDRWRBB20}), enabling  \textit{ab initio} descriptions of clustering, collectivity, and continuum  in \textbf{(b)} exotic nuclei as  $^{10}$He and \textbf{(c)} intermediate-mass nuclei as $^{20}$Ne.  \textbf{d.} Experimental and theoretical $B(E2; 5/2^+ \to 9/2^+)$ values for $^{21}$Mg and $^{21}$F;  the {\it ab initio}  SA-NCSM calculation is shown without the use of effective charges (figure from Ref. \cite{Ruotsalainen19}). \textbf{e.}  Energy and cluster structure of the lowest-lying $0^+$ states in the challenging $^8$Be (figure adapted from Ref. \cite{PhysRevLett.128.202503}). Fig.~(a), (d), and (e) are reproduced with permission.}
    \label{fig:sancsm}
\end{figure}

To capture both static and dynamical correlations in medium-mass \textit{ab initio} calculations, it is advantageous to use a physically motivated, correlated basis to solve the many-body problem. This is the case in the symmetry-adapted no-core shell model (SA-NCSM) (reviewed in \cite{launey16_2403,LauneyMD_ARNPS21}), which exploits symmetries of the nuclear dynamics (Fig. \ref{fig:sancsm}a). Specifically, \textit{ab initio} descriptions of spherical and deformed nuclei up through the calcium region are now possible in the SA-NCSM without the use of interaction renormalization procedures and effective charges. 
Using the SA basis, one can reach higher shells necessary for the accurate description of clustering, collective, and continuum degrees of freedom (dof's), and to reach heavier nuclei such as $^{20}$Ne, $^{21,22}$Mg, $^{28}$Mg,  as well as $^{32}$Ne and $^{48}$Ti \cite{LauneyMD_ARNPS21} (e.g., Fig. \ref{fig:sancsm}b \& c). These studies provide collective observables, such as E2 transition strengths and quadrupole moments, as well as $\alpha$-decay partial widths and ANCs \cite{DreyfussLESBDD20,PhysRevLett.128.202503} (Fig. \ref{fig:sancsm}d \& e).
The SA eigenfunctions are represented in terms of all possible deformed microscopic configurations, including zero deformation and single-particle dof's, and directly reveal the deformation content of a nucleus (Fig. \ref{fig:sancsm}). This, in turn, provides a straightforward way to detect deformation. Indeed, any microscopic model through projections onto the SA basis can identify deformation, even in $0^+$ states.
The critical importance of dynamical correlations is recognized when the SU(3) SA basis \cite{DraayerLPL89,DraayerW83} is further reorganized according to the symplectic $\text{Sp}(3,\mathbf{R})$ group~\cite{RosensteelR77}, and highly regular patterns in wave functions become strikingly clear. In particular, a single nuclear shape (or only a few shapes) is found to dominate low-lying nuclear states (Fig. \ref{fig:sancsm}) \cite{DytrychLDRWRBB20}. Collectivity is almost completely driven by these shapes, and typically only a few hundred shapes or less are needed for convergence of various short- and long-range observables.
Remarkably, it is the dynamical deformations that, when properly developed, drastically decrease the energy of a nuclear shape (the 4p-4h enhanced static deformation of the Hoyle state lies at $\sim 30$ MeV  excitation energy or higher, whereas the vibrations lower it to $\sim 7$ MeV); in addition, as shown in Ref.~\cite{Rowe_book16}, without vibrations the quadrupole moment is underestimated roughly by a factor of two. This factor is often accommodated by various models by an effective charge or starting from an effective static shape.

In short, state-of-the-art many-body methods can provide insight in how deformation emerges in exotic nuclei and how the associated few-body dynamics affect their properties. They can also provide inputs for effective approaches with collective degrees of freedom. However,  difficulties remain in either capturing deformation from first principles, or in dealing with couplings to continuum states near particle emission thresholds.
In this regard, it would be worth exploring new strategies to include static correlations in \textit{ab initio} calculations of deformation-relevant observables, as quadrupole moments and B(E2) transitions, with an eye to the reduction of their computational cost. This effort should be accompanied by robust estimates of theoretical uncertainties, which play a key role in the comparison to experimental data.

\subsection*{Clustering} 


Parallel to collectivity emerging in rotovibrational nuclei is the phenomenon of nuclear clustering. Traditionally, clustering refers to the formation of distinct substructures within atomic nuclei; for example, states in $^{20}$Ne having a $^{16}$O+$^{4}$He character. Naturally, such states are expected to form in the proximity of the respective decay threshold~\cite{ikeda68_2329,oertzen06_1017,freer07_1018,okolowicz13_241,okolowicz12_998}, however they are also observed at higher energies~\cite{johnson09_1479}. An even more extreme example of clustering is seen in halo nuclei, where the clustered state lies slightly below the decay threshold, leading to spatially extended nucleus that can be well described in a core plus particle(s) picture. 

From a practical point of view, clustered states can significantly impact astrophysical reaction rates since clustering typically translates to relatively large decay widths~\cite{Kubono1995,Descouvemont2008,Shen2020,DreyfussLESBDD20}. This is perhaps most famously exemplified in the case of the Hoyle state, which has a pronounced 3$\alpha$ cluster arrangement allowing for the triple-$\alpha$ reaction to proceed in solar environments, bypassing the mass-eight gap. Indeed, the population of nuclear states in, for example, transfer reactions--one of the main tools to be employed at FRIB for the study of exotic nuclei--is expected to be directly proportional to the degree of clustering of a state (typically quantified in terms of spectroscopic factors)~\cite{endt77_2579}. 

Thus, a concrete understanding of how clustering (and other types of collectivity) emerge from single particle degrees of freedom and the nuclear interaction, and how the resulting interplay of the bound and continuum parts of the nuclear wave function affects observables is a key target of nuclear theory in the coming years. However, building such an understanding requires both advances in few- and many-body nuclear theory as well as a significant amount of new experimental data, especially along isotopic chains and near the driplines where cluster phenomena may dominate the spectrum. 

On the theory side, first steps have already been made for the microscopic description of clustered states in light nuclei, both in phenomenological and \textit{ab initio} approaches, for example Refs~\cite{wiringa13_1182,kanadaenyo12_2568,elhatisari15_2324,elhatisari16_2311,kravvaris17_1960,PhysRevLett.128.202503} to name a few, and in intermediate-mass nuclei \cite{LauneyMD_ARNPS21}. Despite this progress, a lot remains to be done. 

In particular, the competition between various forms of clusterization remains a pressing issue, as this type of shape coexistence can serve to quench or amplify clustering effects~\cite{Kravvaris2017}. Furthermore, while first \textit{ab initio} studies of how clustering evolves along an isotopic change have been performed~\cite{Elhatisari2017}, the pertinent question of how composite clusters manifest in the nuclear medium remains. Moreover, as origins of $\alpha$-clustering have been investigated in mean-field approaches~\cite{Ebran2012}, it is relevant to draw parallels to the fundamental nuclear interactions and map out the connection. Last, but definitely not least, the observed signatures of $\alpha$-clustering in heavier nuclei~\cite{vonTresckow2021} suggest that there may still exist new mechanisms, that are significantly different than the ones active in light systems and cannot yet be accessed in \textit{ab initio} approaches, marking the need for further investigations.


As a particular example where collectivity manifests itself one can focus on dipole-excited states. In such collective states the analysis of the so-called “cluster sum rule”~\cite{alhassid1982,hencken2004} can give some insight on how clustering phenomena emerge from first principles. 
This sum rule becomes particularly interesting when looking at halo nuclei, which are usually modeled as a tightly-bound core, surrounded by one or more weakly-bound nucleons. 
The ratio of this cluster sum rule with respect to the full energy-weighted dipole sum rule of the nucleus provides an estimate of how much of the dipole strength is given by the relative motion between the core and the halo. 
Here it is worth pointing out that taking ratios of observables instead of their absolute value allows for the suppression of systematic errors~\cite{PhysRevLett.128.202503,caprio2022, capel2011}, and thus leads to more robust predictions. 
Such an analysis has been done in a recent work on the halo nucleus $^8$He~\cite{bonaiti2022}. 
There, the discretized dipole response of $^8$He was computed in the framework of coupled-cluster theory, merged with the Lorentz Integral Transform (LIT-CC)~\cite{bacca2013,bacca2014}. 
The evaluation of the cluster sum rule in this case pointed out that around 1/3 of the total dipole strength in $^8$He is determined by the relative motion between the $^4$He core and the four-neutron halo. 

\subsection*{Halo structures}

The definition of a halo state, usually accepted as a state in which one or few nucleons have a 50\% probability of being found in the classically forbidden region of the potential generated by the rest of the system, is not without ambiguity \cite{jensen04_233,canham08_2450,frederico12_372,tanihata13_549}. 
First, the classically forbidden region is not so clear for more than one nucleon in the halo, and second, the rest of the system can be itself made of more than one cluster. 
It could be argued that the latter situation is found in the first excited state of $^{11}$Be \cite{fukuda91_1605}, which can be seen as one neutron halo around $^{10}\text{Be} = 2\alpha+2n$. 
Another challenge to the accepted definition of a halo state is the presence of deformation in the core, as mentioned previously. 
While it can be shown that the deformation of the core becomes irrelevant in the limit of a vanishing binding energy between the core and halo nucleon \cite{misu97_1181}, 
in practice many weakly bound nuclei have binding energy sufficiently large for couplings between the deformed core and the halo nucleon to play a role as, for example, in $^{17}$C \cite{suzuki08_1557} or $^{31}$Ne \cite{nakamura09_1513,takechi12_1512}. 
When deformation matters, the deformed Nilsson orbits in which halo nucleons live can be represented as linear combinations of spherical shells, and can include a mix of $\ell=0$ or $\ell = 1$ shells with $\ell > 1$ shells. 
The question then becomes whether or not we still have a halo system if $\ell>1$ shells have a significant occupation and do not extend far into the classically forbidden region. 
In a naive classical picture halo nucleons would be on an elliptical orbit passing close to the core and extending far from it. 
From a theoretical point of view, clarifying the definition of a halo structure is critical as FRIB will allow pursuing the exploration of drip lines into the medium-mass region and is likely to uncover many deformed halo states, potentially involving several neutrons in the halo. 
Clustering is usually expected to play a smaller role in medium-mass nuclei based on the dominance of the mean-field, but this is not necessarily true in excited states and especially if states are found just below decay thresholds as in $^{11}$Be. 
To summarize, testing the boundaries of what is considered a halo state should be an experimental priority to stimulate progress in our understanding of halo physics. 
On the theory side, priority should be given to the development of more comprehensive models of halo structures through the development of effective fields theories going beyond existing halo EFTs \cite{bertulani02_869,ryberg14_997,ji14_1101,Hammer2017}, and of many-body method capable of describing such physics from nucleons as degrees of freedom more efficiently than current approaches.

\subsection*{Making the connection with nuclear forces} 

Besides the challenge of understanding how few-body effects emerge from the many-body dynamics, 
there is the question of which specific components of nuclear forces drive collective phenomena. 
For example, sensitivity studies on the low-energy constants (LECs) of a two-body chiral EFT have been carried out for the quadrupole moments of ${}^6$Li and ${}^{12}$C using the SA-NCSM~\cite{BeckerLED22,Becker23}. This analysis demonstrated, on the one hand, that the quadrupole moments are predominantly sensitive to only two of the LECs parameterizing chiral interactions at NNLO. On the other hand, simultaneous variation of all 14 LECs at this order has little effect on the dominant nuclear shape(s) and the associated symplectic symmetry. It remains to be seen if the extension to other observables, such as quadrupole and dipole transitions, will lead to the same outcome. Some recent  studies on ${}^6$Li based on NCSMC~\cite{PhysRevLett.129.042503} and HH~\cite{PhysRevC.102.014001}  showed the strong dependence of clustering observables to the tensor terms of the two and three-body nuclear interactions.
In Ref.~\cite{PhysRevC.102.014001} it has been shown that the strength of $^6$Li quadrupole moment is directly connected to the strength of the tensor forces in the NN nuclear interaction. A similar effect is expected to be determined also by three-nucleon interactions. Indeed in Ref.~\cite{PhysRevLett.129.042503} the comparison of the ratio of the ANCs obtained with the NN interaction only and NN+3b interaction shows an increase of the strength of the $D$-wave component in ${}^6$Li which suggests that three-nucleon interactions play an important role too in determining cluster and deformation observables. Clearly, these studies are limited only to $^6$Li and should be extended to more systems and chiral interactions to define a general trend. 
Therefore, in the future, it would be interesting to extend the LECs sensitivity studies of the SA-NCSM, which have the possibility to deal with larger and more deformed systems, also to individual terms of the two and three-body nuclear interaction, \textit{i.e.} single diagrams of the chiral expansion, in particular the ones that give rise to tensor forces.  In addition, this type of investigation should be performed on other systems with different properties to better understand how individual interaction terms give rise to collective behavior across the Segre chart.  We note that the problem of the prohibitive computational cost of such calculations can be contained thanks to the reduced basis method known as eigenvector continuation (see for example the symmetry-adapted eigenvector continuation (SA-EVC) case \cite{BeckerLED22}), which provides the basis to build efficient and precise emulators of the many-body method~\cite{Frame:2017fah,Konig:2019adq,Ekstrom:2019lss}.

Another way to study the impact of specific components of nuclear forces on emergent properties in exotic nuclei is to disentangle the role of two- and three-body forces. 
It appears that two-body forces alone seem to be able to generate most emergent phenomena observed in exotic nuclei, 
which is a consequence of the fact that many of these systems live close to the unitary limit and manifest signs of universal behavior (see Sec.~\ref{SecEfimov}). 
However, as discussed earlier, in exotic nuclei where an interplay between continuum couplings and standard phenomena such as deformation and clustering is present, 
the situation becomes more complex (for instance in so-called deformed halo states where one or more nucleons form a halo around a deformed core whose rotational motion can have an impact on the dynamics of the halo, as is the case in $^{40}$Mg \cite{crawford14_1607,crawford19_2359,refId0}). 
Indeed, \textit{ab initio} calculations suggest the need of three-body forces in specific two-body parameterizations to obtain an accurate description of nuclear structure in general, but even more so in nuclei where clustering is present such as $^{10}$B \cite{navratil07_733,mccutchan12_2321,freer18_2138}, where it was found that using ${ (n,\alpha) }$ scattering data to determine three-body forces improved the ordering of the low-lying states \cite{carlson15_1610}. 
Unfortunately the role of three-body forces in collective phenomena has not been fully investigated yet, but this knowledge will be necessary to understand many exotic nuclei at the drip lines. 

Finally, we can also turn the tables and use emergent phenomena in exotic nuclei to constrain nuclear forces. 
For example, widths of resonances depend exponentially on the associated decay thresholds, making them difficult to describe and fairly sensitive to details of the interaction. 
This sensitivity can be exacerbated when considering multi-nucleon resonances, like the famous two-neutron decay of $^{26}$O \cite{guillemaud90_1761,lunderberg12_556,kohley13_1541,caesar13_1765,kondo16_1439}, which only happens because the energy pattern of the ground states of $^{24}$O and $^{25}$O, confined within a few keV, allows it. 
Even more exotic patterns can be observed in neutron-rich helium isotopes where the presence of a large number of neutrons in $^{9}$He \cite{kalanee13_1909} leads to a situation where it is still unclear whether or not the lowest state observed, with $J^\pi = 1/2^+$ and strongly impacted by $s_{1/2}$ partial waves, is a resonance or a virtual state and can thus be considered as the ground state \cite{vorabbi18_1977,fossez18_2171,votaw20_2353}. 
While testing nuclear interactions with precision on such phenomena remains challenging, they can shed some light on which parts of the nuclear forces we have not yet fully under control and give us some hints on how to constrain them. Therefore, when it is possible, we should post-select interactions reproducing data on exotic phenomena.

\clearpage

\section{
Understanding few-body dynamics 
in knockout and transfer reactions}
\textbf{T. Frederico, B. Kay, 
Ch. Elster, L. Hlophe, M. Higgins, and S. K\"onig}\label{SecUnderstandingFewBody}
\smallbreak

\subsection*{Overview and motivation}

Low-energy universal dynamics are present in a nuclear system when the degrees of freedom for the driving physics are configurations characterized by length scales larger than the range of the forces involved in the process~\cite{jensen04_233,frederico12_372}.  In EFT this is associated with long wavelength scales in weakly bound systems and a short wavelength scale that allows a systematic expansion of the scattering amplitude, including in the LECs, for example, effects from the effective ranges, besides the scattering lengths or low-lying resonances taken into account directly at leading order (see e.g.~\cite{Hammer2017,HammerReview2020}).
A manifestation of this universality are so-called ``halo nuclei'' states, which can be thought of one or more valence nucleons loosely bound to a core with lowest excitation energy much smaller than the  separation energy of the core valence nucleons.
The low-energy continuum of  such few-body halo states may exhibit the properties that are model independent and closely related to the bound halo formation, namely excitation energies of few hundreds of keV.  These properties can be probed in $(d,p)$ and $(t,p)$ reactions, and other processes, where the projectile gets one or two neutrons and eventually breaks.  In this section we discuss possibilities to study universality from that angle, suggesting new experiments and perspectives for accompanying theoretical efforts.

The study of $(d,p)$ and $(t,p)$ reactions, and also of nucleon or nucleons removal processes, will allow to learn about the evolution of universality in the formation of neutron halos along isotopic chains and their final state interaction (FSI) as the continuum counterpart to the bound halo state. The outcome of such investigation will be the determination of the minimal conditions from the nuclear force that allows to form the neutron halos and their continuum counterparts. Possibly, much more relying on the on-shell properties of the  neutron-core interaction, which may be determined as essential for the halo formation along the isotopic chain and  for the continuum properties at low excitation energies, as long as universality and model independence holds. We will be able to  provide the conditions for the stability of neutron-halos in the presence of a core, and also  the continuum features, as for example, how the tetraneutron properties manifests in the presence of a core and its relation to the formation of neutron rich nuclei.  At the same time, the involved few-body reaction mechanism will challenge our methods due to the explicit presence of the halo neutrons, which will require further theory developments.

\subsection*{Suggested experimental studies}

Various known halo nuclei are interesting candidates to study final-state universality, like for example $^6$He$(t,p)$$^8$He$^*$, with $^8$He$^* \, = \,\alpha$+4$n$, or other situations as $^9$Li$(t,p)$$^{11}$Li$^*$,
$^{11}$Li$^* =\,^9$Li+2$n$, where we are interested in the physics revealed by the fragments FSI, which comes from the transfer to the continuum. These final states  with low-excitation energies can show up a diverse range of universal phenomena, which we discuss further using some typical examples: (i) $^{11}$Li$^*$, (ii) $^{19}$B$^*$ and (iii) $^8$He$^*$ (the $^*$ indicates excitation to the continuum). 

In the case of $^{11}$Li$^*\,=\,^9$Li+$n$+$n$, the dominant interactions will be on the s-wave. Indeed for $^{10}$Li$^*\,=\,^9$Li+$n$ already the virtual $s_{1/2}$ state has been studied~\cite{BarrancoPRC2020} in a $^{9}$Li$(d,p)$  reaction, where the scattering length of $-8\,$fm was found. The  $^9$Li+$n$+$n$ characterizes a three-body system in the continuum, which is dominated at hundreds of keV  by the s-wave physics, considering in addition to the $^{10}$Li s-wave virtual state the dominant $^1$S$_0$ neutron-neutron channel. It is expected that the final state interaction within the $^9$Li+$n$+$n$ system  is dominated by the Efimov physics~\cite{Efimov:1990rzd}, when the interaction is resonant in the s-wave, namely the scattering lengths have large magnitudes with respect to the interaction range. Although there is not yet a two-neutron halo system, Borromean or Samba ($^{20}$C), which can have an Efimov excited state~\cite{AmorimPRC1997,frederico12_372,Hammer:2022lhx}, still would be possible that Borromean systems  present an Efimov-like continuum  resonance  for large and negative scattering lengths~\cite{Bringas:2004zz,Deltuva:2020sdd,Dietz:2021haj}, which on the theory side has to be extended to mass-imbalanced systems. It could well be that these continuum resonances, associated with Borromean states, would be broad and hard to be seen, although the few-body universal Efimov dynamics will drive the FSI in  core+$n$+$n$ systems with the quantum numbers of the ground state.  We return to discuss in Sect.~\ref{SecEfimov}  the possibility to form an excited Efimov state by removing nucleons from the core, while keeping the neutron halo intact during some time.

In particular, we should mention the $^{17}$B+$n$+$n$ system, which has a shallow Borromean bound state of $^{19}$B~\cite{HiyamaPRC2019} and a large n$-\,^{17}$B scattering length.
The transfer to the continuum in the reaction $^{17}$B$(t,p)$$^{19}$B$^*$ close to the threshold could eventually populate a nearby Efimov continuum resonance. In this case it will be desirable to have measurements for $^{17}$B$(d,p)$$^{18}$B$^*$ to  achieve a better accuracy in the scattering length, as has been done for $^{9}$Li$(d,p)$~\cite{BarrancoPRC2020}. Increasing the number of neutrons in the final state, for example with the transfer reaction $^{19}$B$(t,p)$$^{17}$B+4$n$, at low relative energies,  the FSI in $^{17}$B+4$n$ system, should be determined by $S_{2n}$ in $^{19}$B, the $nn$ and $n-^{17}$B scattering lengths, as suggested by the dominance of the universal Efimov physics.  Curiously enough, when the nn system is at unitarity the neutron cloud turns to be a system within  ``unnuclear physics''~\cite{Hammer:2022lhx}.  Therefore,  it seems quite interesting the situation where the nuclear core is in the presence of a neutron cloud with 2$n$, 3$n$, 4$n$,... dominated by ``unnuclear'' FSI effects at low relative energies, which correspond to small excitation energies in the continuum of the core+$Nn$ $(N\geq 4)$, going beyond the drip line. Indeed the much debated tetraneutron, which is now understood as a strong correlation between the free neutrons~\cite{DuerNat2022} could reveal itself in the presence of the core, when the core+4$n$ is at low excitation energies. The detection of the charged core can provide further information of the interaction of a quasi ``unnuclear'' system, i.e. the neutron cloud, with the core nucleus. Furthermore, it may well be possible  that a core+$n$+$n$ continuum Efimov resonance can be formed in  the presence of the two other neutrons  at low relative energies from the $(t,p)$ reaction when  the n-core scattering length is large and negative. 

Among other possibilities which are not Efimov-physics dominated we mention the $^6$He$(t,p)$$^4$He+4$n$ reaction close to the scattering threshold where, the low-lying p-wave $^5$He resonance  is the main interaction channel of the $\alpha$ with the neutron cloud.
The model independent and universal correlations are enriched with low-lying resonances in the subsystems, like the $^5$He $p$-wave one necessary to establish the properties of $^6$He. Furthermore, such a low-lying p-wave resonance may possibly dominate the $^8$He properties in a $\alpha +$4$n$ description.

\subsection*{Overview of available theory approaches}

The theory should be able  to compute the inclusive  cross-sections  in  $A(d,p)$, $A(d,n)$, $A(t,p)$ and $A(t,n)$, where  for example the final states  $(A+Nn)$$^*$ or $(A+n+p)$$^*$ can be achieved. If the projectile is a weakly bound one-  or two-neutron halo, the  different processes are possible $(A+n)(d,n)(A+2n)$$^*$,  $(A+n)(t,n)(A+3n)$$^*$, $(A+2n)(d,n)(A+3n)$$^*$,  $ (A+2n)(t,n)(A+4n)$$^*$, and eventually it will require up to six-body calculations. The inclusive cross-sections where the proton is observed and the excitation spectrum is obtained, is one of the possibilities, and if in addition  the charged core is detected, further information on the FSI of the core and neutrons can be revealed. The angular distribution of the core as well as the proton at low excitation energies of the system ($A+Nn$), can be a tool to select the angular momentum states of the neutron cloud and the core nucleus. It is interesting to observe that at non-zero total orbital angular momentum of the
core+$Nn$ system the effect of the FSI at low excitation energies is even less sensitive to the details of the potential, beyond the low-energy on-shell pairwise observables, enhancing the universality of the expected results. Developments were done incorporating the halo-EFT in reaction models for one-body halos~\cite{Capel:2018kss,Moschini:2018lwh,Moschini:2019tyx,Capel:2020obz,Hebborn:2021mzi}, which has to be generalized to two or more neutron halos, to allow the investigation of the N-neutron cloud properties in the presence of the core. Effects as few-body forces induced by the core excitation can be for example incorporated by the theory~\cite{Capel:2020obz}, however the systems worked out recently within EFT are restricted to one-body halos, as has also been done in extracting the virtual state of $^{10}$Li from the excitation spectrum in a $(d,p)$ reaction~\cite{BarrancoPRC2020}.

The s-wave interactions in the limit of zero range collapse the three-body system, and therefore a three-body information has  to be included as a short-range scale.  The low-energy observables will be correlated in an universal form and  independent on the details of the short-range interaction once the two-neutron separation energy and the scattering lengths are given.  This, for example, provides the different radii in the core-n-n weakly bound nuclei~\cite{frederico12_372}. Corrections due to the effective range are expected and already studied~\cite{HammerReview2020}. These universal correlations are represented by scaling functions, which are limit cycles from the breaking of the continuous scale symmetry to a discrete one~\cite{frederico12_372,HammerReview2020}. Systems including more neutrons will not exhibit further sensitivity to short-range effects beyond the three-body one, as the Pauli principle does not allow more than two neutrons be on the top of the other. Therefore, at leading-order a system like $^{18}$C+4$n$ for low excitation energies  will only depend on the two neutron separation energy in $^{20}$C and the nn scattering length, in a model independent way. One can be reminded of the Bertsch unitary Fermi gas (see~\cite{BakerPRC1999}), for which the only scale is the Fermi momentum.

Theory has to evolve to treat the five-body system at low excitation energies to unravel the universal physics of the configuration of an unbound cloud of four neutrons and a core nucleus. Of course, this should not be restricted to light nucleus as the core, but to other nuclei, which the low lying excited state can be a MeV or so above the ground state, without being  restrictive. The reaction mechanism has also to be taken care, as the  quasi-free scattering can be distorted by the remaining proton, in $(d,p)$ or $(t,p)$ reactions, as FSI have to be accounted for. The elastic breakup and the incomplete fusion processes may overlap in the kinematical region of low excitation energies. 

\subsubsection*{\it Hyperspherical expansion methods}

Hyperspherical methods have been extensively used in studying few-body physics in a wide variety of atomic and nuclear systems, such as cold and ultracold atomic collisions and low-energy states in light nuclei. Some of the types of systems studied include three-body recombination~\cite{PhysRevA.65.010705}, some four-body recombination rates (i.e. the formation of Efimov trimers in the recombination process $B+B+B+B\rightarrow B_3+B$~\cite{StecherNature,NatureKramer2006}), and even five-body recombination~\cite{Zenesini_2013}, dimer-dimer collisions~\cite{PhysRevA.79.030501}, universality in trapped two-component fermionic systems~\cite{PhysRevA.77.043619}, the BCS-BEC crossover problem~\cite{PhysRevLett.99.090402}, among other physical processes. In the nuclear context, hyperspherical methods have been used to study bound states of light nuclei up to $A=7$~\cite{PhysRevC.102.014001,KievskyARNP2021}, and few--nucleon scattering processes such as $n+d$~\cite{PhysRevA.86.052709}, $t+p$ and $d+d$ elastic and inelastic collisions~\cite{PhysRevC.102.034007,PhysRevC.105.014001}. The hyperspherical approach has even treated scattering in the full $N$-body continuum, as is the case for the three and four neutron systems~\cite{PhysRevLett.125.052501,PhysRevC.103.024004}. Existing formalism can also accommodate systems of a core plus a few neutrons (i.e. processes where the final state includes core+$Nn$), using model core+nucleon interactions.

The adiabatic hyperspherical treatment can treat local two- and three-body interactions including but not limited to central, spin-orbit, and tensor interactions. However, the adiabatic hyperspherical formalism in its current development is incompatible with interactions that are non-local. A Hamiltonian containing non-local interactions would lead to non-locality in the hyperradial wave function and thus in the adiabatic hyperspherical potential curves. It is unclear how to interpret non-local adiabatic hyperspherical potential curves, which would actually be potential surfaces in hyperradii $\rho$ and $\rho^{\prime}$. While the current treatment of the hyperspherical method in an adiabatic representation is incompatible with non-local interactions, semi-local $l$-dependent potentials in a partial-wave expansion can be used, like in the case of the $^{10}\text{Be}+n+n$ system to study states of $^{12}\text{Be}^{*}$~\cite{PhysRevC.77.054313,PhysRevC.86.024310}.

With the adiabatic hyperspherical approach, there are a vast number of nuclear systems and reaction processes that can be studied. A class of reactions that would be of interest to study are knockout reactions such as the four-body processes $^{6}\mathrm{He}(p,p\alpha)nn$ and $^{10}\mathrm{Be}(^{3}\mathrm{H},p){^{12}\mathrm{Be}^{*}}$ to probe two-neutron halo dynamics in $^{6}\mathrm{He}+p\rightarrow({^{4}\mathrm{He}}+n+n)+p$ and $^{10}\mathrm{Be}+{^{3}\mathrm{H}}\rightarrow({^{10}\mathrm{Be}}+n+n)+p$. In principle adiabatic hyperspherical methods can be pushed to more than four particles. In this case, the computational limitations of this method would be put to a rigorous test since the vast majority of studies using this approach are for systems with $N=3$ or $N=4$ \cite{Rittenhouse_2011}. For systems larger than $N=4$, only the lowest hyperradial channel has been studied (i.e. in studies of alpha clusters up to $N=10$ \cite{BlumeGreene2000,PhysRevA.105.022824}). There are many nuclear reactions that are modeled as reactions of nucleons and clusters with $N>4$ that are of experimental and theoretical interest. Such systems include studying five-body systems like $^{8}\mathrm{He}$ to probe the final state interaction of an alpha core interacting with four neutrons, and five- and six-body reaction processes $^8\mathrm{He}(p,p\alpha)4n$, and $^{4}\mathrm{He}(^{8}\mathrm{He},^{8}\mathrm{Be})4n$ (by modeling $^{8}\mathrm{Be}$ as an $\alpha\alpha$ cluster). The reactions $^8\mathrm{He}(p,p\alpha)4n$ and $^{4}\mathrm{He}(^{8}\mathrm{He},^{8}\mathrm{Be})4n$ would be of particular importance to better understand the nature of the low-energy correlation of four--neutrons seen in both the Kisimori et al. experiment in (2016) \cite{PhysRevLett.116.052501} and subsequently the M. Duer et al. experiment in (2022) \cite{DuerNat2022}.

\subsubsection*{\it Faddeev equations framework}

The Faddeev formalism~\cite{faddeev1960mathematical} provides a rigorous mathematical formulation of the quantum mechanical three-body problem in the framework of non-relativistic dynamics. Alt, Grassberger, and Sandhas~\cite{alt1967reduction} (AGS) introduced a momentum space formulation of the Faddeev equations for scattering in terms of three-particle transition amplitudes. The latter are convenient since they are smooth functions of the momentum contrary to the original Faddeev wave function components. For a long time applications of the Faddeev-AGS formalism to the nuclear scattering problem with realistic potentials were limited by the difficulty of treating the long-ranged Coulomb potential. A breakthrough was achieved~\cite{deltuva2005momentum,deltuva2005benchmark} by developing the screening and renormalization procedure for the Coulomb interaction in momentum space using a
smooth but at the same time sufficiently rapid screening. This approach has been successfully applied to obtain ab initio solutions for the three-body scattering problem  (below and above the breakup threshold) for particles with arbitrary masses~\cite{deltuva2006coulomb,deltuva2007three}. Following the success of the Faddeev-AGS approach,  coordinate space techniques based on the Faddeev-Merkuriev~\cite{MERKURIEV1980395:1980} (FM)  were developed and successfully applied to the three-particle scattering problem~\cite{lazauskas2011application,Deltuva2014} using complex scaling~\cite{Nuttall:1969nap,Balslev:1971vb} to handle the complicated boundary conditions above the breakup threshold. 

On one hand, the FM method has been extensively applied to lighter systems (i.e. three nucleons or a light core plus two nucleons). On the other hand, the Faddeev-AGS equations have been successfully used to describe nuclear reactions involving a broader range of nuclei, from $^4$He to $^{58}$Ni. Limitations are primarily due to the utilization of the screening and renormalization procedure needed to treat the long-ranged Coulomb potential. Numerical instabilities limit the applicability of this approach to lighter systems where the Coulomb potential is not too strong. An alternative approach to incorporating the Coulomb potential was already suggested by AGS~\cite{alt1967reduction}, which involved reformulating the Faddeev-AGS equations in a basis of Coulomb scattering wave functions by adopting a separable expansion  of the pairwise potentials. A more rigorous development of the method was presented in Refs.~\cite{Mukhamedzhanov:2000qg,Mukhamedzhanov:2000nt} and was later extended to incorporate core excitations~\cite{Mukhamedzhanov:2012qv}.  This approach has the advantage that the Rutherford scattering amplitudes are isolated and treated analytically so that only the well-behaved short-ranged three-body scattering amplitudes are computed numerically. The need for screening the Coulomb potential is thus eliminated allowing the approach to be applied to heavier systems (and lower energies) where the Coulomb effects are very strong. Recently, a practical implementation of this formulation has been carried out~\cite{Hlophe:2022fdn} and extensions above the breakup threshold are nearing completion. Further, the Faddeev-AGS equations are suitable for generalizations to systems with more than three clusters and a formulation for four clusters was already developed in the seventies~\cite{Alt:1970xe}. The Faddeev-AGS approach is thus suitable for computing the aforementioned deuteron-induced transfer and breakup cross sections for processes, namely, $A(d,p)$,  $A(d,n)$, and $A(d,np)$.    

Following their successful application to three-cluster systems, the Faddeev-AGS equations have been used  to provide an exact description of four nucleon reactions~\cite{Deltuva:2007xv,Deltuva:2006sz}. Application of the Faddeev-AGS formalism to systems with four clusters, e.g., $A$+$n$+$n$+$n$, is thus feasible in principle. Since the number of two-body and three-body partial waves increases with the size of the system, the feasibility of computing four-cluster reactions involving heavy nuclei still needs to be tested. An implementation of the four-particle Faddeev-AGS equations would enable the description of reactions involving, e.g., one-neutron halos such as $(A+n)+d\longrightarrow A+2n+p$. The Faddeev-AGS equations can in principle be generalized to an even larger number of clusters (e.g., five or six clusters). Such calculations maybe feasible especially for systems of the type $A$+$3n$+$p$ and core+$3n$+p, where the number of coupled equations are reduced considerably due to the identical nature of the external nucleons.

The generalization of Faddeev-AGS to more complex systems, four and five-particles, is crucial to understand/describe exotic nuclear structures close to the driplines and central to FRIB science. 
However, such developments have not yet occurred  and the Faddeev-AGS description of reactions involving five and six clusters remains a subject for future investigations.
We note that a four and five-particle generalization of the FM equations, the coordinate counterpart of the Faddeev-AGS equations, has been successfully applied to four- and five-nucleon scattering~\cite{Lazauskas:2012jc,Lazauskas:2017paz}, although the five-nucleon scattering has been limited to low energy elastic scattering.           


\clearpage
\section{
Continuum and cluster effects in reaction theory }
\textbf{G. Potel, L. Hlophe, and J. Lubian Rios} \label{SecReactionTheory}
\smallbreak
Not surprisingly, our experimental understanding of the physics of clusters and halos in nuclei derives mostly from nuclear reactions. Clustering effects can be  probed, for example, in cluster transfer reactions, resonant scattering, and through the interplay of incomplete fusion (ICF) and complete fusion (CF).

When a clustered nucleus collides with a target, the fusion process can proceed both through CF, where the whole projectile fuses with the target, or through ICF, where only one of the fragments undergo fusion, while the rest of the projectile flies away. The relative importance of these two contributions depend strongly on the cluster dynamics of the process, as well as on the degree of clustering of  the projectile. This problem gathered attention in the 80's, when
the Ichimura Austern Vincent (IAV)~\cite{IAV85} model was developed. This model was recently revived by the Sevilla group to determine the CF for the $^{6,7}$Li +$^{209}$Bi  ~\cite{LMo15,LMo17,LMo19},  and has also been applied to the study inclusive breakup cross sections in reactions induced by $^6$He and  $^{6,7}$Li in Ref.~\cite{SCC21} and by $^{17}$F in Ref.~\cite{YLY21}.

The Rio group was also able to compute the CF as well as the ICF of each fragment for the case of projectiles breaking into two fragments. This model has been applied to the study
of fusion of  $^{6,7}$Li with heavy targets ($^{209}$Bi, $^{197}$Au, $^{198}$Pt, and $^{124}$Sn) with success~\cite{RCL20,CRF20,LFR22}. The effect of the breakup channels (continuum) on the fusion cross section has been discussed in detail, and  it has been shown that the CF is hindered above the barrier due to the dynamic effects of the couplings of the elastic channel with the continuum, that produces a repulsive polarization. It was also shown that the cluster structure of these projectiles produces a static effect that reduces the Coulomb barrier (V$_B$) and makes it thinner, producing an enhancement at energies above and below the barrier. The net effect is a competition between static and dynamic effects. The total fusion (defined as the sum of CF and ICF) is enhanced at energies below V$_B$ because the ICF becomes dominant and larger than the CF. However, it is essential to have more exclusive and inclusive cross sections for CF, ICF, TF, elastic breakup and nonelastic breakup that would put strong constrains on the theoretical calculations.
 
Breakup effects are particularly important for weakly bound nuclei, to the extent that clustered configurations are often present in  weakly bound and halo nuclei. However, present theoretical models for the derivation of CF and ICF were developed for two-body projectiles, and the extension to  projectiles that break into more than two fragments is still pending.  

The effects of the breakup channel on other reaction mechanisms are not the same at different energy regimes and mass ranges. In the case of light projectiles the CF and ICF cannot usually be separated because they have similar decay modes, calling, therefore, for the development of new detection techniques.
In a more general case, where the target is deformed, the models have to be improved in order to account for inelastic channels. The competition of the breakup channel with transfer reactions, as well as core excitation and sequential processes (breakup triggered by transfer) might also be relevant and should be included in the present models.

These reaction calculations rely on a variety of nuclear structure  ingredients. Optical potentials are needed to describe the interaction between the fragments and the target, and the double-folding São Paulo has been successfully used in this context  (SP)~\cite{CCG02}. Its first version uses a systematic parametrization for matter density distributions for the projectile and target,  based on experimental data and Dirac-Hartree-Bogolyubov calculations. In order to extend the applicability away from stability, a new version of the SP potential was published~\cite{CCG21} where the specific matter densities for each specific nucleus near the drip-line were included. Within this context, fragment-target elastic scattering angular distributions could be useful to constrain nucleus-nucleus optical potentials.
In order to generate the cluster-cluster and cluster-particle continuum states, Woods-Saxon potentials are usually used. The parameters of these simple potentials (depth, reduced radius, and diffuseness) are adjusted to describe experimental binding energies, position and width of the resonances, root mean square radius of the composite nucleus, reduced transition probabilities, multipole moments, etc. The use of more microscopic wavefunctions would provide further tests to the nuclear structure models  in  reaction calculations, as they can be introduced as external information in some reaction codes. 
In order to implement the aforementioned coupled channel calculations for deformed nuclei, deformation parameters of collective motions are usually taken from systematic such as Ref.~\cite{RNT01} for $\beta_2$ and Ref.~\cite{KiS02} for $\beta_3$. Transfer reactions are also good candidates to study the cluster configurations of exotic nuclei and pairing correlations, and the main ingredients and experimental inputs needed for the corresponding theoretical description are similar to the ones mentioned above. 

When one or two neutrons are transferred to/from a one- or two- neutron halo system, one has to add to the above ingredients a good description of the continuum. In particular, an accurate quantitative account of the two-neutron transfer cross section associated with Borromean two-neutron halo systems requires the consideration of the successive transfer contribution through the intermediate, one-neutron transfer, unbound channel \cite{Potel:10,Descouvemont:21}. The role of continuum is, unarguably, even more conspicuous when the one- or two-neutron state populated in the final channel is unbound \cite{BarrancoPRC2020,Cappuzzello:15,Barranco:17b}. Among the techniques employed to describe virtual or real states in the continuum in the contexts describe above, we mention dicretization in a finite spherical box \cite{Potel:10},  Continuum Discretized Coupled Channels (CDCC) calculations \cite{Descouvemont:21,Cappuzzello:15}, and the Green's Function Transfer (GFT) formalism \cite{Barranco:17b,BarrancoPRC2020}. Let us finally mention that the accuracy of reaction calculations in these contexts relies on a good description of the structure associated with the relevant unbound states. In particular, one needs to describe one- and two-neutron resonances, and describe pairing correlations in the continuum. An excellent example of such combined structure and reaction challenges is given by the quest for the Giant Pairing Vibration (GPV) \cite{Cappuzzello:15,Cavallaro:19,Assie:19}. 

\clearpage
\section{Spectroscopic factors, experimental probes and perspectives }
\textbf{C. Hebborn, G. Potel, D. Bazin, and A. Macchiavelli  } \label{SecQuenching}
\smallbreak
The quantitative signature of clustering in nuclei is often expressed making use of the concept of spectroscopic factors, which are also widely applied to the study of single-particle degrees of freedom~\cite{AUMANN2021103847}. Within this context, transfer processes, quasi-free scattering, and knockout reactions, which either add or remove one nucleon, play a prominent role in experiments~\cite{GM18,LeePRC06,Tetal09,PhysRevC.103.054610}.  The standard theoretical approach to describe these processes consists in combining the overlaps computed within some structure formalism of choice with a description of the collision process making use of appropriate reaction theory.     

\begin{figure}[h]
    \centering
    \includegraphics[clip,trim=1.6cm 3cm 2cm 0cm,width=\linewidth]{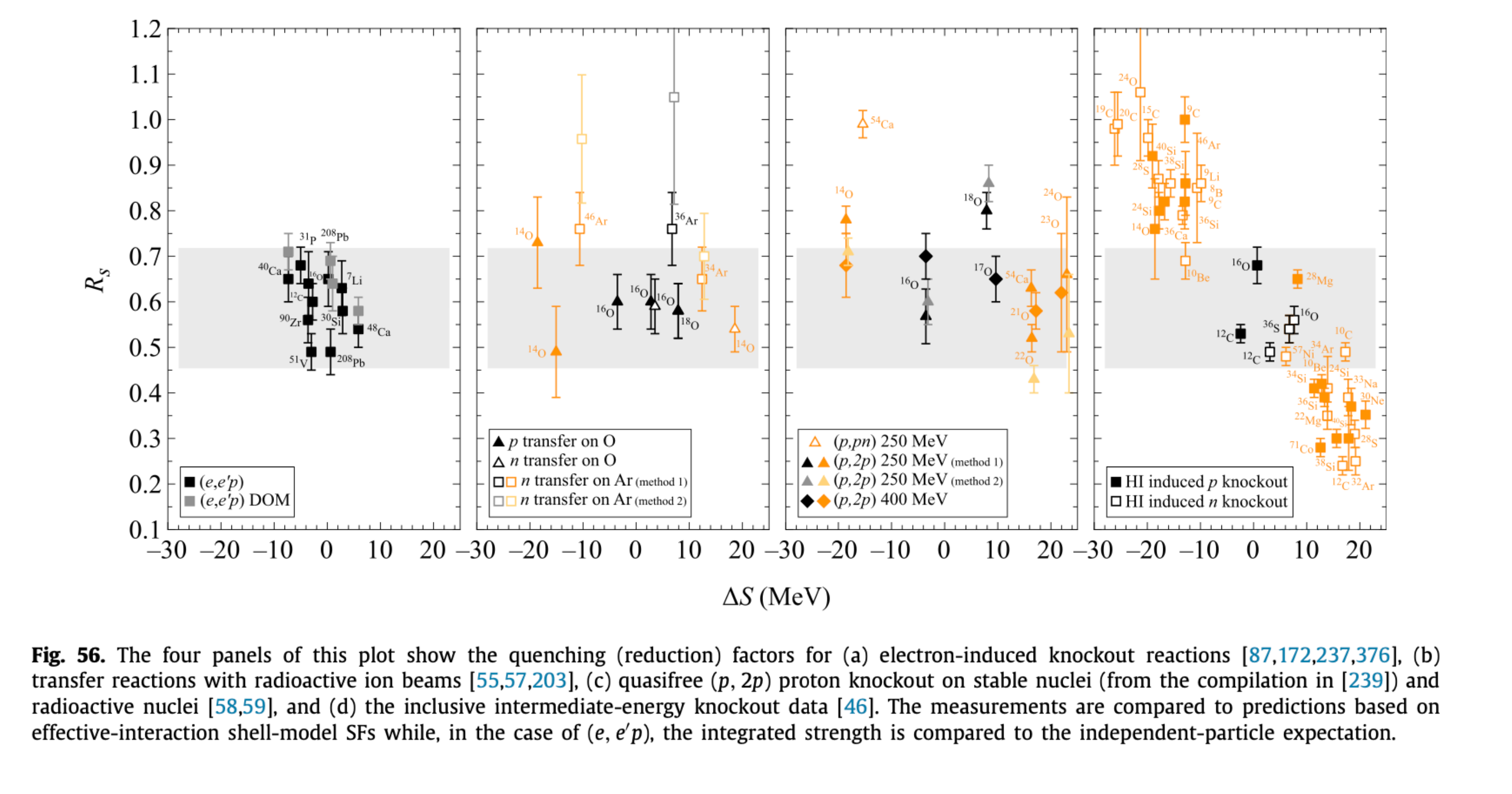}
    \caption{Quenching of the spectroscopic factor for (from left to right): electron-induced knockout reactions, transfer reactions, quasifree  knockout reactions, and    
knockout reaction  on composite target (typically $^9\rm Be$ or $^{12}\rm C$). This
figure is taken from Ref.~\cite{AUMANN2021103847}.}
    \label{GT21_Fig2}
\end{figure}

Systematic measurements of quasi-free scattering, transfer and knockout have been done for nuclei exhibiting different binding energies from the valley of stability to extremely exotic regions of the nuclear chart. The difference between the theoretical cross section and the experimental one has been interpreted as a quantitative measure of the missing many-body correlations in the structure model. Arguably, the most striking feature of this  analysis is that the ratio between the experimental and theoretical knockout cross sections exhibits a systematic trend as a function of the neutron-proton asymmetry of the studied nucleus (see right panel of Fig.~\ref{GT21_Fig2}), which is still not well understood~\cite{PhysRevC.103.054610,AUMANN2021103847}. Since similar analyses on electron-induced~\cite{KRAMER2001267}, quasi-free~\cite{GM18} and transfer reactions~\cite{Tetal09,LeePRC06,PhysRevLett.129.152501} do not show this trend (see three left panels of Fig.~\ref{GT21_Fig2}), a possible deficiency in the  theoretical description of knockout reactions has been put forward~\cite{AUMANN2021103847}.
The analysis of knockout reactions relies on shell model spectroscopic factors and  the high-energy eikonal approximation~\cite{Hansen03,PhysRevC.103.054610}. In order to understand the asymmetry dependence of the knockout cross section, the validity of these approaches have been thoroughly analyzed for the last two decades. 

In the shell model, nucleons filling completely shells are treated as an inert core, the rest of the nucleons are considered as active, and excited states are obtained by elevating a nucleon to a higher shell~\cite{B88}. With this approximation, the nuclear structure is approximated by Slater determinants built from various single-particle states, describing the active nucleons as independent particles subject to a mean field generated by the inert core.
This simplistic view therefore approximates the correlations associated with coupling of nucleons to high-momentum states arising from short-range nucleon-nucleon interactions and long-range couplings involving collective excitations, through the use of effective shell-model interactions. 

Different studies have quantified correlations beyond the independent-particle motion. In particular,  microscopic structure studies~\cite{PhysRevLett.107.032501,PhysRevLett.103.202502,PhysRevC.92.014306,PhysRevC.104.L061301} showed that many-body correlations arising from a combination of configuration mixing at the Fermi surface and coupling to the continuum induce a reduction---so called quenching---of the spectroscopic strength for deficient nucleon species.
 Other studies~\cite{PhysRevC.92.034313,PhysRevC.104.034311,PhysRevC.106.024324,PhysRevC.106.024616} have also pointed out that spectroscopic factors are not observables as they run with the  resolution scale of the interaction. Moreover, systematic studies of knockout cross sections have shown that these observables are not sensitive to the interior of the single-particle wave functions, and  scale with the square of the asymptotic normalization constants for halo nuclei and mean square radii for more bound systems~\cite{PhysRevC.100.054607,Hebborn:2021mzi,HCinprep,PhysRevC.77.044306,PhysRevLett.119.262501}.
 Both the scheme dependence and these sensitivity analyses  highlight  the need of a consistent description  of structure and reaction observables. 
 
 Other groups using  phenomenological approaches have also studied the isospin dependency of the spectroscopic strengths. In particular, the Dispersive Optical Model (DOM), which  treats consistently structure and reaction by constraining an optical potential with both  bound and scattering data,  also observes a quenching of the spectroscopic strengths for more deeply bound nucleons~\cite{ATKINSON2019135027}. Another phenomenological study~\cite{PASCHALIS2020135110} showed that the quenching of the spectroscopic factor in the knockout of a deficient species is consistent with the recent results on short-range correlations from high-energy electron scattering experiment.
However, all of the microscopic and phenomenological studies predict a rather mild dependence of the quenching of the spectroscopic factor for different proton-neutron asymmetry of nuclei, suggesting that the reaction theory involved in the analysis of knockout data are also at stake.

The typical observables for knockout reactions  are inclusive with respect to the target states, i.e., they are summed over all possible target-nucleon final states. The experimental cross  sections are usually analyzed with the  eikonal approximation, which assumes that the projectile is only slightly deflected by the
target and thus that the projectile-target relative motion does not differ much from the
initial plane wave~\cite{G59,Hansen03}. This model is  valid only at high enough energy,
i.e., beam energy above 50 MeV/nucleon. We want to stress here that two subsequent approximations  are made in the traditional eikonal description of knockout reactions~\cite{Hansen03}: the adiabatic  and the core-spectator approximations.

The adiabatic approximation implies that the projectile’s intrinsic energy and its velocity  
stay constant during the collision and, as a consequence, violates energy      
conservation. The resulting parallel momentum distributions are symmetric with respect to the maximum at zero momentum transfer, while experimental ones exhibit some degree of asymmetry that becomes very apparent for knockout of more deeply bound nucleon, for which the ratio between the single-nucleon separation energy and the beam energy is smaller~\cite{PhysRevLett.108.252501}. Some works including in their analysis this mismatch between initial and final energies obtained theoretical asymmetric momentum distributions which reproduce very well the experimental shapes~\cite{PhysRevLett.108.252501}. However, recent knockout measurements done at higher energies, at which the adiabatic approximation is expected to be more accurate, exhibit the same quenching of the ratio of the experimental to theoretical cross sections~\cite{PhysRevC.103.054610}. This suggests that the cross section integrated over the whole distorted momentum distribution is essentially identical to the one obtained from the adiabatic, symmetric one. 

The core-spectator approximation assumes that the core degrees of freedom are ``frozen'' during the collision process, thus ignoring dissipation effects associated with the extraction of the nucleon. A recent work~\cite{PhysRevC.83.011601}  suggests that these dissipation effects might populate excited states above the particle emission thresholds, leading to particle evaporation and a loss of flux that could explain part of the quenching of spectroscopic factors. This has stimulated efforts in the theory community to revisit the reaction theory at the basis of the description of knockout experiments. Recently,  Gómez Ramos and Moro studied the effects of absorption in the final channel~\cite{GOMEZRAMOS2022137252}, while Hebborn and Potel went also beyond the core-spectator approximation within the Green's Function Knockout (GFK) formalism~\cite{HP2022}, in order to account for dissipative effects in the nucleon extraction process. None of these approaches use the adiabatic approximation.  A systematic revisiting of the available knockout data  making use of these and other improved reaction formalisms is desirable and might contribute to elucidate the quenching puzzle. 

The above parlance highlights the general need to integrate in a consistent theoretical framework the structure and dynamics of many nucleons in a nucleus, and the description of reactions used to study them in an experimental context. A possible approach is suggested in the GFK, where the dynamics (driven by the optical potential) and the structure (embodied in the Green's function) are consistently integrated making use of the DOM.

Experimentally, the possibility to disentangle the reaction model and structure contributions to the quenching trend observed with knockout reactions has been put forward using mirror reactions. In this type of study, knockout cross sections between mirror initial and final states are compared to each other, with the assumption that the relevant spectroscopic factors are identical due to isospin symmetry. Pioneering works have been performed on some p-shell nuclei~\cite{GRI11, KUC22}, but need to be extended to more cases and to heavier nuclei. 

It is clear that more experimental data is needed to improve the quenching comparisons between different reaction probes. This is particularly true for transfer and quasi-free scattering for which the few existing data sets only cover a limited range of the neutron-proton asymmetry energy. Progress is being made in that direction, for instance using high luminosity detectors such as the AT-TPC~\cite{AT-TPC}, that enable to perform transfer reactions with lower beam intensities. Reactions such as (d,p) on proton-rich nuclei or ($^3$He,d) on neutron-rich nuclei would probe the quenching effect on deeply bound nucleons, for which the knockout data shows largest effect. 

Finally,   theoretical uncertainties, coming mainly from the choice of the optical potentials (discussed more extensively in Sec. \ref{SecUQ}), are usually not quantified. Recent works on transfer reactions have shown that these uncertainties introduce an error  from $\sim 20\%$ to $100\%$ on the spectroscopic strength inferred from transfer data (see Ref.~\cite{Lovell_2020} for a recent review). To be able to make a robust comparison between  different experimental probes, all the theoretical uncertainties should be quantified. A first step in that direction would be to quantify the errors introduced by the choice of optical potentials, which strongly influence the magnitude of these cross sections and hence the extracted spectroscopic factors.

Ultimately, the quantification of clustering or halo effects in nuclei relies on the determination of spectroscopic factors of the corresponding states or resonances. As these can be best populated via reactions, it becomes essential to have the most accurate theoretical tools to extract the spectroscopic factors from measured cross sections. The understanding of the different quenching trends observed between various reaction probes is therefore paramount to achieve this objective.

\clearpage
\section{
Universal corrections for few-body reaction models}
\textbf{L. Hlophe and  Ch. Elster}
\label{SecUniversalCorr}

\smallbreak

Experiments at FRIB are poised to produce and study the properties of a wide variety of exotic isotopes, often using light probes such as the deuteron or the triton. Such nuclear reactions generally constitute a complex many-body scattering problem for which exact solutions are typically not feasible. Traditional models rely on identifying the relevant degrees of freedom and projecting the many-body system onto the few-body space, leading to a system of few clusters interacting via pairwise forces. The latter are usually constructed by fitting phenomenological (Woods-Saxon) forms to (elastic) scattering data. A few-body Hamiltonian ($H_{FB}$) is then constructed using the effective pairwise potentials and observables are computed from eigenstates of $H_{FB}$. However, a formal reduction of the many-body problem onto the few body space generally leads to `irreducible" few-body forces beyond the direct sum of the pairwise potentials. A very common case is that of three-body reaction model which is employed to described processes such as the transfer of a cluster from one nucleus to another as well as three-body breakup processes. In this case, the irreducible effective three-body potential arises from antisymmetrization of the many-body problem and excitation of the clusters. In the past, attempts based on simplified nuclear structure and reaction models~\cite{Ioannides19781331,Johnson1981750,Johnson1982348,TOSTEVIN198783} were made to quantify the effects of the irreducible three-body force (3BF) on deuteron-induced reaction observables. Such works already gave indications that the impact of the 3BF on reaction observables was significant, necessitating more thorough investigations using modern techniques for describing nuclear structure and reactions. Recently a study of the effects of the 3BF arising from core excitations based in deuteron-induced reactions has been carried out~\cite{Dinmore:2019ouu}. The study utilizes multiple scattering theory to estimate leading order contribution to the 3BF, which is expressed in terms of the phenomenological nucleon-nucleus potentials. This work indicates that the impact of the 3BF on transfer cross sections can be substantial and should be accounted in order to reliably extract structure information from reaction measurements. 

\begin{figure}[ht]
\begin{center}
\includegraphics[width=10cm]{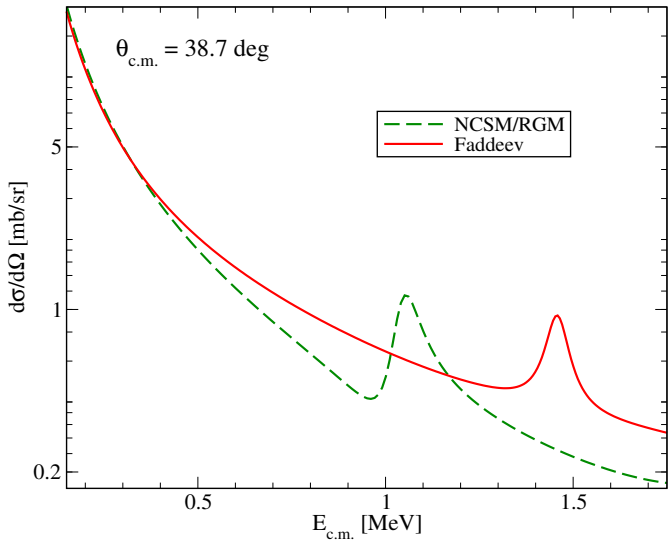}
\vspace{3mm}
\caption{The differential cross section  for elastic $d+\alpha$ scattering as a function of the center-of-mass energy $E_{\rm c.m.}$ at the scattering angle $\theta_{\rm c.m.}=38.7$~deg. The dashed lines shows phase shifts computed using the NCSM/RGM while the Faddeev results are depicted by solid lines. The
model space for the Faddeev calculation is restricted to a total two-body angular momentum of ${\cal I}_{n+p}\le3$ and ${\cal I}_{N{\rm +}\alpha}\le9/2$ for the $n$+$p$ and $N$-$\alpha$ subsystems. The NCSM/RGM model
space is truncated at $N_{\rm max}=12/13$ for the positive/negative parity partial waves and $\hbar\Omega=14$~MeV. 
}
\label{fig:dsigmae_da}
\end{center}
\end{figure}

Quantifying effects of the 3BF arising from exchange of nucleons between clusters requires a microscopic theory that is valid for the description of both structure and reaction dynamics. In Ref.~\cite{Theeten:2007} several bound systems ($^6$He=$\alpha$+$n$+$n$, $^{9}$Be=$\alpha$+$\alpha$+$n$, $^{12}$C=$\alpha$+$\alpha$+$\alpha$) were studied using the no-core shell model coupled with the resonating group method  (NCSM/RGM) starting from phenomenological nucleon-nucleon (NN) forces. The bound state energies for each nucleus were computed using two approaches, namely, (1)  a fully microscopic NCSM/RGM calculation and (2) a three-body calculation using effective nucleon-nucleus potentials evaluated using the NCSM/RGM starting from the same NN potentials used in (1). To ensure that differences only reflect the effects of nucleon exchanges between clusters, all clusters were fixed in their ground states. The result showed that the three-body model consistently leads to smaller binding energies when compared to the fully microscopic calculation. The work of Ref.~\cite{Hlophe:2022fdn} investigated the effects of the 3BF arising from antisymmetrization in deuteron ($d$)+nucleus systems, using as a test case the $d$+$\alpha$. Using a similar strategy as in Ref.~\cite{Theeten:2007}, fully microscopic NCSM/RGM calculations for the $^{6}$Li=$\alpha$+$p$+$n$ bound state were compared to three-body Faddeev calculations based on microscopic NCSM/RGM nucleon-$\alpha$ potentials. Both the fully microscopic NCSM/RGM calculation and the NCSM/RGM nucleon-$\alpha$ were based on the same chiral effective theory ($\chi$EFT) NN potentials. The three-body model was found to be underbound by approximately $600$keV, consistent with the observations in Ref.~\cite{Theeten:2007}. In addition to comparing the bound state energies, the study of Ref.~\cite{Hlophe:2022fdn}  was extended to the scattering regime. The fully microscopic NCM/RGM was used to compute phase shifts and cross sections for deuteron-$\alpha$  scattering below the deuteron breakup threshold, starting from the same $\chi$EFT NN potential adopted for the bound-state calculations. The results were compared against three-body Faddeev calculations using the aforementioned NCSM/RGM microscopic nucleon-$\alpha$ potentials. It was observed that the two approaches yield significantly different cross sections. Particularly, the $3^+$ resonance peak is found shifted to a higher energy in the three-body (see Fig.~\ref{fig:dsigmae_da}). Both the bound state and scattering results suggest that the contribution to the 3BF due to antisymmetrization is attractive. Owing primarily to the difference in the resonance position, the $d$+$\alpha$ angular distributions computed using the two methods also differ considerably. While the contribution to the binding energy was shown to increase with system size in Ref.~\cite{Theeten:2007}, a similar investigation of the effects on reactions observables is still outstanding and should be a subject for future studies. In addition, a study of the full effects of the 3BF (antisymmetrization and core excitations) can be quantified by adopting the state-of-the-art ab initio reaction theories such as the no-core shell model with continuum in place of the NCSM/RGM. %

To take full advantage of the capabilities at FRIB, accurate few-body reaction models are necessary as they provide means to translate observed experimental yields into useful structure information. In particular, single nucleon transfer reactions are typically described within a three-body model where the primary input are the nucleon-nucleus and nucleon-nucleon potentials. While a lot of progress has been made in ensuring that the three-body dynamics in these reactions are treated correctly, questions pertaining to the three-body Hamiltonian remain. Currently there is a considerable effort being dedicated to improving the effective nucleon-nucleus potentials (optical potentials) with a particular focus on exotic nuclear systems. While this is an essential step, it should be coupled with universal corrections to the three-body model itself through the quantification of irreducible three-body effects. As already indicated above, such corrections alter the theoretical energy and angular distributions and can thus potentially distort the extracted structure quantities such as spectroscopic factors. Moreover, some of the FRIB transfer measurements will use heavier probes such as $^3$H, $^3$He, and $^4$He. Although approximate solutions of the few-body problem provide means for analyzing those experiments, exact methods will need to be extended to enable a treatment of more degrees of freedom. This is necessary not only for verifying the accuracy of the approximate methods but also make reliable predictions for other exotic nuclei.

\newpage

\section{
Remarks on experimental and theoretical uncertainties in reaction studies}\label{SecUQ}

\textbf{X. Zhang, F. Nunes, and B. P. Kay}

\smallbreak

Reactions are often used to populate, dissect or modify few-body cluster states in nuclei. The understanding from a measurement requires the confluence of a detailed account on the experimental errors but also quantification of model uncertainties used in the interpretation. Arguably as compared to the experimental counterpart, the theoretical community has lagged behind in scrutinizing theory uncertainties. It is very important to make a significant investment to first understand all sources of uncertainty and then devise strategies to minimize the uncertainties so that meaningful physics can be extracted. It is common for experimental data to be plotted including  statistical uncertainties only, while the information regarding systematic uncertainties are mentioned in a less prominent way in the text of a publication. Theorists have traditionally published predictions without uncertainties (recent theory efforts on uncertainty quantification are mentioned briefly in  the last section of this contribution). Estimates of variability are provided by showing results with two different models (or parameterizations). Only recently has reaction theory begun to perform a quantitative evaluation of  uncertainties, particularly due to the effective interactions used in models. Ultimately, final results should contain the full uncertainty, combining experimental and theoretical contributions.

\subsection*{Experimental uncertainties}

Experimentally, there are many parameters to consider when assessing the uncertainties of, or in designing, a measurement~\cite{AUMANN2021103847}. Relevant to the studies of halo and cluster states is being able to infer the relative and absolute cross sections, to assign spin and parity of the final state and to compare the cross section to a reaction and nuclear--structure model. Each of these can be impacted by the experimental approach, such as detector system, the reaction used, the incident beam energy, and the angular coverage available. Some of these factors are commented on below in a general manner.

In some measurements the experimental circumstances are such that relative cross sections are obtained with good accuracy over the excitation energy range of interest, while the absolute uncertainties are substantially greater. Relative cross sections for a broad class of spectrometer techniques are often better than a few percent, though caution has to be taken when probing the unbound region in coincidence with recoil fragments, which may have different acceptances. Such data can be valuable in exploring a broad range of nuclear structure properties, but to make meaningful comparisons to theory predictions, absolute cross sections are necessary.

The absolute value of the cross section is most reliably determined by minimizing systematic errors, for instance using the same apparatus for multiple similar measurements. Accurate determination of the number of  incident beam ions on target, precise definitions of apertures and acceptance, intrinsic detector efficiency, etc., are all needed. Challenges often arise in determining the target thickness precisely. Ideally, but not always possible, is measuring the elastic scattering cross section on the same target as the direct reaction of interest but with a beam energy well below the Coulomb barrier, where the scattering is Rutherford. Other factors to consider are target uniformity and target degradation, which can be concerns when using heavier beams on thin plastic targets. Some detector systems, such as active targets can have the advantage of being self-normalizing in that the beam dose is determined absolutely, in the same way that the reaction yield is (for example, Ref.~\cite{Avila17}).

In the treatment of cross sections, once the quantum numbers for a transition are well established, it is apparent that the comparison between a reaction model and the experimental data is best done in the angular region where the cross section is at its maximum. The shape of angular distributions may not be well described at larger angles, where cross sections can be orders of magnitude lower than the primary maxima.

The range of uncertainties in absolute cross sections cannot be easily generalized as it depends on many experimental parameters. With that said, it is not uncommon for measurements with stable light-ion beams to be better than 10\% and many radioactive ion beam measurements are now around 20-30\%, though in some cases much better.

\subsection*{Interface between experimental data and model interpretation}

The choice of beam energies can have a significant impact not on the uncertainty of the measured cross sections (they are what they are), but in the reliability of the model used in the interpretation of the cross sections. It has been well-documented~\cite{Schiffer12,Schiffer13,Wimmer18,AUMANN2021103847} that for direct reaction studies, modeled as peripheral, single-step reactions, energies a few MeV/u above the Coulomb barrier in both the incident and outgoing channels are the most reliable regime (largest cross sections) in which to determine absolute cross sections for comparison to models like the distorted-wave Born approximation.

However, experiments have different goals. If the objective is to determine the asymptotic normalization of the wave function, which is needed in order to estimate astrophysical cross section for astrophysical purposes (mostly for unbound resonances), then sub-Coulomb energies are generally the most reliable. 
At these sub-Coulomb energies, the incoming wave is affected mostly by the Coulomb barrier and the transfer cross sections become proportional to the square magnitude of the tail of the wave function, the Asymptotic Normalization Coefficient (ANC) (a quantity that is relevant for both bound and resonance states). Note that the ANC can also be extracted using breakup or knockout reactions at high energy (e.g. \cite{capel2007}).

Of course, there are a whole class of intermediate and high-energy reactions, where different models have been developed and again, in those cases, strong arguments are made for specific energy ranges, such as quasi-free scattering and so on. The choice of beam energy and reaction then depends on which quantity is of primary interest.

Related to the topic above is that of momentum matching, which is modified by the choice of reaction used. For the study of halo states and many clusters, often these are low angular momentum states, being $\ell=0$, 1, etc. Reactions such as ($d$,$p$) and ($t$,$p$) are well suited to probing such excitations, in the nominal energy range of 10~MeV/u. It is worth remembering that these reactions do not reveal the parity of the final state necessarily, though it can often be easy to infer using the shell-model or from complementary experimental probes.

Other reactions, such as ($\alpha$,$^3$He) will add a neutron much like the ($d$,$p$) reaction, but are not suitable for low-$\ell$ transfer due to the often large $Q$ values involved.

Each detector set up and experimental arrangement, will have a certain angular coverage. A challenge with some systems commonly used for the study of e.g., transfer reactions in inverse kinematics, is angular coverage which can impact  the uncertainties associated with fitting of angular distributions and thus the physics interpretation.

Typically, one would like to cover the primary maximum and minima of angular distributions. When considering the canonical 10~MeV/u, and reactions like ($d$,$p$) and ($t$,$p$), one often sees these distinct features in the 0-50$^{\circ}$ center of mass range. Of particular importance, but also often difficult to access, are the most forward center-of-mass angles where $\ell=0$ angular distributions are peaked. Particular effort in the experiment community  to design detector systems that can explore this region more reliably is ongoing.

\subsection*{Theory uncertainty quantification}

\begin{figure}
\centering
\includegraphics[width=0.8\textwidth]{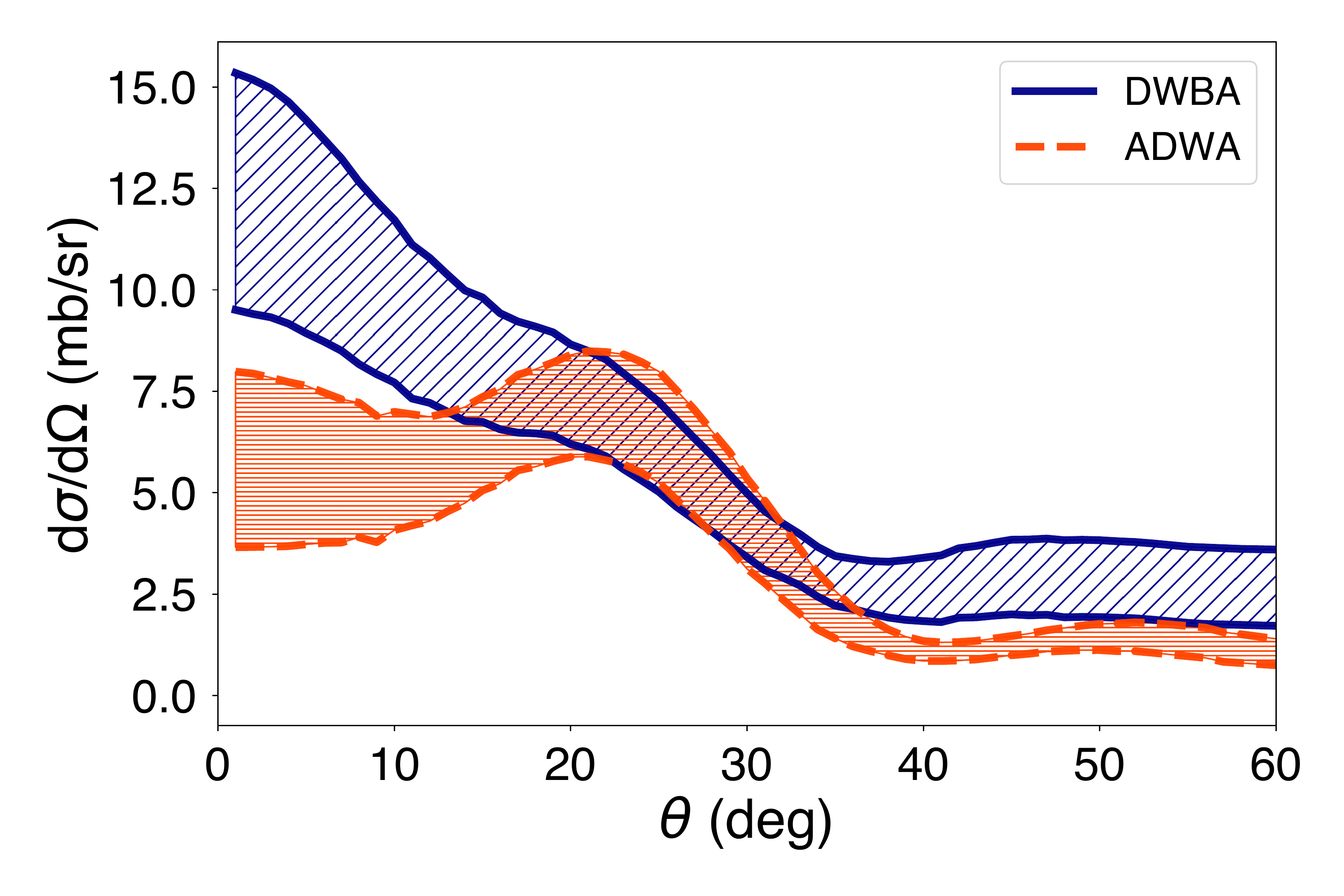}
\caption{Comparison between the 95\% confidence intervals for $^{40}$Ca(d,p)$^{41}$Ca(g.s.) cross section using two standard reaction models: DWBA (blue) and ADWA (red). Figure taken from \cite{Lovell_2020}.}
\label{fig:uqtransfer}
\end{figure}

Bayesian statistics has attracted much attention in theoretical low-energy nuclear physics
~\cite{Phillips:2020dmw, Zhang:2015ajn, Iliadis:2016vkw, Acharya:2016kfl, Wesolowski:2018lzj, Zhang:2019odg,Premarathna:2019tup, Acharya:2019fij, Melendez:2019izc, Maris:2020qne, Higa:2020pvj, Drischler:2020hwi, Drischler:2020yad, Poudel:2021mii, Acharya:2021lrv, Odell:2021nmp, Wesolowski:2021cni, Acharya:2022drl}. Recently, Bayesian analyses of the optical model have been performed \cite{lovell2018,king2018,catacora2019,Lovell_2020,catacora2021}. In these studies, nucleon elastic scattering data is used to  calibrate the nucleon optical potential, which is then propagated to transfer (d,p) reactions. An illustration is shown in Fig. \ref{fig:uqtransfer} where two standard reaction models for (d,p) transfer are compared, with the shaded areas corresponding to the 95\% confidence interval resulting from the uncertainty in the optical potentials in each model.

A detailed comparison between the Bayesian approach and the traditional frequentist approach \cite{king2019} demonstrated that the traditional approach underestimates uncertainties and that uncertainties from the optical model are in general larger than previously thought. New statistical tools \cite{catacora2021} can help in identifying which measurements are more promising to reduce these uncertainties. 

The optical models considered in the aforementioned studies are the simplest available. That choice was specifically due to computational limitations.
For a Bayesian analysis, large number of samples are needed to explore the model's parameter space. A full continuum state solution of a few-body Hamiltonian at a single point in parameter space can be expensive, e.g. the Continuum Discretized Coupled Channel (CDCC) Method, the method of choice when studying breakup reactions with halos. Repeating such calculations for thousands or even millions of times, as required in Bayesian analysis, becomes computationally infeasible.

Emulators are developed to solve  this bottleneck issue. They can be considered as fast and accurate ways to interpolate and extrapolate  either the full solutions or a particular set of observable predictions of a given complex model. Generally speaking, there are  data-driven and model-driven emulators~\cite{Melendez:2022kid,Bonilla:2022rph}. 
An example of a data-driven emulator is the recent work on quantifying the uncertainty in the breakup of $^8$B \cite{surer2022}. In that work, off-the-shelf emulators were trained on cross section breakup data to replace the full computationally intensive CDCC calculations.

For model-driven emulators, the so-called  eigenvector continuation (EC) method\footnote{The EC method was recently recognized as one of the  reduced basis methods (RBM) developed in the field of model order reduction~\cite{Melendez:2022kid,Bonilla:2022rph}.} was initially developed for  many-body bound state (discrete level) calculations~\cite{Frame:2017fah,Konig:2019adq,Ekstrom:2019lss},  and later generalized to two-body scattering~\cite{Furnstahl:2020abp,Drischler:2021qoy, Melendez:2021lyq,Bai:2021xok}, and lately to three-body scattering~\cite{Zhang:2021jmi}.  Compared to the data-driven emulators which learn physics  from data, the model-driven ones incorporate the relevant physics and therefore require generally less training data and  have better emulation accuracy. Some exemplary comparisons can be found in \cite{Konig:2019adq}. 

\begin{figure}
\centering
\includegraphics[width=0.7\textwidth]{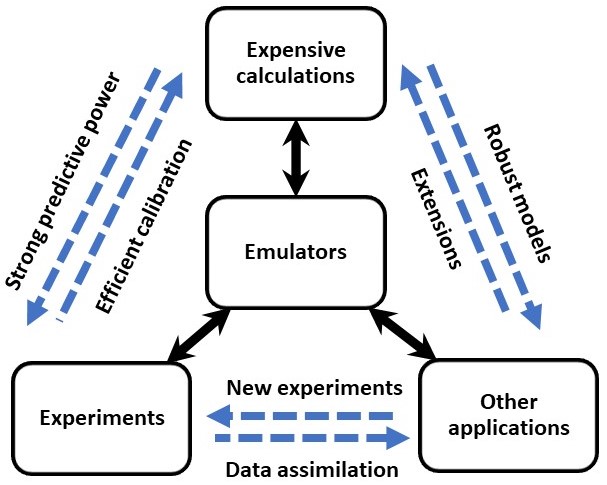}
\caption{The emulator’s role in coupling different research areas. The dashed arrows indicate
desirable connections enabled by the emulators. Figure taken from \cite{Zhang:2021jmi}.}
\label{fig:emulators}
\end{figure}

Effort is ongoing to further generalize the RBM/EC emulation to higher-energies  and N-body systems (with N larger than 3), and its implementation in a Bayesian framework for uncertainty quantification. It is worth noting that the applications of these emulators are not limited to uncertainty quantification. A scheme depicting their roles in research workflow is presented in Fig.~\ref{fig:emulators}. They can serve effectively as the interfaces for expensive calculations. Through these interfaces, users can easily access  the expensive calculations without spending significant computing resources. As a result, new interactions between different research areas can be created, and novel studies might then emerge.

\clearpage
\section{
New opportunities for nuclear Efimov states}
\textbf{S. K\"onig, T. Frederico, and F. Nunes}\label{SecEfimov}
\smallbreak

\begin{figure}
    \centering
    \includegraphics[width=0.8\textwidth]{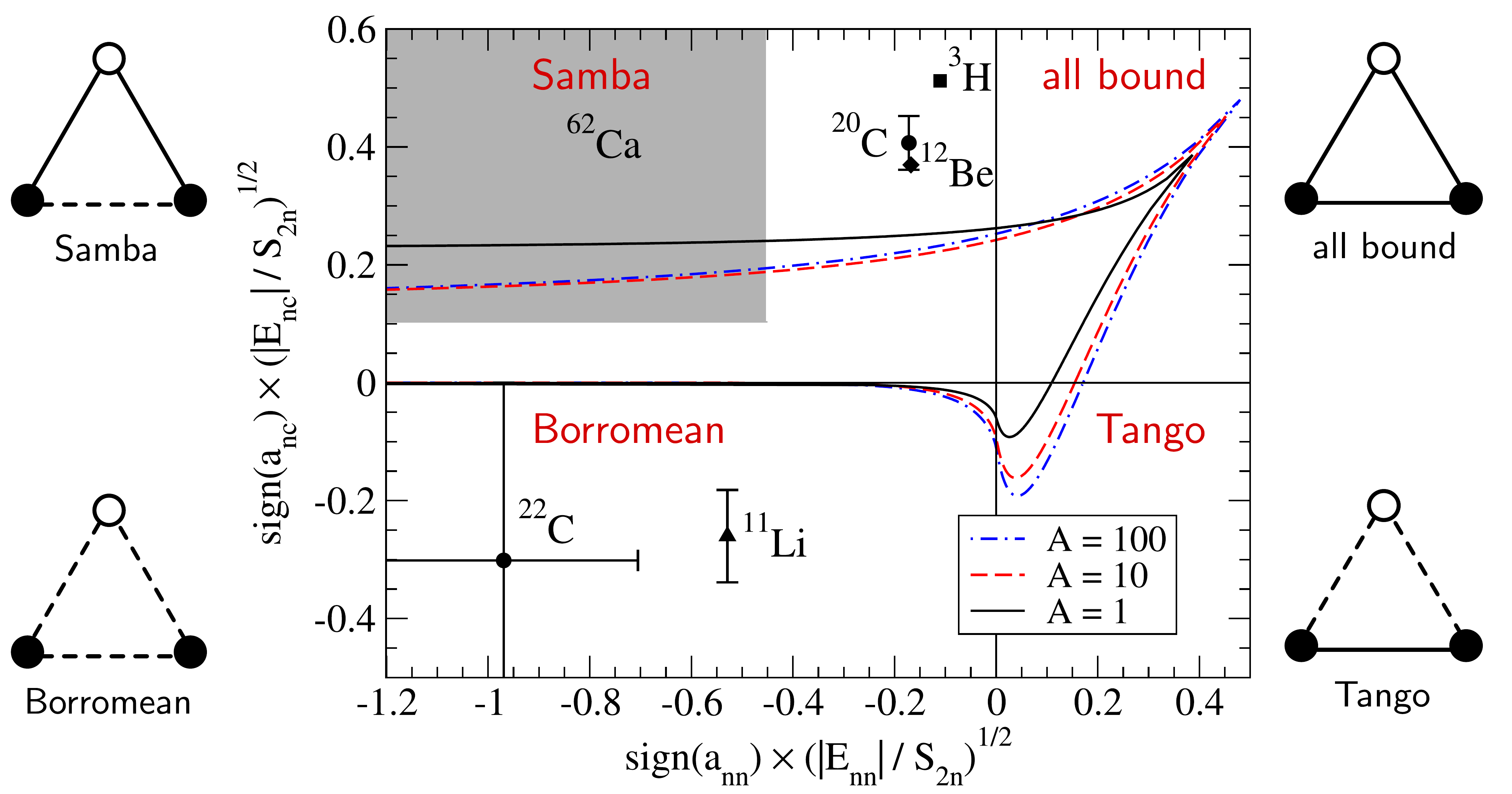}
    \caption{Critical region for the Efimov excited state above a bound one. The axis are the ratios of the neutron-core energy $(S_{1n})$ and the two-neutron  $(S_{2n})$ separation energies. $E_{nn}$ the neutron-neutron virtual or bound energies. The positive axis means a bound n-core s-wave system and the negative one a virtual s-wave state. $A$ is the core mass number. The physical region of the two-neutron halo nuclei are the left-upper quadrant (samba nuclei) and the  left-lower quadrant (Borromean nuclei).
    The triangles illustrate the possible configurations, all-bound, tango, samba and Borromean, where the continuous line corresponds to a bound s-wave state of the pair and the dashed a virtual state. The quoted nuclei ground states with a core+$n$+$n$ structure are $^{62}$Ca, $^{20}$C, $^{12}$Be, $^3$H in the samba configuration and  $^{22}$C and $^{11}$Li in the Borromean one.  Thus figure has been adapted from a similar one that appeared in Ref.~\cite{Hammer2017}.  We
    thank Hans-Werner Hammer for making the original material available to us and for giving us permission to adapt it.}
    \label{fig:EfimovRegion}
\end{figure}

Searches for the Efimov effect in atomic nuclei have by now a relatively long history (see e.g.~\cite{jensen04_233,Macchiavelli:2015xml}), and yet these states remain elusive in nuclear systems.  Weakly bound two-neutron halo nuclei are in principle excellent candidates to observe this effect, the minimum signature of which would be two levels apart in energy by a universal geometric scaling factor.  The scattering length between the neutrons and the core $a_{nc}$ and the neutron-neutron scattering length $a_{nn}$ determine the Efimov spectrum: in the unitarity limit, where both these scattering lengths are infinite, the Efimov effect in its purest form implies the existence of infinitely many geometrically spaced energy levels, that is, the ratio of energies of two successive states is a constant.  This has been presented in Efimov's pioneering work for three identical bosons~\cite{Efimov:1970zz,Efimov:1971zz}, where he showed that the reason for such a remarkable phenomenon is the emergence of an effective hyper-radial long-range inverse square potential, with strength large enough to allow Landau's ``fall to the center''~\cite{Landau:1991wop}.
For finite scattering lengths, only a limited number of bound states exists, the precise number of which depends on the actual values of the scattering lengths.\footnote{There is additionally a dependence on the mass ratio between the halo constituents and the core, but its influence is rather weak.}  Alternatively, one can express the Efimov criterion in terms of one- and two-neutron separation energies, and in this parameter space one can map out a critical region for the existence of a bound excited Efimov state above the halo ground state.  This is shown in Fig.~\ref{fig:EfimovRegion} and, as can be seen in this figure, essentially all candidates considered to date elude the critical region.

Efimov states move when the system moves away from unitarity limit to finite scattering lengths, and they gradually disappear one by one beyond the lowest scattering threshold~\cite{Efimov:1970zz,Efimov:1971zz}.  In the situation where at least one of the subsystems becomes bound, an Efimov state approaches the continuum threshold and eventually becomes a virtual state by decreasing the scattering length or increasing the subsystem binding energy~\cite{Adhikari:1982zz}.  In other words, an Efimov state migrates to the second complex energy sheet of the S-matrix, moving through the lowest two-body cut as the corresponding subsystem binding energy increases. 
This is the case, for example, for the triton~\cite{Adhikari:1982zz} (more recently studied in EFT~\cite{Rupak:2018gnc}), $^{20}$C (see e.g.~\cite{frederico12_372}), and possibly for $^{12}$Be and $^{62}$Ca~\cite{Hammer2017}, all of which are located in the second quadrant of Fig.~\ref{fig:EfimovRegion}.  If we had a knob to play with the strength of the $n$-core interaction, the virtual states of these nuclei might be pulled down into the Efimov region, by increasing the neutron-core scattering length.  We propose to explore this scenario experimentally by suddenly stripping out nucleons from the core of a halo nucleus.

All the candidates included in Fig.~\ref{fig:EfimovRegion} are bound nuclei, classified as halos in their ground states.  Rare isotope facilities provide a fascinating opportunity to look at more exotic scenarios where the halo configuration occurs as an excited state.  If the lifetime is sufficiently long compared to the typical time scale of the halo dynamics, these states may be present as resonances.  For example, one might start with a beam of known halo states, falling short of exhibiting the Efimov effect, and use reactions that remove one or more nucleons \emph{from the core}, leaving, potentially, a new halo state where the parameters are shifted in such a way that it is nudged inside the critical region shown in Fig.~\ref{fig:EfimovRegion}.  Very recently, an experiment performed at TRIUMF~\cite{Kuhn:2021dvu} achieved exactly such a stripping process to study excited states in $^{10}${Be} through the $^{11}$Be$(p,d)$ reaction at low energies. In that work, the authors showed that the data is consistent with a one-step process that removes a neutron from the core while leaving the one-neutron halo in the original nucleus intact.

Due to the decoupling of core and halo degrees of freedom, we consider that, for a short period of time after removing the core nucleons, the valence nucleons remain in their extended halo configuration (where the fall-off is still ruled by the same three-body binding energy).
Naively, we can assume that immediately after the removal of the nucleons from the core, the two-body scattering length increases due to the change in mass as $a'_{nc}\approx \sqrt{M_{\text{final}}/M_{\text{initial}}}\, a_{nc}$, reducing the corresponding two-body separation energy, and decreasing the ratio $S_{1n}/S_{2n}$ towards the Efimov region, particularly in the second quadrant shown in Fig.~\ref{fig:EfimovRegion}. 
Based on this expectation, we suggest that new experiments utilize the technique of removing multiple neutrons/protons from the core, with the goal of modifying the neutron-excited-core scattering length so that it is more favorable to populate excited Efimov states in nuclei.
\clearpage 
\section*{Acknowledgements}
\addcontentsline{toc}{section}{Acknowledgements}
This work is supported by the U.S. Department of Energy, Office of Science, Office of Nuclear Physics, under the FRIB Theory Alliance award no.\ DE-SC0013617. 
This work is supported by the National Science Foundation under Grant Nos. PHY-1555030, PHY-2111426, PHY-1913728, PHY-2209060, PHY-2044632, PHY-1912350, OAC-2004601 and the U.S. Department of Energy, Office of Science, Office of
Nuclear Physics under Contract Nos.  DE-AC52-07NA27344, DE-AC05-00OR22725, DE-SC0021422, DE-AC02-06CH1135, DE-FG02-93ER40756, DE-SC0020451, and DE-AC05-06OR23177.
GP's  work is  supported by the LLNL-LDRD Program under Project No. 21-ERD-006. 
KSB greatly appreciates the financial support of a research fellowship from the Louisiana Board of Regents; it benefited from computing resources provided by the National Energy Research Scientific Computing Center NERSC (under Contract No. DE-AC02-05CH11231), Frontera computing project at the Texas Advanced Computing Center (under National Science Foundation award OAC-1818253) and LSU ({\tt www.hpc.lsu.edu}).
FB's work is supported by the Deutsche Forschungsgemeinschaft (DFG, German Research Foundation) through Project-ID 279384907 - SFB 1245. FB would like to acknowledge Sonia Bacca for useful discussions. 
TF's work is partially supported by  Funda\c{c}\~{a}o de Amparo \`{a} Pesquisa do Estado de S\~{a}o Paulo  (2017/05660-0, 2019/07767-1), Conselho Nacional de Desenvolvimento Cient\'{i}fico e Tecnol\'{o}gico  (308486/2015-3) and the INCT-FNA project No.~464898/2014-5.


\begin{thebibliography}{308}
\ifx \bisbn   \undefined \def \bisbn  #1{ISBN #1}\fi
\ifx \binits  \undefined \def \binits#1{#1}\fi
\ifx \bauthor  \undefined \def \bauthor#1{#1}\fi
\ifx \batitle  \undefined \def \batitle#1{#1}\fi
\ifx \bjtitle  \undefined \def \bjtitle#1{#1}\fi
\ifx \bvolume  \undefined \def \bvolume#1{\textbf{#1}}\fi
\ifx \byear  \undefined \def \byear#1{#1}\fi
\ifx \bissue  \undefined \def \bissue#1{#1}\fi
\ifx \bfpage  \undefined \def \bfpage#1{#1}\fi
\ifx \blpage  \undefined \def \blpage #1{#1}\fi
\ifx \burl  \undefined \def \burl#1{\textsf{#1}}\fi
\ifx \doiurl  \undefined \def \doiurl#1{\url{https://doi.org/#1}}\fi
\ifx \betal  \undefined \def \betal{\textit{et al.}}\fi
\ifx \binstitute  \undefined \def \binstitute#1{#1}\fi
\ifx \binstitutionaled  \undefined \def \binstitutionaled#1{#1}\fi
\ifx \bctitle  \undefined \def \bctitle#1{#1}\fi
\ifx \beditor  \undefined \def \beditor#1{#1}\fi
\ifx \bpublisher  \undefined \def \bpublisher#1{#1}\fi
\ifx \bbtitle  \undefined \def \bbtitle#1{#1}\fi
\ifx \bedition  \undefined \def \bedition#1{#1}\fi
\ifx \bseriesno  \undefined \def \bseriesno#1{#1}\fi
\ifx \blocation  \undefined \def \blocation#1{#1}\fi
\ifx \bsertitle  \undefined \def \bsertitle#1{#1}\fi
\ifx \bsnm \undefined \def \bsnm#1{#1}\fi
\ifx \bsuffix \undefined \def \bsuffix#1{#1}\fi
\ifx \bparticle \undefined \def \bparticle#1{#1}\fi
\ifx \barticle \undefined \def \barticle#1{#1}\fi
\bibcommenthead
\ifx \bconfdate \undefined \def \bconfdate #1{#1}\fi
\ifx \botherref \undefined \def \botherref #1{#1}\fi
\ifx \url \undefined \def \url#1{\textsf{#1}}\fi
\ifx \bchapter \undefined \def \bchapter#1{#1}\fi
\ifx \bbook \undefined \def \bbook#1{#1}\fi
\ifx \bcomment \undefined \def \bcomment#1{#1}\fi
\ifx \oauthor \undefined \def \oauthor#1{#1}\fi
\ifx \citeauthoryear \undefined \def \citeauthoryear#1{#1}\fi
\ifx \endbibitem  \undefined \def \endbibitem {}\fi
\ifx \bconflocation  \undefined \def \bconflocation#1{#1}\fi
\ifx \arxivurl  \undefined \def \arxivurl#1{\textsf{#1}}\fi
\csname PreBibitemsHook\endcsname

\bibitem{FRIBupgrade}
\begin{botherref}
{FRIB400 The Scientific Case for the 400 {MeV}/u Energy Upgrade of FRIB}.
\url{https://frib.msu.edu/_files/pdfs/frib400_final.pdf}
\end{botherref}
\endbibitem

\bibitem{Equ}
\begin{botherref}
Experimental Equipment Needs for The Facility for Rare Isotope Beams (FRIB).
\url{https://fribusers.org/documents/2014/FRIB_EQUIPMENT_whitepaper.pdf}
\end{botherref}
\endbibitem

\bibitem{NAS}
\begin{bbook}
\bauthor{\bsnm{{National Research Council}}}:
\bbtitle{Nuclear {P}hysics: {E}xploring the {H}eart Of {M}atter}.
\bpublisher{The National Academies Press},
\blocation{Washington, DC}
(\byear{2013}).
\doiurl{10.17226/13438}
\end{bbook}
\endbibitem

\bibitem{LRP1}
\begin{botherref}
\oauthor{\bsnm{{The 2015 Nuclear Science Advisory Committee}}}:
Reaching for the horizon: The 2015 long range plan for nuclear science
(2015)
\end{botherref}
\endbibitem

\bibitem{AT-TPC}
\begin{barticle}
\bauthor{\bsnm{Bradt}, \binits{J.}},
\bauthor{\bsnm{Bazin}, \binits{D.}},
\bauthor{\bsnm{Abu-Nimeh}, \binits{F.}},
\bauthor{\bsnm{Ahn}, \binits{T.}},
\bauthor{\bsnm{Ayyad}, \binits{Y.}},
\bauthor{\bsnm{Novo}, \binits{S.B.}},
\bauthor{\bsnm{Carpenter}, \binits{L.}},
\bauthor{\bsnm{Cortesi}, \binits{M.}},
\bauthor{\bsnm{Kuchera}, \binits{M.P.}},
\bauthor{\bsnm{Lynch}, \binits{W.G.}},
\bauthor{\bsnm{Mittig}, \binits{W.}},
\bauthor{\bsnm{Rost}, \binits{S.}},
\bauthor{\bsnm{Watwood}, \binits{N.}},
\bauthor{\bsnm{Yurkon}, \binits{J.}}:
\batitle{Commissioning of the active-target time projection chamber}.
\bjtitle{Nucl. Instrum. Methods. Phys. Res. A}
\bvolume{875},
\bfpage{9}
(\byear{2017})
\end{barticle}
\endbibitem

\bibitem{hergert20_2412}
\begin{barticle}
\bauthor{\bsnm{Hergert}, \binits{H.}}:
\batitle{A guided tour of \textit{ab initio} nuclear many-body theory}.
\bjtitle{Front. Phys.}
\bvolume{8},
\bfpage{379}
(\byear{2020})
\end{barticle}
\endbibitem

\bibitem{HammerReview2020}
\begin{barticle}
\bauthor{\bsnm{Hammer}, \binits{H.-W.}},
\bauthor{\bsnm{K\"onig}, \binits{S.}},
\bauthor{\bparticle{van} \bsnm{Kolck}, \binits{U.}}:
\batitle{Nuclear effective field theory: Status and perspectives}.
\bjtitle{Rev. Mod. Phys.}
\bvolume{92},
\bfpage{66}
(\byear{2020})
\end{barticle}
\endbibitem

\bibitem{Navratil_2016}
\begin{barticle}
\bauthor{\bsnm{Navr\'{a}til}, \binits{P.}},
\bauthor{\bsnm{Quaglioni}, \binits{S.}},
\bauthor{\bsnm{Hupin}, \binits{G.}},
\bauthor{\bsnm{Romero-Redondo}, \binits{C.}},
\bauthor{\bsnm{Calci}, \binits{A.}}:
\batitle{Unified \textit{ab initio} approaches to nuclear structure and
  reactions}.
\bjtitle{Phys. Scr}
\bvolume{91},
\bfpage{053002}
(\byear{2016})
\end{barticle}
\endbibitem

\bibitem{quaglioni08_755}
\begin{barticle}
\bauthor{\bsnm{Quaglioni}, \binits{S.}},
\bauthor{\bsnm{Navr\'atil}, \binits{P.}}:
\batitle{\textit{Ab initio} many-body calculations of ${ n - {}^{3}\text{H} , n
  - {}^{4}\text{He}, p - {}^{3,4}\text{He} }$, and ${ n - {}^{10}\text{Be} }$
  scattering}.
\bjtitle{Phys. Rev. Lett.}
\bvolume{101},
\bfpage{092501}
(\byear{2008})
\end{barticle}
\endbibitem

\bibitem{quaglioni09_756}
\begin{barticle}
\bauthor{\bsnm{Quaglioni}, \binits{S.}},
\bauthor{\bsnm{Navr\'atil}, \binits{P.}}:
\batitle{\textit{Ab initio} many-body calculations of nucleon-nucleus
  scattering}.
\bjtitle{Phys. Rev. C.}
\bvolume{79},
\bfpage{044606}
(\byear{2009})
\end{barticle}
\endbibitem

\bibitem{lee09_949}
\begin{barticle}
\bauthor{\bsnm{Lee}, \binits{D.}}:
\batitle{Lattice simulations for few- and many-body systems}.
\bjtitle{Prog. Part. Nucl. Phys.}
\bvolume{63},
\bfpage{117}
(\byear{2009})
\end{barticle}
\endbibitem

\bibitem{nollett12_835}
\begin{barticle}
\bauthor{\bsnm{Nollett}, \binits{K.M.}}:
\batitle{\textit{Ab initio} calculations of nuclear widths via an integral
  relation}.
\bjtitle{Phys. Rev. C}
\bvolume{86},
\bfpage{044330}
(\byear{2012})
\end{barticle}
\endbibitem

\bibitem{flores22}
\begin{botherref}
\oauthor{\bsnm{Flores}, \binits{A.R.}},
\oauthor{\bsnm{Nollett}, \binits{K.M.}}:
Variational {M}onte {C}arlo calculations of ${ n + {}^{3}\text{H} }$
  scattering.
ArXiv
(2022)
\end{botherref}
\endbibitem

\bibitem{stroberg21_2483}
\begin{barticle}
\bauthor{\bsnm{Stroberg}, \binits{S.R.}},
\bauthor{\bsnm{Holt}, \binits{J.D.}},
\bauthor{\bsnm{Schwenk}, \binits{A.}},
\bauthor{\bsnm{Simonis}, \binits{J.}}:
\batitle{\textit{Ab initio} limits of atomic nuclei}.
\bjtitle{Phys. Rev. Lett.}
\bvolume{126},
\bfpage{022501}
(\byear{2021})
\end{barticle}
\endbibitem

\bibitem{caurier05_424}
\begin{barticle}
\bauthor{\bsnm{Caurier}, \binits{E.}},
\bauthor{\bsnm{{Martinez-Pinedo}}, \binits{G.}},
\bauthor{\bsnm{Nowacki}, \binits{F.}},
\bauthor{\bsnm{Poves}, \binits{A.}},
\bauthor{\bsnm{Zuker}, \binits{A.P.}}:
\batitle{The shell model as a unified view of nuclear structure}.
\bjtitle{Rev. Mod. Phys.}
\bvolume{77},
\bfpage{427}
(\byear{2005})
\end{barticle}
\endbibitem

\bibitem{rotter91_448}
\begin{barticle}
\bauthor{\bsnm{Rotter}, \binits{I.}}:
\batitle{A continuum shell model for the open quantum mechanical nuclear
  system}.
\bjtitle{Rep. Prog. Phys.}
\bvolume{54},
\bfpage{635}
(\byear{1991})
\end{barticle}
\endbibitem

\bibitem{michel09_2}
\begin{barticle}
\bauthor{\bsnm{Michel}, \binits{N.}},
\bauthor{\bsnm{Nazarewicz}, \binits{W.}},
\bauthor{\bsnm{P{\l}oszajczak}, \binits{M.}},
\bauthor{\bsnm{Vertse}, \binits{T.}}:
\batitle{Shell model in the complex energy plane}.
\bjtitle{J. Phys. G: Nucl. Part. Phys.}
\bvolume{36},
\bfpage{013101}
(\byear{2009})
\end{barticle}
\endbibitem

\bibitem{volya05_470}
\begin{barticle}
\bauthor{\bsnm{Volya}, \binits{A.}},
\bauthor{\bsnm{Zelevinsky}, \binits{V.}}:
\batitle{Discrete and continuum spectra in the unified shell model approach}.
\bjtitle{Phys. Rev. Lett.}
\bvolume{94},
\bfpage{052501}
(\byear{2005})
\end{barticle}
\endbibitem

\bibitem{baroni13_385}
\begin{barticle}
\bauthor{\bsnm{Baroni}, \binits{S.}},
\bauthor{\bsnm{Navr\'atil}, \binits{P.}},
\bauthor{\bsnm{Quaglioni}, \binits{S.}}:
\batitle{Unified \textit{ab initio} approach to bound and unbound states:
  {N}o-core shell model with continuum and its application to ${
  {}^{7}\text{He} }$}.
\bjtitle{Phys. Rev. C}
\bvolume{87},
\bfpage{034326}
(\byear{2013})
\end{barticle}
\endbibitem

\bibitem{matsuo10_1238}
\begin{barticle}
\bauthor{\bsnm{Matsuo}, \binits{M.}},
\bauthor{\bsnm{Nakatsukasa}, \binits{T.}}:
\batitle{Open problems in nuclear structure near drip lines}.
\bjtitle{J. Phys. G: Nucl. Part. Phys.}
\bvolume{37},
\bfpage{064017}
(\byear{2010})
\end{barticle}
\endbibitem

\bibitem{michel10_4}
\begin{barticle}
\bauthor{\bsnm{Michel}, \binits{N.}},
\bauthor{\bsnm{Nazarewicz}, \binits{W.}},
\bauthor{\bsnm{Oko{\l}owicz}, \binits{J.}},
\bauthor{\bsnm{P{\l}oszajczak}, \binits{M.}}:
\batitle{Open problems in the theory of nuclear open quantum systems}.
\bjtitle{J. Phys. G: Nucl. Part. Phys.}
\bvolume{37},
\bfpage{064042}
(\byear{2010})
\end{barticle}
\endbibitem

\bibitem{okolowicz03_21}
\begin{barticle}
\bauthor{\bsnm{Oko{\l}owicz}, \binits{J.}},
\bauthor{\bsnm{P{\l}oszajczak}, \binits{M.}},
\bauthor{\bsnm{Rotter}, \binits{I.}}:
\batitle{Dynamics of quantum systems embedded in a continuum}.
\bjtitle{Phys. Rep.}
\bvolume{374},
\bfpage{271}
(\byear{2003})
\end{barticle}
\endbibitem

\bibitem{otsuka20_2383}
\begin{barticle}
\bauthor{\bsnm{Otsuka}, \binits{T.}},
\bauthor{\bsnm{Gade}, \binits{A.}},
\bauthor{\bsnm{Sorlin}, \binits{O.}},
\bauthor{\bsnm{Suzuki}, \binits{T.}},
\bauthor{\bsnm{Utsuno}, \binits{Y.}}:
\batitle{Evolution of shell structure in exotic nuclei}.
\bjtitle{Rev. Mod. Phys.}
\bvolume{92},
\bfpage{015002}
(\byear{2020})
\end{barticle}
\endbibitem

\bibitem{fossez16_1335}
\begin{barticle}
\bauthor{\bsnm{Fossez}, \binits{K.}},
\bauthor{\bsnm{Nazarewicz}, \binits{W.}},
\bauthor{\bsnm{Jaganathen}, \binits{Y.}},
\bauthor{\bsnm{Michel}, \binits{N.}},
\bauthor{\bsnm{P{\l}oszajczak}, \binits{M.}}:
\batitle{Nuclear rotation in the continuum}.
\bjtitle{Phys. Rev. C}
\bvolume{93},
\bfpage{011305}
(\byear{2016})
\end{barticle}
\endbibitem

\bibitem{fossez16_1793}
\begin{barticle}
\bauthor{\bsnm{Fossez}, \binits{K.}},
\bauthor{\bsnm{Rotureau}, \binits{J.}},
\bauthor{\bsnm{Michel}, \binits{N.}},
\bauthor{\bsnm{Liu}, \binits{Q.}},
\bauthor{\bsnm{Nazarewicz}, \binits{W.}}:
\batitle{Single-particle and collective motion in unbound deformed ${
  {}^{39}\text{Mg} }$}.
\bjtitle{Phys. Rev. C}
\bvolume{94},
\bfpage{054302}
(\byear{2016})
\end{barticle}
\endbibitem

\bibitem{fossez18_2171}
\begin{barticle}
\bauthor{\bsnm{Fossez}, \binits{K.}},
\bauthor{\bsnm{Rotureau}, \binits{J.}},
\bauthor{\bsnm{Nazarewicz}, \binits{W.}}:
\batitle{Energy spectrum of neutron-rich helium isotopes: {C}omplex made
  simple}.
\bjtitle{Phys. Rev. C}
\bvolume{98},
\bfpage{061302}
(\byear{2018})
\end{barticle}
\endbibitem

\bibitem{kravvaris17_1960}
\begin{barticle}
\bauthor{\bsnm{Kravvaris}, \binits{K.}},
\bauthor{\bsnm{Volya}, \binits{A.}}:
\batitle{Study of nuclear clustering from an \textit{ab initio} perspective}.
\bjtitle{Phys. Rev. Lett.}
\bvolume{119},
\bfpage{062501}
(\byear{2017})
\end{barticle}
\endbibitem

\bibitem{wang18_2144}
\begin{barticle}
\bauthor{\bsnm{Wang}, \binits{S.M.}},
\bauthor{\bsnm{Nazarewicz}, \binits{W.}}:
\batitle{Puzzling two-proton decay of ${ {}^{67}\text{Kr} }$}.
\bjtitle{Phys. Rev. Lett.}
\bvolume{120},
\bfpage{212502}
(\byear{2018})
\end{barticle}
\endbibitem

\bibitem{fossez22_2540}
\begin{barticle}
\bauthor{\bsnm{Fossez}, \binits{K.}},
\bauthor{\bsnm{Rotureau}, \binits{J.}}:
\batitle{Density matrix renormalization group description of the island of
  inversion isotopes ${ {}^{28-33}\text{F} }$}.
\bjtitle{Phys. Rev. C}
\bvolume{106},
\bfpage{034312}
(\byear{2022})
\end{barticle}
\endbibitem

\bibitem{jensen04_233}
\begin{barticle}
\bauthor{\bsnm{Jensen}, \binits{A.S.}},
\bauthor{\bsnm{Riisager}, \binits{K.}},
\bauthor{\bsnm{Fedorov}, \binits{D.V.}},
\bauthor{\bsnm{Garrido}, \binits{E.}}:
\batitle{Structure and reactions of quantum halos}.
\bjtitle{Rev. Mod. Phys.}
\bvolume{76},
\bfpage{215}
(\byear{2004})
\end{barticle}
\endbibitem

\bibitem{frederico12_372}
\begin{barticle}
\bauthor{\bsnm{Frederico}, \binits{T.}},
\bauthor{\bsnm{Delfino}, \binits{A.}},
\bauthor{\bsnm{Tomio}, \binits{L.}},
\bauthor{\bsnm{Yamashita}, \binits{M.T.}}:
\batitle{Universal aspects of light halo nuclei}.
\bjtitle{Prog. Part. Nucl. Phys.}
\bvolume{67},
\bfpage{939}
(\byear{2012})
\end{barticle}
\endbibitem

\bibitem{tanihata13_549}
\begin{barticle}
\bauthor{\bsnm{Tanihata}, \binits{I.}},
\bauthor{\bsnm{Savajols}, \binits{H.}},
\bauthor{\bsnm{Kanungo}, \binits{R.}}:
\batitle{Recent experimental progress in nuclear halo structure studies}.
\bjtitle{Prog. Part. Nucl. Phys.}
\bvolume{68},
\bfpage{215}
(\byear{2013})
\end{barticle}
\endbibitem

\bibitem{freer18_2138}
\begin{barticle}
\bauthor{\bsnm{Freer}, \binits{M.}},
\bauthor{\bsnm{Horiuchi}, \binits{H.}},
\bauthor{\bsnm{{Kanada-En'yo}}, \binits{Y.}},
\bauthor{\bsnm{Lee}, \binits{D.}},
\bauthor{\bsnm{Mei{\ss}ner}, \binits{U.}}:
\batitle{Microscopic clustering in light nuclei}.
\bjtitle{Rev. Mod. Phys.}
\bvolume{90},
\bfpage{035004}
(\byear{2018})
\end{barticle}
\endbibitem

\bibitem{ikeda68_2329}
\begin{barticle}
\bauthor{\bsnm{Ikeda}, \binits{K.}},
\bauthor{\bsnm{Takigawa}, \binits{N.}},
\bauthor{\bsnm{Horiuchi}, \binits{H.}}:
\batitle{The systematic structure-change into the molecule-like structures in
  the self-conjugate ${ 4n }$ nuclei}.
\bjtitle{Prog. Theor. Phys. Suppl.}
\bvolume{E68},
\bfpage{464}
(\byear{1968})
\end{barticle}
\endbibitem

\bibitem{oertzen06_1017}
\begin{barticle}
\bauthor{\bsnm{{von Oertzen}}, \binits{W.}},
\bauthor{\bsnm{Freer}, \binits{M.}},
\bauthor{\bsnm{{Kanada-En'yo}}, \binits{Y.}}:
\batitle{Nuclear clusters and nuclear molecules}.
\bjtitle{Phys. Rep.}
\bvolume{432},
\bfpage{43}
(\byear{2006})
\end{barticle}
\endbibitem

\bibitem{freer07_1018}
\begin{barticle}
\bauthor{\bsnm{Freer}, \binits{M.}}:
\batitle{The clustered nucleus--cluster structures in stable and unstable
  nuclei}.
\bjtitle{Rep. Prog. Phys.}
\bvolume{70},
\bfpage{2149}
(\byear{2007})
\end{barticle}
\endbibitem

\bibitem{okolowicz13_241}
\begin{barticle}
\bauthor{\bsnm{Oko{\l}owicz}, \binits{J.}},
\bauthor{\bsnm{Nazarewicz}, \binits{W.}},
\bauthor{\bsnm{P{\l}oszajczak}, \binits{M.}}:
\batitle{Toward understanding the microscopic origin of nuclear clustering}.
\bjtitle{Fortschr. Phys.}
\bvolume{61},
\bfpage{66}
(\byear{2013})
\end{barticle}
\endbibitem

\bibitem{okolowicz12_998}
\begin{barticle}
\bauthor{\bsnm{Oko{\l}owicz}, \binits{J.}},
\bauthor{\bsnm{P{\l}oszajczak}, \binits{M.}},
\bauthor{\bsnm{Nazarewicz}, \binits{W.}}:
\batitle{On the origin of nuclear clustering}.
\bjtitle{Prog. Theor. Phys. Supp.}
\bvolume{196},
\bfpage{230}
(\byear{2012})
\end{barticle}
\endbibitem

\bibitem{guillemaud90_1761}
\begin{barticle}
\bauthor{\bsnm{{Guillemaud-Mueller}}, \binits{D.}},
\bauthor{\bsnm{Jacmart}, \binits{J.C.}},
\bauthor{\bsnm{Kashy}, \binits{E.}},
\bauthor{\bsnm{Latimier}, \binits{A.}},
\bauthor{\bsnm{Mueller}, \binits{A.C.}},
\bauthor{\bsnm{Pougheon}, \binits{F.}},
\bauthor{\bsnm{Richard}, \binits{A.}},
\bauthor{\bsnm{Penionzhkevich}, \binits{Y.E.}},
\bauthor{\bsnm{Artuhk}, \binits{A.G.}},
\bauthor{\bsnm{Belozyorov}, \binits{A.V.}},
\bauthor{\bsnm{Lukyanov}, \binits{S.M.}},
\bauthor{\bsnm{Anne}, \binits{R.}},
\bauthor{\bsnm{Bricault}, \binits{P.}},
\bauthor{\bsnm{D\'etraz}, \binits{C.}},
\bauthor{\bsnm{Lewitowicz}, \binits{M.}},
\bauthor{\bsnm{Zhang}, \binits{Y.}},
\bauthor{\bsnm{Lyutostansky}, \binits{Y.S.}},
\bauthor{\bsnm{Zverev}, \binits{M.V.}},
\bauthor{\bsnm{Bazin}, \binits{D.}},
\bauthor{\bsnm{{Schmidt-Ott}}, \binits{W.D.}}:
\batitle{Particle stability of the isotopes ${ {}^{26}\text{O} }$ and ${
  {}^{32}\text{Ne} }$ in the reaction 44 {MeV}/nucleon ${ {}^{48}\text{Ca} +
  \text{Ta} }$}.
\bjtitle{Phys. Rev. C}
\bvolume{41},
\bfpage{937}
(\byear{1990})
\end{barticle}
\endbibitem

\bibitem{blank08_449}
\begin{barticle}
\bauthor{\bsnm{Blank}, \binits{B.}},
\bauthor{\bsnm{P{\l}oszajczak}, \binits{M.}}:
\batitle{Two-proton radioactivity}.
\bjtitle{Rep. Prog. Phys.}
\bvolume{71},
\bfpage{046301}
(\byear{2008})
\end{barticle}
\endbibitem

\bibitem{pfutzner12_1169}
\begin{barticle}
\bauthor{\bsnm{Pf\"utzner}, \binits{M.}},
\bauthor{\bsnm{Karny}, \binits{M.}},
\bauthor{\bsnm{Grigorenko}, \binits{L.V.}},
\bauthor{\bsnm{Riisager}, \binits{K.}}:
\batitle{Radioactive decays at limits of nuclear stability}.
\bjtitle{Rev. Mod. Phys.}
\bvolume{84},
\bfpage{567}
(\byear{2012})
\end{barticle}
\endbibitem

\bibitem{spyrou12_1216}
\begin{barticle}
\bauthor{\bsnm{Spyrou}, \binits{A.}},
\bauthor{\bsnm{Kohley}, \binits{Z.}},
\bauthor{\bsnm{Baumann}, \binits{T.}},
\bauthor{\bsnm{Bazin}, \binits{D.}},
\bauthor{\bsnm{Brown}, \binits{B.A.}},
\bauthor{\bsnm{Christian}, \binits{G.}},
\bauthor{\bsnm{{DeYoung}}, \binits{P.A.}},
\bauthor{\bsnm{Finck}, \binits{J.E.}},
\bauthor{\bsnm{Frank}, \binits{N.}},
\bauthor{\bsnm{Lunderberg}, \binits{E.}},
\bauthor{\bsnm{Mosby}, \binits{S.}},
\bauthor{\bsnm{Peters}, \binits{W.A.}},
\bauthor{\bsnm{Schiller}, \binits{S.}},
\bauthor{\bsnm{Smith}, \binits{J.K.}},
\bauthor{\bsnm{Snyder}, \binits{J.}},
\bauthor{\bsnm{Strongman}, \binits{M.J.}},
\bauthor{\bsnm{Thoennessen}, \binits{M.}},
\bauthor{\bsnm{Volya}, \binits{A.}}:
\batitle{First observation of ground state dineutron decay: ${ {}^{16}\text{Be}
  }$}.
\bjtitle{Phys. Rev. Lett.}
\bvolume{108},
\bfpage{102501}
(\byear{2012})
\end{barticle}
\endbibitem

\bibitem{thoennessen13_1776}
\begin{barticle}
\bauthor{\bsnm{Thoennessen}, \binits{M.}},
\bauthor{\bsnm{Kohley}, \binits{Z.}},
\bauthor{\bsnm{Spyrou}, \binits{A.}},
\bauthor{\bsnm{Lunderberg}, \binits{E.}},
\bauthor{\bsnm{{DeYoung}}, \binits{P.A.}},
\bauthor{\bsnm{Attanayake}, \binits{H.}},
\bauthor{\bsnm{Baumann}, \binits{T.}},
\bauthor{\bsnm{Bazin}, \binits{D.}},
\bauthor{\bsnm{Brown}, \binits{B.A.}},
\bauthor{\bsnm{Christian}, \binits{G.}},
\bauthor{\bsnm{Divaratne}, \binits{D.}},
\bauthor{\bsnm{Grimes}, \binits{S.M.}},
\bauthor{\bsnm{Haagsma}, \binits{A.}},
\bauthor{\bsnm{Finck}, \binits{J.E.}},
\bauthor{\bsnm{Frank}, \binits{N.}},
\bauthor{\bsnm{Luther}, \binits{B.}},
\bauthor{\bsnm{Mosby}, \binits{S.}},
\bauthor{\bsnm{Nagi}, \binits{T.}},
\bauthor{\bsnm{Peaslee}, \binits{G.F.}},
\bauthor{\bsnm{Peters}, \binits{W.A.}},
\bauthor{\bsnm{Schiller}, \binits{A.}},
\bauthor{\bsnm{Smith}, \binits{J.K.}},
\bauthor{\bsnm{Snyder}, \binits{J.}},
\bauthor{\bsnm{Strongman}, \binits{M.}},
\bauthor{\bsnm{Volya}, \binits{A.}}:
\batitle{Observation of ground-state two-neutron decay}.
\bjtitle{Acta Phys. Pol.}
\bvolume{44},
\bfpage{543}
(\byear{2013})
\end{barticle}
\endbibitem

\bibitem{johnson20_2389}
\begin{barticle}
\bauthor{\bsnm{Johnson}, \binits{C.W.}},
\bauthor{\bsnm{Launey}, \binits{K.D.}},
\bauthor{\bsnm{Auerbach}, \binits{N.}},
\bauthor{\bsnm{Bacca}, \binits{S.}},
\bauthor{\bsnm{Barrett}, \binits{B.R.}},
\bauthor{\bsnm{Brune}, \binits{C.}},
\bauthor{\bsnm{Caprio}, \binits{M.A.}},
\bauthor{\bsnm{Descouvemont}, \binits{P.}},
\bauthor{\bsnm{Dickhoff}, \binits{W.H.}},
\bauthor{\bsnm{Elster}, \binits{C.}},
\bauthor{\bsnm{Fasano}, \binits{P.J.}},
\bauthor{\bsnm{Fossez}, \binits{K.}},
\bauthor{\bsnm{Hergert}, \binits{H.}},
\bauthor{\bsnm{{Hjorth-Jensen}}, \binits{M.}},
\bauthor{\bsnm{Hlophe}, \binits{L.}},
\bauthor{\bsnm{Hu}, \binits{B.}},
\bauthor{\bsnm{{Id Betan}}, \binits{R.M.}},
\bauthor{\bsnm{Idini}, \binits{A.}},
\bauthor{\bsnm{K\"onig}, \binits{S.}},
\bauthor{\bsnm{Kravvaris}, \binits{K.}},
\bauthor{\bsnm{Lee}, \binits{D.}},
\bauthor{\bsnm{Lei}, \binits{J.}},
\bauthor{\bsnm{Maris}, \binits{P.}},
\bauthor{\bsnm{Mercenne}, \binits{A.}},
\bauthor{\bsnm{Minomo}, \binits{K.}},
\bauthor{\bsnm{{Navarro P\'erez}}, \binits{R.}},
\bauthor{\bsnm{Nazarewicz}, \binits{W.}},
\bauthor{\bsnm{Nunes}, \binits{F.M.}},
\bauthor{\bsnm{P{\l}oszajczak}, \binits{M.}},
\bauthor{\bsnm{Quaglioni}, \binits{S.}},
\bauthor{\bsnm{Rotureau}, \binits{J.}},
\bauthor{\bsnm{Rupak}, \binits{G.}},
\bauthor{\bsnm{Shirokov}, \binits{A.M.}},
\bauthor{\bsnm{Thompson}, \binits{I.}},
\bauthor{\bsnm{Vary}, \binits{J.P.}},
\bauthor{\bsnm{Volya}, \binits{A.}},
\bauthor{\bsnm{Xu}, \binits{F.}},
\bauthor{\bsnm{Zelevinsky}, \binits{V.}},
\bauthor{\bsnm{Zhang}, \binits{X.}}:
\batitle{White paper: {F}rom bound states to the continuum}.
\bjtitle{J. Phys. G: Nucl. Part. Phys.}
\bvolume{47},
\bfpage{123001}
(\byear{2020})
\end{barticle}
\endbibitem

\bibitem{volya06_94}
\begin{barticle}
\bauthor{\bsnm{Volya}, \binits{A.}},
\bauthor{\bsnm{Zelevinsky}, \binits{V.}}:
\batitle{Continuum shell model}.
\bjtitle{Phys. Rev. C}
\bvolume{74},
\bfpage{064314}
(\byear{2006})
\end{barticle}
\endbibitem

\bibitem{rotureau06_15}
\begin{barticle}
\bauthor{\bsnm{Rotureau}, \binits{J.}},
\bauthor{\bsnm{Michel}, \binits{N.}},
\bauthor{\bsnm{Nazarewicz}, \binits{W.}},
\bauthor{\bsnm{P{\l}oszajczak}, \binits{M.}},
\bauthor{\bsnm{Dukelsky}, \binits{J.}}:
\batitle{Density matrix renormalisation group approach for many-body open
  quantum systems}.
\bjtitle{Phys. Rev. Lett.}
\bvolume{97},
\bfpage{110603}
(\byear{2006})
\end{barticle}
\endbibitem

\bibitem{hagen07_976}
\begin{barticle}
\bauthor{\bsnm{Hagen}, \binits{G.}},
\bauthor{\bsnm{Dean}, \binits{D.J.}},
\bauthor{\bsnm{{Hjorth-Jensen}}, \binits{M.}},
\bauthor{\bsnm{Papenbrock}, \binits{T.}}:
\batitle{Complex coupled-cluster approach to an \textit{ab initio} description
  of open quantum systems}.
\bjtitle{Phys. Lett. B}
\bvolume{656},
\bfpage{169}
(\byear{2007})
\end{barticle}
\endbibitem

\bibitem{carbonell14_1240}
\begin{barticle}
\bauthor{\bsnm{Carbonell}, \binits{J.}},
\bauthor{\bsnm{Deltuva}, \binits{A.}},
\bauthor{\bsnm{Fonseca}, \binits{A.C.}},
\bauthor{\bsnm{Lazauskas}, \binits{R.}}:
\batitle{Bound state techniques to solve the multiparticle scattering problem}.
\bjtitle{Prog. Part. Nucl. Phys.}
\bvolume{74},
\bfpage{55}
(\byear{2014})
\end{barticle}
\endbibitem

\bibitem{jaganathen14_988}
\begin{barticle}
\bauthor{\bsnm{Jaganathen}, \binits{Y.}},
\bauthor{\bsnm{Michel}, \binits{N.}},
\bauthor{\bsnm{P{\l}oszajczak}, \binits{M.}}:
\batitle{Gamow shell model description of proton scattering on ${
  {}^{18}\text{Ne} }$}.
\bjtitle{Phys. Rev. C}
\bvolume{89},
\bfpage{034624}
(\byear{2014})
\end{barticle}
\endbibitem

\bibitem{fossez15_1119}
\begin{barticle}
\bauthor{\bsnm{Fossez}, \binits{K.}},
\bauthor{\bsnm{Michel}, \binits{N.}},
\bauthor{\bsnm{P{\l}oszajczak}, \binits{M.}},
\bauthor{\bsnm{Jaganathen}, \binits{Y.}},
\bauthor{\bsnm{{Id Betan}}, \binits{R.M.}}:
\batitle{Description of the proton and neutron radiative capture reactions in
  the gamow shell model}.
\bjtitle{Phys. Rev. C}
\bvolume{91},
\bfpage{034609}
(\byear{2015})
\end{barticle}
\endbibitem

\bibitem{ono92_2334}
\begin{barticle}
\bauthor{\bsnm{Ono}, \binits{A.}},
\bauthor{\bsnm{Horiuchi}, \binits{H.}},
\bauthor{\bsnm{Maruyama}, \binits{T.}},
\bauthor{\bsnm{Ohnishi}, \binits{A.}}:
\batitle{Fragment formation studied with antisymmetrized version of molecular
  dynamics with two-nucleon collisions}.
\bjtitle{Phys. Rev. Lett.}
\bvolume{68},
\bfpage{2898}
(\byear{1992})
\end{barticle}
\endbibitem

\bibitem{enyo01_2335}
\begin{barticle}
\bauthor{\bsnm{{Kanada-En'yo}}, \binits{Y.}},
\bauthor{\bsnm{Horiuchi}, \binits{H.}}:
\batitle{Structure of light unstable nuclei studied with antisymmetrized
  molecular dynamics}.
\bjtitle{Prog. Theor. Phys. Suppl.}
\bvolume{142},
\bfpage{205}
(\byear{2001})
\end{barticle}
\endbibitem

\bibitem{feldmeier00_2337}
\begin{barticle}
\bauthor{\bsnm{Feldmeier}, \binits{H.}},
\bauthor{\bsnm{Schnack}, \binits{J.}}:
\batitle{Molecular dynamics for fermions}.
\bjtitle{Rev. Mod. Phys.}
\bvolume{72},
\bfpage{655}
(\byear{2000})
\end{barticle}
\endbibitem

\bibitem{launey16_2403}
\begin{barticle}
\bauthor{\bsnm{Launey}, \binits{K.D.}},
\bauthor{\bsnm{Dytrych}, \binits{T.}},
\bauthor{\bsnm{Draayer}, \binits{J.P.}}:
\batitle{Symmetry-guided large-scale shell-model theory}.
\bjtitle{Prog. Part. Nucl. Phys.}
\bvolume{89},
\bfpage{101}
(\byear{2016})
\end{barticle}
\endbibitem

\bibitem{feshbach1958unified}
\begin{barticle}
\bauthor{\bsnm{Feshbach}, \binits{H.}}:
\batitle{Unified theory of nuclear reactions}.
\bjtitle{Ann. Phys.}
\bvolume{5},
\bfpage{390}
(\byear{1958})
\end{barticle}
\endbibitem

\bibitem{TN09}
\begin{bbook}
\bauthor{\bsnm{Thompson}, \binits{I.J.}},
\bauthor{\bsnm{Nunes}, \binits{F.M.}}:
\bbtitle{Nuclear Reactions for Astrophysics : Principles, Calculation and
  Applications of Low-energy Reactions}.
\bpublisher{Cambridge University Press},
\blocation{Cambridge}
(\byear{2009})
\end{bbook}
\endbibitem

\bibitem{BC12}
\begin{bchapter}
\bauthor{\bsnm{Baye}, \binits{D.}},
\bauthor{\bsnm{Capel}, \binits{P.}}:
\bctitle{Breakup reaction models for two- and three-cluster projectiles}.
In: \beditor{\bsnm{Beck}, \binits{C.}} (ed.)
\bbtitle{Clusters in Nuclei, Vol. 2}
vol. \bseriesno{848}.
\bpublisher{Springer},
\blocation{Berlin, Heidelberg}
(\byear{2012}).
\doiurl{10.1007/978-3-642-24707-1_3}
\end{bchapter}
\endbibitem

\bibitem{Hlophe:2022fdn}
\begin{botherref}
\oauthor{\bsnm{Hlophe}, \binits{L.}},
\oauthor{\bsnm{Kravvaris}, \binits{K.}},
\oauthor{\bsnm{Quaglioni}, \binits{S.}}:
Quantifying uncertainties due to irreducible three-body forces in
  deuteron-nucleus reactions.
arXiv:2208.10568
(2022)
\end{botherref}
\endbibitem

\bibitem{PhysRevC.74.044304}
\begin{barticle}
\bauthor{\bsnm{Theeten}, \binits{M.}},
\bauthor{\bsnm{Baye}, \binits{D.}},
\bauthor{\bsnm{Descouvemont}, \binits{P.}}:
\batitle{Comparison of local, semi-microscopic, and microscopic three-cluster
  models}.
\bjtitle{Phys. Rev. C}
\bvolume{74},
\bfpage{14}
(\byear{2006})
\end{barticle}
\endbibitem

\bibitem{PhysRevC.104.034614}
\begin{barticle}
\bauthor{\bsnm{Dinmore}, \binits{M.J.}},
\bauthor{\bsnm{Timofeyuk}, \binits{N.K.}},
\bauthor{\bsnm{Al-Khalili}, \binits{J.S.}}:
\batitle{Three-body optical potentials in ${ (d,p) }$ reactions and their
  influence on indirect study of stellar nucleosynthesis}.
\bjtitle{Phys. Rev. C}
\bvolume{104},
\bfpage{12}
(\byear{2021})
\end{barticle}
\endbibitem

\bibitem{PhysRevC.104.024612}
\begin{barticle}
\bauthor{\bsnm{Johnson}, \binits{R.C.}}:
\batitle{Three-body model of the ${ d+A }$ system in an antisymmetrized,
  translationally invariant many nucleon theory}.
\bjtitle{Phys. Rev. C}
\bvolume{104},
\bfpage{18}
(\byear{2021})
\end{barticle}
\endbibitem

\bibitem{AUSTERN1987125}
\begin{barticle}
\bauthor{\bsnm{Austern}, \binits{N.}},
\bauthor{\bsnm{Iseri}, \binits{Y.}},
\bauthor{\bsnm{Kamimura}, \binits{M.}},
\bauthor{\bsnm{Kawai}, \binits{M.}},
\bauthor{\bsnm{Rawitscher}, \binits{G.}},
\bauthor{\bsnm{Yahiro}, \binits{M.}}:
\batitle{Continuum-discretized coupled-channels calculations for three-body
  models of deuteron-nucleus reactions}.
\bjtitle{Phys. Rep.}
\bvolume{154},
\bfpage{04}
(\byear{1987})
\end{barticle}
\endbibitem

\bibitem{Yetal86}
\begin{barticle}
\bauthor{\bsnm{Yahiro}, \binits{M.}},
\bauthor{\bsnm{Iseri}, \binits{Y.}},
\bauthor{\bsnm{Kameyama}, \binits{H.}},
\bauthor{\bsnm{Kamimura}, \binits{M.}},
\bauthor{\bsnm{Kawai}, \binits{M.}}:
\batitle{Chapter {III}. {E}ffects of deuteron virtual breakup on deuteron
  elastic and inelastic scattering}.
\bjtitle{Prog. Theor. Phys. Supp.}
\bvolume{89},
\bfpage{32}--\blpage{8}
(\byear{1986})
\end{barticle}
\endbibitem

\bibitem{Kawai86}
\begin{barticle}
\bauthor{\bsnm{Kawai}, \binits{M.}}:
\batitle{Chapter {II}. {F}ormalism of the method of coupled discretized
  continuum channel}.
\bjtitle{Prog. Theor. Phys. Supp.}
\bvolume{89},
\bfpage{11}
(\byear{1986})
\end{barticle}
\endbibitem

\bibitem{PhysRevC.68.064609}
\begin{barticle}
\bauthor{\bsnm{Ogata}, \binits{K.}},
\bauthor{\bsnm{Yahiro}, \binits{M.}},
\bauthor{\bsnm{Iseri}, \binits{Y.}},
\bauthor{\bsnm{Matsumoto}, \binits{T.}},
\bauthor{\bsnm{Kamimura}, \binits{M.}}:
\batitle{New coupled-channel approach to nuclear and {C}oulomb breakup
  reactions}.
\bjtitle{Phys. Rev. C}
\bvolume{68},
\bfpage{7}
(\byear{2003})
\end{barticle}
\endbibitem

\bibitem{G59}
\begin{bchapter}
\bauthor{\bsnm{Glauber}, \binits{R.J.}}:
\bctitle{High energy collision theory}.
In: \beditor{\bsnm{Brittin}, \binits{W.E.}},
\beditor{\bsnm{Dunham}, \binits{L.G.}} (eds.)
\bbtitle{Lecture in Theoretical Physics}
vol. \bseriesno{1},
p. \bfpage{315}.
\bpublisher{Interscience},
\blocation{New York}
(\byear{1959})
\end{bchapter}
\endbibitem

\bibitem{Hansen03}
\begin{barticle}
\bauthor{\bsnm{Hansen}, \binits{P.G.}},
\bauthor{\bsnm{Tostevin}, \binits{J.A.}}:
\batitle{Direct reactions with exotic nuclei}.
\bjtitle{Annu. Rev. Nucl. Part. Sci.}
\bvolume{53},
\bfpage{61}
(\byear{2003})
\end{barticle}
\endbibitem

\bibitem{PhysRevLett.95.082502}
\begin{barticle}
\bauthor{\bsnm{Baye}, \binits{D.}},
\bauthor{\bsnm{Capel}, \binits{P.}},
\bauthor{\bsnm{Goldstein}, \binits{G.}}:
\batitle{Collisions of halo nuclei within a dynamical eikonal approximation}.
\bjtitle{Phys. Rev. Lett.}
\bvolume{95},
\bfpage{4}
(\byear{2005})
\end{barticle}
\endbibitem

\bibitem{JOHNSON197456}
\begin{barticle}
\bauthor{\bsnm{Johnson}, \binits{R.C.}},
\bauthor{\bsnm{Tandy}, \binits{P.C.}}:
\batitle{An approximate three-body theory of deuteron stripping}.
\bjtitle{Nucl. Phys. A}
\bvolume{235},
\bfpage{4}
(\byear{1974})
\end{barticle}
\endbibitem

\bibitem{TIMOFEYUK2020103738}
\begin{barticle}
\bauthor{\bsnm{Timofeyuk}, \binits{N.K.}},
\bauthor{\bsnm{Johnson}, \binits{R.C.}}:
\batitle{Theory of deuteron stripping and pick-up reactions for nuclear
  structure studies}.
\bjtitle{Prog. Part. Nucl. Phys.}
\bvolume{111},
\bfpage{103738}
(\byear{2020})
\end{barticle}
\endbibitem

\bibitem{PhysRevC.101.064611}
\begin{barticle}
\bauthor{\bsnm{Descouvemont}, \binits{P.}}:
\batitle{Low-energy ${ ^{11}\text{Li}+p }$ and ${ ^{11}\text{Li}+d }$
  scattering in a multicluster model}.
\bjtitle{Phys. Rev. C}
\bvolume{101},
\bfpage{11}
(\byear{2020})
\end{barticle}
\endbibitem

\bibitem{PhysRevC.78.054602}
\begin{barticle}
\bauthor{\bsnm{Capel}, \binits{P.}},
\bauthor{\bsnm{Baye}, \binits{D.}},
\bauthor{\bsnm{Suzuki}, \binits{Y.}}:
\batitle{Coulomb-corrected eikonal description of the breakup of halo nuclei}.
\bjtitle{Phys. Rev. C}
\bvolume{78},
\bfpage{10}
(\byear{2008})
\end{barticle}
\endbibitem

\bibitem{CooperPairPotel}
\begin{barticle}
\bauthor{\bsnm{Potel}, \binits{G.}},
\bauthor{\bsnm{Idini}, \binits{A.}},
\bauthor{\bsnm{Barranco}, \binits{F.}},
\bauthor{\bsnm{Vigezzi}, \binits{E.}},
\bauthor{\bsnm{Broglia}, \binits{R.A.}}:
\batitle{Cooper pair transfer in nucl}.
\bjtitle{Rep. Prog. Phys.}
\bvolume{76},
\bfpage{106301}
(\byear{2013})
\end{barticle}
\endbibitem

\bibitem{PhysRevLett.109.232502}
\begin{barticle}
\bauthor{\bsnm{Moro}, \binits{A.M.}},
\bauthor{\bsnm{Lay}, \binits{J.A.}}:
\batitle{Interplay between valence and core excitation mechanisms in the
  breakup of halo nuclei}.
\bjtitle{Phys. Rev. Lett.}
\bvolume{109},
\bfpage{5}
(\byear{2012})
\end{barticle}
\endbibitem

\bibitem{koning03}
\begin{barticle}
\bauthor{\bsnm{Koning}, \binits{A.J.}},
\bauthor{\bsnm{Delaroche}, \binits{J.P.}}:
\batitle{Local and global nucleon optical models from 1 {keV} to 200 {MeV}}.
\bjtitle{Nucl. Phys.}
\bvolume{713},
\bfpage{231}
(\byear{2003})
\end{barticle}
\endbibitem

\bibitem{becchetti69}
\begin{barticle}
\bauthor{\bsnm{Becchetti}, \binits{F.D.}},
\bauthor{\bsnm{Greenlees}, \binits{G.W.}}:
\batitle{Nucleon-nucleus optical-model parameters, ${ A>40 }$, ${ E<50 }$
  {MeV}}.
\bjtitle{Phys. Rev.}
\bvolume{182},
\bfpage{1190}
(\byear{1969})
\end{barticle}
\endbibitem

\bibitem{bertulani02_869}
\begin{barticle}
\bauthor{\bsnm{Bertulani}, \binits{C.A.}},
\bauthor{\bsnm{Hammer}, \binits{H.-W.}},
\bauthor{\bsnm{{van Kolck}}, \binits{U.}}:
\batitle{Effective field theory for halo nuclei: shallow ${ p }$-wave states}.
\bjtitle{Nucl. Phys. A}
\bvolume{712},
\bfpage{37}
(\byear{2002})
\end{barticle}
\endbibitem

\bibitem{ryberg14_997}
\begin{barticle}
\bauthor{\bsnm{Ryberg}, \binits{E.}},
\bauthor{\bsnm{Forss\'en}, \binits{C.}},
\bauthor{\bsnm{Hammer}, \binits{H.-W.}},
\bauthor{\bsnm{Platter}, \binits{L.}}:
\batitle{Effective field theory for proton halo nuclei}.
\bjtitle{Phys. Rev. C}
\bvolume{89},
\bfpage{014325}
(\byear{2014})
\end{barticle}
\endbibitem

\bibitem{ji14_1101}
\begin{barticle}
\bauthor{\bsnm{Ji}, \binits{C.}},
\bauthor{\bsnm{Elster}, \binits{C.}},
\bauthor{\bsnm{Phillips}, \binits{D.R.}}:
\batitle{${ {}^{6}\text{He} }$ nucleus in halo effective field theory}.
\bjtitle{Phys. Rev. C}
\bvolume{90},
\bfpage{044004}
(\byear{2014})
\end{barticle}
\endbibitem

\bibitem{Hammer2017}
\begin{barticle}
\bauthor{\bsnm{Hammer}, \binits{H.-W.}},
\bauthor{\bsnm{Ji}, \binits{C.}},
\bauthor{\bsnm{Phillips}, \binits{D.R.}}:
\batitle{Effective field theory description of halo nuclei}.
\bjtitle{J. Phys. G: Nucl. Part. Phys.}
\bvolume{44},
\bfpage{103002}
(\byear{2017})
\end{barticle}
\endbibitem

\bibitem{Capel:2018kss}
\begin{barticle}
\bauthor{\bsnm{Capel}, \binits{P.}},
\bauthor{\bsnm{Phillips}, \binits{D.R.}},
\bauthor{\bsnm{Hammer}, \binits{H.-W.}}:
\batitle{Dissecting reaction calculations using halo effective field theory and
  \textit{ab initio} input}.
\bjtitle{Phys. Rev. C}
\bvolume{98},
\bfpage{034610}
(\byear{2018})
\end{barticle}
\endbibitem

\bibitem{papenbrock11_1310}
\begin{barticle}
\bauthor{\bsnm{Papenbrock}, \binits{T.}}:
\batitle{Effective theory for deformed nuclei}.
\bjtitle{Nucl. Phys. A}
\bvolume{852},
\bfpage{36}
(\byear{2011})
\end{barticle}
\endbibitem

\bibitem{papenbrock14_1312}
\begin{barticle}
\bauthor{\bsnm{Papenbrock}, \binits{T.}},
\bauthor{\bsnm{Weidenm\"uller}, \binits{H.A.}}:
\batitle{Effective field theory for finite systems with spontaneously broken
  symmetry}.
\bjtitle{Phys. Rev. C}
\bvolume{89},
\bfpage{014334}
(\byear{2014})
\end{barticle}
\endbibitem

\bibitem{papenbrock15_1311}
\begin{barticle}
\bauthor{\bsnm{Papenbrock}, \binits{T.}},
\bauthor{\bsnm{Weidenm\"uller}, \binits{H.A.}}:
\batitle{Effective field theory of emergent symmetry breaking in deformed
  atomic nuclei}.
\bjtitle{J. Phys. G: Nucl. Part. Phys.}
\bvolume{42},
\bfpage{106103}
(\byear{2015})
\end{barticle}
\endbibitem

\bibitem{WhitePaperOptPot22}
\begin{barticle}
\bauthor{\bsnm{Hebborn}, \binits{C.}},
\bauthor{\bsnm{Nunes}, \binits{F.M.}},
\bauthor{\bsnm{Potel}, \binits{G.}},
\bauthor{\bsnm{Dickhoff}, \binits{W.H.}},
\bauthor{\bsnm{Holt}, \binits{J.W.}},
\bauthor{\bsnm{Atkinson}, \binits{M.C.}},
\bauthor{\bsnm{Baker}, \binits{R.B.}},
\bauthor{\bsnm{Barbieri}, \binits{C.}},
\bauthor{\bsnm{Blanchon}, \binits{G.}},
\bauthor{\bsnm{Burrows}, \binits{M.}},
\bauthor{\bsnm{Capote}, \binits{R.}},
\bauthor{\bsnm{Danielewicz}, \binits{P.}},
\bauthor{\bsnm{Dupuis}, \binits{M.}},
\bauthor{\bsnm{Elster}, \binits{C.}},
\bauthor{\bsnm{Escher}, \binits{J.E.}},
\bauthor{\bsnm{Hlophe}, \binits{L.}},
\bauthor{\bsnm{Idini}, \binits{A.}},
\bauthor{\bsnm{Jayatissa}, \binits{H.}},
\bauthor{\bsnm{Kay}, \binits{B.P.}},
\bauthor{\bsnm{Kravvaris}, \binits{K.}},
\bauthor{\bsnm{Manfredi}, \binits{J.J.}},
\bauthor{\bsnm{Mercenne}, \binits{A.}},
\bauthor{\bsnm{Morillon}, \binits{B.}},
\bauthor{\bsnm{Perdikakis}, \binits{G.}},
\bauthor{\bsnm{Pruitt}, \binits{C.D.}},
\bauthor{\bsnm{Sargsyan}, \binits{G.H.}},
\bauthor{\bsnm{Thompson}, \binits{I.J.}},
\bauthor{\bsnm{Vorabbi}, \binits{M.}},
\bauthor{\bsnm{Whitehead}, \binits{T.R.}}:
\batitle{{Optical potentials for the rare-isotope beam era}}.
\bjtitle{J. Phys. G}
\bvolume{50}(\bissue{6}),
\bfpage{060501}
(\byear{2023})
{\href{https://arxiv.org/abs/2210.07293}{{arXiv:2210.07293}}}
{[nucl-th]}.
\doiurl{10.1088/1361-6471/acc348}
\end{barticle}
\endbibitem

\bibitem{volya03_858}
\begin{barticle}
\bauthor{\bsnm{Volya}, \binits{A.}},
\bauthor{\bsnm{Zelevinsky}, \binits{V.}}:
\batitle{Exploring quantum dynamics in an open many-body system: transition to
  superradiance}.
\bjtitle{J. Opt. B: Quantum Semiclass. Opt.}
\bvolume{5},
\bfpage{450}
(\byear{2003})
\end{barticle}
\endbibitem

\bibitem{garrido13_1171}
\begin{barticle}
\bauthor{\bsnm{Garrido}, \binits{E.}},
\bauthor{\bsnm{Jensen}, \binits{A.S.}},
\bauthor{\bsnm{Fedorov}, \binits{D.V.}}:
\batitle{Rotational bands in the continuum illustrated by ${ {}^{8}\text{Be} }$
  results}.
\bjtitle{Phys. Rev. C}
\bvolume{88},
\bfpage{024001}
(\byear{2013})
\end{barticle}
\endbibitem

\bibitem{DytrychLDRWRBB20}
\begin{barticle}
\bauthor{\bsnm{Dytrych}, \binits{T.}},
\bauthor{\bsnm{Launey}, \binits{K.D.}},
\bauthor{\bsnm{Draayer}, \binits{J.P.}},
\bauthor{\bsnm{Rowe}, \binits{D.J.}},
\bauthor{\bsnm{Wood}, \binits{J.L.}},
\bauthor{\bsnm{Rosensteel}, \binits{G.}},
\bauthor{\bsnm{Bahri}, \binits{C.}},
\bauthor{\bsnm{Langr}, \binits{D.}},
\bauthor{\bsnm{Baker}, \binits{R.B.}}:
\batitle{Physics of nuclei: Key role of an emergent symmetry}.
\bjtitle{Phys. Rev. Lett.}
\bvolume{124},
\bfpage{6}
(\byear{2020})
\end{barticle}
\endbibitem

\bibitem{Mottelson_NP}
\begin{botherref}
\oauthor{\bsnm{Mottelson}, \binits{B.R.}}:
Nobel lectures.
Physics 1971-1980
(1992)
\end{botherref}
\endbibitem

\bibitem{Elliott58}
\begin{barticle}
\bauthor{\bsnm{Elliott}, \binits{J.P.}}:
\batitle{Collective motion in the nuclear shell model. i. classification
  schemes for states of mixed configurations}.
\bjtitle{Proc. Roy. Soc. A}
\bvolume{245},
\bfpage{128}
(\byear{1958})
\end{barticle}
\endbibitem

\bibitem{HeydeW11}
\begin{barticle}
\bauthor{\bsnm{Heyde}, \binits{K.}},
\bauthor{\bsnm{Wood}, \binits{J.L.}}:
\batitle{Shape coexistence in atomic nuclei}.
\bjtitle{Rev. Mod. Phys.}
\bvolume{83},
\bfpage{1467}
(\byear{2011})
\end{barticle}
\endbibitem

\bibitem{Wood16}
\begin{botherref}
\oauthor{\bsnm{Wood}, \binits{J.L.}}:
1.
Nuclear Collectivity — Its Emergent Nature Viewed from Phenomenology and
  Spectroscopy,
p. 3
\end{botherref}
\endbibitem

\bibitem{RoweW2010book}
\begin{bbook}
\bauthor{\bsnm{Rowe}, \binits{D.J.}},
\bauthor{\bsnm{Wood}, \binits{J.L.}}:
\bbtitle{Fundamentals of Nuclear Models: Foundational Models}.
\bpublisher{World Scientific},
\blocation{Singapore}
(\byear{2010})
\end{bbook}
\endbibitem

\bibitem{physics4030048}
\begin{barticle}
\bauthor{\bsnm{Stuchbery}, \binits{A.E.}},
\bauthor{\bsnm{Wood}, \binits{J.L.}}:
\batitle{To shell model, or not to shell model, that is the question}.
\bjtitle{Physics}
\bvolume{4},
\bfpage{773}
(\byear{2022})
\end{barticle}
\endbibitem

\bibitem{weinberg1990}
\begin{barticle}
\bauthor{\bsnm{Weinberg}, \binits{S.}}:
\batitle{Nuclear forces from chiral lagrangians}.
\bjtitle{Phys. Lett. B}
\bvolume{251},
\bfpage{92}
(\byear{1990})
\end{barticle}
\endbibitem

\bibitem{ordonez1992}
\begin{barticle}
\bauthor{\bsnm{Ordóñez}, \binits{C.}},
\bauthor{\bsnm{{van Kolck}}, \binits{U.}}:
\batitle{Chiral lagrangians and nuclear forces}.
\bjtitle{Phys. Lett. B}
\bvolume{291},
\bfpage{64}
(\byear{1992})
\end{barticle}
\endbibitem

\bibitem{papenbrock2016}
\begin{barticle}
\bauthor{\bsnm{Papenbrock}, \binits{T.}},
\bauthor{\bsnm{Weidenmüller}, \binits{H.A.}}:
\batitle{Effective field theory for deformed atomic nuclei}.
\bjtitle{Phys. Scr.}
\bvolume{91},
\bfpage{053004}
(\byear{2016})
\end{barticle}
\endbibitem

\bibitem{luo21_2394}
\begin{barticle}
\bauthor{\bsnm{{Yu-Xuan Luo}}},
\bauthor{\bsnm{Fossez}, \binits{K.}},
\bauthor{\bsnm{{Quan Liu}}},
\bauthor{\bsnm{{Jian-You Guo}}}:
\batitle{Role of quadrupole deformation and continuum effects in the "island of
  inversion" nuclei ${ {}^{28,29,31}\text{F} }$}.
\bjtitle{Phys. Rev. C}
\bvolume{104},
\bfpage{014307}
(\byear{2021})
\end{barticle}
\endbibitem

\bibitem{eisenberg1976}
\begin{botherref}
\oauthor{\bsnm{Eisenberg}, \binits{J.M.}},
\oauthor{\bsnm{Greiner}, \binits{W.}}:
Nuclear Theory. Excitation Mechanisms of the Nucleus,
Netherlands
(1976)
\end{botherref}
\endbibitem

\bibitem{stroberg2017}
\begin{barticle}
\bauthor{\bsnm{Stroberg}, \binits{S.R.}},
\bauthor{\bsnm{Calci}, \binits{A.}},
\bauthor{\bsnm{Hergert}, \binits{H.}},
\bauthor{\bsnm{Holt}, \binits{J.D.}},
\bauthor{\bsnm{Bogner}, \binits{S.K.}},
\bauthor{\bsnm{Roth}, \binits{R.}},
\bauthor{\bsnm{Schwenk}, \binits{A.}}:
\batitle{Nucleus-dependent valence-space approach to nuclear structure}.
\bjtitle{Phys. Rev. Lett.}
\bvolume{118},
\bfpage{6}
(\byear{2017})
\end{barticle}
\endbibitem

\bibitem{hagen2014}
\begin{barticle}
\bauthor{\bsnm{Hagen}, \binits{G.}},
\bauthor{\bsnm{Papenbrock}, \binits{T.}},
\bauthor{\bsnm{Hjorth-Jensen}, \binits{M.}},
\bauthor{\bsnm{Dean}, \binits{D.J.}}:
\batitle{Coupled-cluster computations of atomic nuclei}.
\bjtitle{Reports on Progress in Physic}
\bvolume{77},
\bfpage{096302}
(\byear{2014})
\end{barticle}
\endbibitem

\bibitem{stroberg2022}
\begin{barticle}
\bauthor{\bsnm{Stroberg}, \binits{S.R.}},
\bauthor{\bsnm{Henderson}, \binits{J.}},
\bauthor{\bsnm{Hackman}, \binits{G.}},
\bauthor{\bsnm{Ruotsalainen}, \binits{P.}},
\bauthor{\bsnm{Hagen}, \binits{G.}},
\bauthor{\bsnm{Holt}, \binits{J.D.}}:
\batitle{Systematics of ${ E2 }$ strength in the ${ sd }$ shell with the
  valence-space in-medium similarity renormalization group}.
\bjtitle{Phys. Rev. C}
\bvolume{105},
\bfpage{10}
(\byear{2022})
\end{barticle}
\endbibitem

\bibitem{cordoba2016}
\begin{barticle}
\bauthor{\bsnm{Ramos-Cordoba}, \binits{E.}},
\bauthor{\bsnm{Salvador}, \binits{P.}},
\bauthor{\bsnm{Matito}, \binits{E.}}:
\batitle{Separation of dynamic and nondynamic correlation}.
\bjtitle{Phys. Chem. Chem. Phys.}
\bvolume{18},
\bfpage{4023}
(\byear{2016})
\end{barticle}
\endbibitem

\bibitem{yao2018}
\begin{barticle}
\bauthor{\bsnm{Yao}, \binits{J.M.}},
\bauthor{\bsnm{Engel}, \binits{J.}},
\bauthor{\bsnm{Wang}, \binits{L.J.}},
\bauthor{\bsnm{Jiao}, \binits{C.F.}},
\bauthor{\bsnm{Hergert}, \binits{H.}}:
\batitle{Generator-coordinate reference states for spectra and ${
  0\nu\beta\beta }$ decay in the in-medium similarity renormalization group}.
\bjtitle{Phys. Rev. C}
\bvolume{98},
\bfpage{11}
(\byear{2018})
\end{barticle}
\endbibitem

\bibitem{hagen2022}
\begin{barticle}
\bauthor{\bsnm{Hagen}, \binits{G.}},
\bauthor{\bsnm{Novario}, \binits{S.J.}},
\bauthor{\bsnm{Sun}, \binits{Z.H.}},
\bauthor{\bsnm{Papenbrock}, \binits{T.}},
\bauthor{\bsnm{Jansen}, \binits{G.R.}},
\bauthor{\bsnm{Lietz}, \binits{J.G.}},
\bauthor{\bsnm{Duguet}, \binits{T.}},
\bauthor{\bsnm{Tichai}, \binits{A.}}:
\batitle{Angular-momentum projection in coupled-cluster theory: Structure of ${
  ^{34}\text{Mg} }$}.
\bjtitle{Phys. Rev. C}
\bvolume{105},
\bfpage{23}
(\byear{2022})
\end{barticle}
\endbibitem

\bibitem{LauneyMD_ARNPS21}
\begin{barticle}
\bauthor{\bsnm{Launey}, \binits{K.D.}},
\bauthor{\bsnm{Mercenne}, \binits{A.}},
\bauthor{\bsnm{Dytrych}, \binits{T.}}:
\batitle{Nuclear dynamics and reactions in the ab initio symmetry-adapted
  framework}.
\bjtitle{Annu. Rev. Nucl. Part. Sci.}
\bvolume{71},
\bfpage{253}
(\byear{2021})
\end{barticle}
\endbibitem

\bibitem{Ruotsalainen19}
\begin{barticle}
\bauthor{\bsnm{Ruotsalainen}, \binits{P.}},
\bauthor{\bsnm{Henderson}, \binits{J.}},
\bauthor{\bsnm{Hackman}, \binits{G.}},
\bauthor{\bsnm{Sargsyan}, \binits{G.H.}},
\bauthor{\bsnm{Launey}, \binits{K.D.}},
\bauthor{\bsnm{Saxena}, \binits{A.}},
\bauthor{\bsnm{Srivastava}, \binits{P.C.}},
\bauthor{\bsnm{Stroberg}, \binits{S.R.}},
\bauthor{\bsnm{Grahn}, \binits{T.}},
\bauthor{\bsnm{Pakarinen}, \binits{J.}},
\bauthor{\bsnm{Ball}, \binits{G.C.}},
\bauthor{\bsnm{Julin}, \binits{R.}},
\bauthor{\bsnm{Greenlees}, \binits{P.T.}},
\bauthor{\bsnm{Smallcombe}, \binits{J.}},
\bauthor{\bsnm{Andreoiu}, \binits{C.}},
\bauthor{\bsnm{Bernier}, \binits{N.}},
\bauthor{\bsnm{Bowry}, \binits{M.}},
\bauthor{\bsnm{Buckner}, \binits{M.}},
\bauthor{\bsnm{Caballero-Folch}, \binits{R.}},
\bauthor{\bsnm{Chester}, \binits{A.}},
\bauthor{\bsnm{Cruz}, \binits{S.}},
\bauthor{\bsnm{Evitts}, \binits{L.J.}},
\bauthor{\bsnm{Frederick}, \binits{R.}},
\bauthor{\bsnm{Garnsworthy}, \binits{A.B.}},
\bauthor{\bsnm{Holl}, \binits{M.}},
\bauthor{\bsnm{Kurkjian}, \binits{A.}},
\bauthor{\bsnm{Kisliuk}, \binits{D.}},
\bauthor{\bsnm{Leach}, \binits{K.G.}},
\bauthor{\bsnm{McGee}, \binits{E.}},
\bauthor{\bsnm{Measures}, \binits{J.}},
\bauthor{\bsnm{M\"ucher}, \binits{D.}},
\bauthor{\bsnm{Park}, \binits{J.}},
\bauthor{\bsnm{Sarazin}, \binits{F.}},
\bauthor{\bsnm{Smith}, \binits{J.K.}},
\bauthor{\bsnm{Southall}, \binits{D.}},
\bauthor{\bsnm{Starosta}, \binits{K.}},
\bauthor{\bsnm{Svensson}, \binits{C.E.}},
\bauthor{\bsnm{Whitmore}, \binits{K.}},
\bauthor{\bsnm{Williams}, \binits{M.}},
\bauthor{\bsnm{Wu}, \binits{C.Y.}}:
\batitle{Isospin symmetry in ${ B(E2) }$ values: {C}oulomb excitation study of
  ${ ^{21}\text{Mg} }$}.
\bjtitle{Phys. Rev. C}
\bvolume{99},
\bfpage{7}
(\byear{2019})
\end{barticle}
\endbibitem

\bibitem{PhysRevLett.128.202503}
\begin{barticle}
\bauthor{\bsnm{Sargsyan}, \binits{G.H.}},
\bauthor{\bsnm{Launey}, \binits{K.D.}},
\bauthor{\bsnm{Burkey}, \binits{M.T.}},
\bauthor{\bsnm{Gallant}, \binits{A.T.}},
\bauthor{\bsnm{Scielzo}, \binits{N.D.}},
\bauthor{\bsnm{Savard}, \binits{G.}},
\bauthor{\bsnm{Mercenne}, \binits{A.}},
\bauthor{\bsnm{Dytrych}, \binits{T.}},
\bauthor{\bsnm{Langr}, \binits{D.}},
\bauthor{\bsnm{Varriano}, \binits{L.}},
\bauthor{\bsnm{Longfellow}, \binits{B.}},
\bauthor{\bsnm{Hirsh}, \binits{T.Y.}},
\bauthor{\bsnm{Draayer}, \binits{J.P.}}:
\batitle{Impact of clustering on the ${ ^{8}\text{Li} }$ ${ \beta }$ decay and
  recoil form factors}.
\bjtitle{Phys. Rev. Lett.}
\bvolume{128},
\bfpage{7}
(\byear{2022})
\end{barticle}
\endbibitem

\bibitem{DreyfussLESBDD20}
\begin{barticle}
\bauthor{\bsnm{Dreyfuss}, \binits{A.C.}},
\bauthor{\bsnm{Launey}, \binits{K.D.}},
\bauthor{\bsnm{Escher}, \binits{J.E.}},
\bauthor{\bsnm{Sargsyan}, \binits{G.H.}},
\bauthor{\bsnm{Baker}, \binits{R.B.}},
\bauthor{\bsnm{Dytrych}, \binits{T.}},
\bauthor{\bsnm{Draayer}, \binits{J.P.}}:
\batitle{Clustering and ${ \alpha }$-capture reaction rate from \textit{ab
  initio} symmetry-adapted descriptions of ${ ^{20}\text{Ne} }$}.
\bjtitle{Phys. Rev. C}
\bvolume{102},
\bfpage{14}
(\byear{2020})
\end{barticle}
\endbibitem

\bibitem{DraayerLPL89}
\begin{barticle}
\bauthor{\bsnm{Draayer}, \binits{J.P.}},
\bauthor{\bsnm{Leschber}, \binits{Y.}},
\bauthor{\bsnm{Park}, \binits{S.C.}},
\bauthor{\bsnm{Lopez}, \binits{R.}}:
\batitle{Representations of {U}(3) in {U}(${ N }$)}.
\bjtitle{Comput. Phys. Commun.}
\bvolume{56},
\bfpage{279}
(\byear{1989})
\end{barticle}
\endbibitem

\bibitem{DraayerW83}
\begin{barticle}
\bauthor{\bsnm{Draayer}, \binits{J.R.}},
\bauthor{\bsnm{Weeks}, \binits{K.J.}}:
\batitle{Shell-model description of the low-energy structure of strongly
  deformed nuclei}.
\bjtitle{Phys. Rev. Lett.}
\bvolume{51},
\bfpage{1422}
(\byear{198})
\end{barticle}
\endbibitem

\bibitem{RosensteelR77}
\begin{barticle}
\bauthor{\bsnm{Rosensteel}, \binits{G.}},
\bauthor{\bsnm{Rowe}, \binits{D.J.}}:
\batitle{"nuclear sp(3,r) model"}.
\bjtitle{Phys. Rev. Lett.}
\bvolume{38},
\bfpage{10}
(\byear{1977})
\end{barticle}
\endbibitem

\bibitem{Rowe_book16}
\begin{bchapter}
\bauthor{\bsnm{Rowe}, \binits{D.J.}}:
\bctitle{The emergence and use of symmetry in the many-nucleon model of atomic
  nuclei}.
In: \bbtitle{Emergent Phenomena in Atomic Nuclei from Large-scale Modeling: a
  Symmetry-guided Perspective},
p. \bfpage{65}.
\bpublisher{World Scientific Publishing Co.},
\blocation{\,}
(\byear{2017})
\end{bchapter}
\endbibitem

\bibitem{johnson09_1479}
\begin{barticle}
\bauthor{\bsnm{Johnson}, \binits{E.D.}},
\bauthor{\bsnm{Rogachev}, \binits{G.V.}},
\bauthor{\bsnm{Goldberg}, \binits{V.Z.}},
\bauthor{\bsnm{Brown}, \binits{S.}},
\bauthor{\bsnm{Robson}, \binits{D.}},
\bauthor{\bsnm{Crisp}, \binits{A.M.}},
\bauthor{\bsnm{Cottle}, \binits{P.D.}},
\bauthor{\bsnm{Fu}, \binits{C.}},
\bauthor{\bsnm{Giles}, \binits{J.}},
\bauthor{\bsnm{Green}, \binits{B.W.}},
\bauthor{\bsnm{Kemper}, \binits{K.W.}},
\bauthor{\bsnm{Lee}, \binits{K.}},
\bauthor{\bsnm{Roeder}, \binits{B.T.}},
\bauthor{\bsnm{Tribble}, \binits{R.E.}}:
\batitle{Extreme ${ \alpha }$-clustering in the ${ {}^{18}\text{O} }$ nucleus}.
\bjtitle{Eur. Phys. J. A}
\bvolume{42},
\bfpage{135}
(\byear{2009})
\end{barticle}
\endbibitem

\bibitem{Kubono1995}
\begin{bchapter}
\bauthor{\bsnm{Kubono}, \binits{S.}}:
\bctitle{Nuclear clustering aspects in astrophysics}.
In: \beditor{\bsnm{Anagnostatos}, \binits{G.S.}},
\beditor{\bsnm{{von Oertzen}}, \binits{W.}} (eds.)
\bbtitle{Atomic and Nuclear Clusters},
p. \bfpage{73}.
\bpublisher{Springer},
\blocation{Berlin, Heidelberg}
(\byear{1995})
\end{bchapter}
\endbibitem

\bibitem{Descouvemont2008}
\begin{barticle}
\bauthor{\bsnm{Descouvemont}, \binits{P.}}:
\batitle{Cluster models in nuclear astrophysics}.
\bjtitle{J. Phys. G: Nucl. Part. Phys.}
\bvolume{35},
\bfpage{014006}
(\byear{2007})
\end{barticle}
\endbibitem

\bibitem{Shen2020}
\begin{barticle}
\bauthor{\bsnm{Shen}, \binits{Y.P.}},
\bauthor{\bsnm{Guo}, \binits{B.}},
\bauthor{\bsnm{Liu}, \binits{W.P.}}:
\batitle{Alpha-cluster transfer reactions: A tool for understanding stellar
  helium burning}.
\bjtitle{Prog. Part. Nucl. Phys.}
\bvolume{119},
\bfpage{103857}
(\byear{2021})
\end{barticle}
\endbibitem

\bibitem{endt77_2579}
\begin{barticle}
\bauthor{\bsnm{Endt}, \binits{P.M.}}:
\batitle{Spectroscopic factors for single-nucleon transfer in the ${ A = 21-44
  }$ region}.
\bjtitle{Atom. Data Nucl. Data Tab.}
\bvolume{19},
\bfpage{23}
(\byear{1977})
\end{barticle}
\endbibitem

\bibitem{wiringa13_1182}
\begin{barticle}
\bauthor{\bsnm{Wiringa}, \binits{R.B.}},
\bauthor{\bsnm{Pastore}, \binits{S.}},
\bauthor{\bsnm{Pieper}, \binits{S.C.}},
\bauthor{\bsnm{Miller}, \binits{G.A.}}:
\batitle{Charge-symmetry breaking forces and isospin mixing in ${
  {}^{8}\text{Be} }$}.
\bjtitle{Phys. Rev. C}
\bvolume{88},
\bfpage{044333}
(\byear{2013})
\end{barticle}
\endbibitem

\bibitem{kanadaenyo12_2568}
\begin{barticle}
\bauthor{\bsnm{{Kanada-En'yo}}, \binits{Y.}},
\bauthor{\bsnm{Kimura}, \binits{M.}},
\bauthor{\bsnm{Ono}, \binits{A.}}:
\batitle{Antisymmetrized molecular dynamics and its applications to cluster
  phenomena}.
\bjtitle{Prog. Theor. Exp. Phys.}
\bvolume{2012},
\bfpage{01}--\blpage{202}
(\byear{2012})
\end{barticle}
\endbibitem

\bibitem{elhatisari15_2324}
\begin{barticle}
\bauthor{\bsnm{Elhatisari}, \binits{S.}},
\bauthor{\bsnm{Lee}, \binits{D.}},
\bauthor{\bsnm{Rupak}, \binits{G.}},
\bauthor{\bsnm{Epelbaum}, \binits{E.}},
\bauthor{\bsnm{Krebs}, \binits{H.}},
\bauthor{\bsnm{L\"ahde}, \binits{T.A.}},
\bauthor{\bsnm{Luu}, \binits{T.}},
\bauthor{\bsnm{Mei{\ss}ner}, \binits{U.}}:
\batitle{\textit{Ab initio} alpha-alpha scattering}.
\bjtitle{Nature}
\bvolume{528},
\bfpage{111}
(\byear{2015})
\end{barticle}
\endbibitem

\bibitem{elhatisari16_2311}
\begin{barticle}
\bauthor{\bsnm{Elhatisari}, \binits{S.}},
\bauthor{\bsnm{Li}, \binits{N.}},
\bauthor{\bsnm{Rokash}, \binits{A.}},
\bauthor{\bsnm{Alarc\'on}, \binits{J.M.}},
\bauthor{\bsnm{Du}, \binits{D.}},
\bauthor{\bsnm{Klein}, \binits{N.}},
\bauthor{\bsnm{{Bing-nan Lu}}},
\bauthor{\bsnm{Mei{\ss}ner}, \binits{U.}},
\bauthor{\bsnm{Epelbaum}, \binits{E.}},
\bauthor{\bsnm{Krebs}, \binits{H.}},
\bauthor{\bsnm{L\"ahde}, \binits{T.A.}},
\bauthor{\bsnm{Lee}, \binits{D.}},
\bauthor{\bsnm{Rupak}, \binits{G.}}:
\batitle{Nuclear binding near a quantum phase transition}.
\bjtitle{Phys. Rev. Lett.}
\bvolume{117},
\bfpage{132501}
(\byear{2016})
\end{barticle}
\endbibitem

\bibitem{Kravvaris2017}
\begin{barticle}
\bauthor{\bsnm{Kravvaris}, \binits{K.}},
\bauthor{\bsnm{Volya}, \binits{A.}}:
\batitle{Quest for superradiance in atomic nuclei}.
\bjtitle{AIP Conf. Proc.}
\bvolume{1912},
\bfpage{020010}
(\byear{2017})
\end{barticle}
\endbibitem

\bibitem{Elhatisari2017}
\begin{barticle}
\bauthor{\bsnm{Elhatisari}, \binits{S.}},
\bauthor{\bsnm{Epelbaum}, \binits{E.}},
\bauthor{\bsnm{Krebs}, \binits{H.}},
\bauthor{\bsnm{L\"ahde}, \binits{T.A.}},
\bauthor{\bsnm{Lee}, \binits{D.}},
\bauthor{\bsnm{Li}, \binits{N.}},
\bauthor{\bsnm{Lu}, \binits{B.}},
\bauthor{\bsnm{Mei\ss{}ner}, \binits{U.}},
\bauthor{\bsnm{Rupak}, \binits{G.}}:
\batitle{\textit{Ab initio} calculations of the isotopic dependence of nuclear
  clustering}.
\bjtitle{Phys. Rev. Lett.}
\bvolume{119},
\bfpage{6}
(\byear{2017})
\end{barticle}
\endbibitem

\bibitem{Ebran2012}
\begin{barticle}
\bauthor{\bsnm{Ebran}, \binits{J.-P.}},
\bauthor{\bsnm{Khan}, \binits{E.}},
\bauthor{\bsnm{Nik{\v s}i{\'c}}, \binits{T.}},
\bauthor{\bsnm{Vretenar}, \binits{D.}}:
\batitle{How atomic nuclei cluster}.
\bjtitle{Nature}
\bvolume{487},
\bfpage{344}
(\byear{2012})
\end{barticle}
\endbibitem

\bibitem{vonTresckow2021}
\begin{barticle}
\bauthor{\bsnm{{von Tresckow}}, \binits{M.}},
\bauthor{\bsnm{Rudigier}, \binits{M.}},
\bauthor{\bsnm{Shneidman}, \binits{T.M.}},
\bauthor{\bsnm{Kröll}, \binits{T.}},
\bauthor{\bsnm{Boromiza}, \binits{M.}},
\bauthor{\bsnm{Clisu}, \binits{C.}},
\bauthor{\bsnm{Costache}, \binits{C.}},
\bauthor{\bsnm{Filipescu}, \binits{D.}},
\bauthor{\bsnm{Florea}, \binits{N..M.}},
\bauthor{\bsnm{Gheorghe}, \binits{I.}},
\bauthor{\bsnm{Gladnishki}, \binits{K.}},
\bauthor{\bsnm{Ionescu}, \binits{A.}},
\bauthor{\bsnm{Kocheva}, \binits{D.}},
\bauthor{\bsnm{Lică}, \binits{R.}},
\bauthor{\bsnm{Mărginean}, \binits{N.}},
\bauthor{\bsnm{Mărginean}, \binits{R.}},
\bauthor{\bsnm{Mashtakov}, \binits{K.R.}},
\bauthor{\bsnm{Mihai}, \binits{C.}},
\bauthor{\bsnm{Mihai}, \binits{R.E.}},
\bauthor{\bsnm{Negret}, \binits{A.}},
\bauthor{\bsnm{Nita}, \binits{C.R.}},
\bauthor{\bsnm{Olacel}, \binits{A.}},
\bauthor{\bsnm{Oprea}, \binits{A.}},
\bauthor{\bsnm{Pascu}, \binits{S.}},
\bauthor{\bsnm{Rainovski}, \binits{G.}},
\bauthor{\bsnm{Sava}, \binits{T.}},
\bauthor{\bsnm{Scheck}, \binits{M.}},
\bauthor{\bsnm{Spagnoletti}, \binits{P.}},
\bauthor{\bsnm{Sotty}, \binits{C.}},
\bauthor{\bsnm{Stan}, \binits{L.}},
\bauthor{\bsnm{Stiru}, \binits{I.}},
\bauthor{\bsnm{Toma}, \binits{S.}},
\bauthor{\bsnm{Turturică}, \binits{A.}},
\bauthor{\bsnm{Ujeniuc}, \binits{S.}}:
\batitle{New evidence for alpha clustering structure in the ground state band
  of ${}^{212}$po}.
\bjtitle{Phys. Lett. B}
\bvolume{821},
\bfpage{136624}
(\byear{2021})
\end{barticle}
\endbibitem

\bibitem{alhassid1982}
\begin{barticle}
\bauthor{\bsnm{Alhassid}, \binits{Y.}},
\bauthor{\bsnm{Gai}, \binits{M.}},
\bauthor{\bsnm{Bertsch}, \binits{G.F.}}:
\batitle{Radiative width of molecular-cluster states}.
\bjtitle{Phys. Rev. Lett.}
\bvolume{49},
\bfpage{0}
(\byear{1982})
\end{barticle}
\endbibitem

\bibitem{hencken2004}
\begin{barticle}
\bauthor{\bsnm{Hencken}, \binits{K.}},
\bauthor{\bsnm{Baur}, \binits{G.}},
\bauthor{\bsnm{Trautmann}, \binits{D.}}:
\batitle{A cluster version of the ggt sum rule}.
\bjtitle{Nucl. Phys. A}
\bvolume{733},
\bfpage{10}
(\byear{2004})
\end{barticle}
\endbibitem

\bibitem{caprio2022}
\begin{barticle}
\bauthor{\bsnm{Caprio}, \binits{M.A.}},
\bauthor{\bsnm{Fasano}, \binits{P.J.}},
\bauthor{\bsnm{Maris}, \binits{P.}}:
\batitle{Robust \textit{ab initio} prediction of nuclear electric quadrupole
  observables by scaling to the charge radius}.
\bjtitle{Phys. Rev. C}
\bvolume{105},
\bfpage{7}
(\byear{2022})
\end{barticle}
\endbibitem

\bibitem{capel2011}
\begin{barticle}
\bauthor{\bsnm{Capel}, \binits{P.}},
\bauthor{\bsnm{Johnson}, \binits{R.C.}},
\bauthor{\bsnm{Nunes}, \binits{F.M.}}:
\batitle{One-neutron halo structure by the ratio method}.
\bjtitle{Phys. Lett. B}
\bvolume{705},
\bfpage{15}
(\byear{2011})
\end{barticle}
\endbibitem

\bibitem{bonaiti2022}
\begin{barticle}
\bauthor{\bsnm{Bonaiti}, \binits{F.}},
\bauthor{\bsnm{Bacca}, \binits{S.}},
\bauthor{\bsnm{Hagen}, \binits{G.}}:
\batitle{\textit{Ab initio} coupled-cluster calculations of ground and dipole
  excited states in ${ ^{8}\text{He} }$}.
\bjtitle{Phys. Rev. C}
\bvolume{105},
\bfpage{9}
(\byear{2022})
\end{barticle}
\endbibitem

\bibitem{bacca2013}
\begin{barticle}
\bauthor{\bsnm{Bacca}, \binits{S.}},
\bauthor{\bsnm{Barnea}, \binits{N.}},
\bauthor{\bsnm{Hagen}, \binits{G.}},
\bauthor{\bsnm{Orlandini}, \binits{G.}},
\bauthor{\bsnm{Papenbrock}, \binits{T.}}:
\batitle{First principles description of the giant dipole resonance in ${
  ^{16}\text{O} }$}.
\bjtitle{Phys. Rev. Lett.}
\bvolume{111},
\bfpage{5}
(\byear{2013})
\end{barticle}
\endbibitem

\bibitem{bacca2014}
\begin{barticle}
\bauthor{\bsnm{Bacca}, \binits{S.}},
\bauthor{\bsnm{Barnea}, \binits{N.}},
\bauthor{\bsnm{Hagen}, \binits{G.}},
\bauthor{\bsnm{Miorelli}, \binits{M.}},
\bauthor{\bsnm{Orlandini}, \binits{G.}},
\bauthor{\bsnm{Papenbrock}, \binits{T.}}:
\batitle{Giant and pigmy dipole resonances in ${ ^{4}\text{He} }$, ${
  ^{16,22}\text{O} }$, and ${ ^{40}\text{Ca} }$ from chiral nucleon-nucleon
  interactions}.
\bjtitle{Phys. Rev. C}
\bvolume{90},
\bfpage{12}
(\byear{2014})
\end{barticle}
\endbibitem

\bibitem{canham08_2450}
\begin{barticle}
\bauthor{\bsnm{Canham}, \binits{D.L.}},
\bauthor{\bsnm{Hammer}, \binits{H.-W.}}:
\batitle{Universal properties and structure of halo nuclei}.
\bjtitle{Eur. Phys. J. A}
\bvolume{37},
\bfpage{367}
(\byear{2008})
\end{barticle}
\endbibitem

\bibitem{fukuda91_1605}
\begin{barticle}
\bauthor{\bsnm{Fukuda}, \binits{M.}},
\bauthor{\bsnm{Ichihara}, \binits{T.}},
\bauthor{\bsnm{Inabe}, \binits{N.}},
\bauthor{\bsnm{Kubo}, \binits{T.}},
\bauthor{\bsnm{Kumagai}, \binits{H.}},
\bauthor{\bsnm{Nakagawa}, \binits{T.}},
\bauthor{\bsnm{Yano}, \binits{Y.}},
\bauthor{\bsnm{Tanihata}, \binits{I.}},
\bauthor{\bsnm{Adachi}, \binits{M.}},
\bauthor{\bsnm{Asahi}, \binits{K.}},
\bauthor{\bsnm{Kouguchi}, \binits{M.}},
\bauthor{\bsnm{Ishihara}, \binits{M.}},
\bauthor{\bsnm{Sagawa}, \binits{H.}},
\bauthor{\bsnm{Shimoura}, \binits{S.}}:
\batitle{Neutron halo in ${ {}^{11}\text{Be} }$ studied \textit{via} reaction
  cross section}.
\bjtitle{Phys. Lett. B}
\bvolume{268},
\bfpage{339}
(\byear{1991})
\end{barticle}
\endbibitem

\bibitem{misu97_1181}
\begin{barticle}
\bauthor{\bsnm{Misu}, \binits{T.}},
\bauthor{\bsnm{Nazarewicz}, \binits{W.}},
\bauthor{\bsnm{{\AA}berg}, \binits{S.}}:
\batitle{Deformed nuclear halos}.
\bjtitle{Nucl. Phys. A}
\bvolume{614},
\bfpage{44}
(\byear{1997})
\end{barticle}
\endbibitem

\bibitem{suzuki08_1557}
\begin{barticle}
\bauthor{\bsnm{Suzuki}, \binits{D.}},
\bauthor{\bsnm{Iwasaki}, \binits{H.}},
\bauthor{\bsnm{Ong}, \binits{H.J.}},
\bauthor{\bsnm{Imai}, \binits{N.}},
\bauthor{\bsnm{Sakurai}, \binits{H.}},
\bauthor{\bsnm{Nakao}, \binits{T.}},
\bauthor{\bsnm{Aoi}, \binits{N.}},
\bauthor{\bsnm{Baba}, \binits{H.}},
\bauthor{\bsnm{Bishop}, \binits{S.}},
\bauthor{\bsnm{Ichikawa}, \binits{Y.}},
\bauthor{\bsnm{Ishihara}, \binits{M.}},
\bauthor{\bsnm{Kondo}, \binits{Y.}},
\bauthor{\bsnm{Kubo}, \binits{T.}},
\bauthor{\bsnm{Kurita}, \binits{K.}},
\bauthor{\bsnm{Motobayashi}, \binits{T.}},
\bauthor{\bsnm{Nakamura}, \binits{T.}},
\bauthor{\bsnm{Okumura}, \binits{T.}},
\bauthor{\bsnm{Onishi}, \binits{T.K.}},
\bauthor{\bsnm{Ota}, \binits{S.}},
\bauthor{\bsnm{Suzuki}, \binits{M.K.}},
\bauthor{\bsnm{Takeuchi}, \binits{S.}},
\bauthor{\bsnm{Togano}, \binits{Y.}},
\bauthor{\bsnm{Yanagisawa}, \binits{Y.}}:
\batitle{Lifetime measurements of excited states in ${ {}^{17}\text{C} }$:
  {P}ossible interplay between collectivity and halo effects}.
\bjtitle{Phys. Lett. B}
\bvolume{666},
\bfpage{222}
(\byear{2008})
\end{barticle}
\endbibitem

\bibitem{nakamura09_1513}
\begin{barticle}
\bauthor{\bsnm{Nakamura}, \binits{T.}},
\bauthor{\bsnm{Kobayashi}, \binits{N.}},
\bauthor{\bsnm{Kondo}, \binits{Y.}},
\bauthor{\bsnm{Satou}, \binits{Y.}},
\bauthor{\bsnm{Aoi}, \binits{N.}},
\bauthor{\bsnm{Baba}, \binits{H.}},
\bauthor{\bsnm{Deguchi}, \binits{S.}},
\bauthor{\bsnm{Fukuda}, \binits{N.}},
\bauthor{\bsnm{Gibelin}, \binits{J.}},
\bauthor{\bsnm{Inabe}, \binits{N.}},
\bauthor{\bsnm{Ishihara}, \binits{M.}},
\bauthor{\bsnm{Kameda}, \binits{D.}},
\bauthor{\bsnm{Kawada}, \binits{Y.}},
\bauthor{\bsnm{Kubo}, \binits{T.}},
\bauthor{\bsnm{Kusaka}, \binits{K.}},
\bauthor{\bsnm{Mengoni}, \binits{A.}},
\bauthor{\bsnm{Motobayashi}, \binits{T.}},
\bauthor{\bsnm{Ohnishi}, \binits{T.}},
\bauthor{\bsnm{Ohtake}, \binits{M.}},
\bauthor{\bsnm{Orr}, \binits{N.A.}},
\bauthor{\bsnm{Otsu}, \binits{H.}},
\bauthor{\bsnm{Otsuka}, \binits{T.}},
\bauthor{\bsnm{Saito}, \binits{A.}},
\bauthor{\bsnm{Sakurai}, \binits{H.}},
\bauthor{\bsnm{Shimoura}, \binits{S.}},
\bauthor{\bsnm{Sumikama}, \binits{T.}},
\bauthor{\bsnm{Takeda}, \binits{H.}},
\bauthor{\bsnm{Takeshita}, \binits{E.}},
\bauthor{\bsnm{Takechi}, \binits{M.}},
\bauthor{\bsnm{Takeuchi}, \binits{S.}},
\bauthor{\bsnm{Tanaka}, \binits{K.}},
\bauthor{\bsnm{Tanaka}, \binits{K.N.}},
\bauthor{\bsnm{Tanaka}, \binits{N.}},
\bauthor{\bsnm{Togano}, \binits{Y.}},
\bauthor{\bsnm{Utsuno}, \binits{Y.}},
\bauthor{\bsnm{Yoneda}, \binits{K.}},
\bauthor{\bsnm{Yoshida}, \binits{A.}},
\bauthor{\bsnm{Yoshida}, \binits{K.}}:
\batitle{Halo structure of the island of inversion nucleus ${ {}^{31}\text{Ne}
  }$}.
\bjtitle{Phys. Rev. Lett.}
\bvolume{103},
\bfpage{262501}
(\byear{2009})
\end{barticle}
\endbibitem

\bibitem{takechi12_1512}
\begin{barticle}
\bauthor{\bsnm{Takechi}, \binits{M.}},
\bauthor{\bsnm{Ohtsubo}, \binits{T.}},
\bauthor{\bsnm{Fukuda}, \binits{M.}},
\bauthor{\bsnm{Nishimura}, \binits{D.}},
\bauthor{\bsnm{Kuboki}, \binits{T.}},
\bauthor{\bsnm{Suzuki}, \binits{T.}},
\bauthor{\bsnm{Yamaguchi}, \binits{T.}},
\bauthor{\bsnm{Ozawa}, \binits{A.}},
\bauthor{\bsnm{Moriguchi}, \binits{T.}},
\bauthor{\bsnm{Ooishi}, \binits{H.}},
\bauthor{\bsnm{Nagae}, \binits{D.}},
\bauthor{\bsnm{Suzuki}, \binits{H.}},
\bauthor{\bsnm{Suzuki}, \binits{S.}},
\bauthor{\bsnm{Izumikawa}, \binits{T.}},
\bauthor{\bsnm{Sumikama}, \binits{T.}},
\bauthor{\bsnm{Ishihara}, \binits{M.}},
\bauthor{\bsnm{Geissel}, \binits{H.}},
\bauthor{\bsnm{Aoi}, \binits{N.}},
\bauthor{\bsnm{Chen}, \binits{R.}},
\bauthor{\bsnm{Fang}, \binits{D.}},
\bauthor{\bsnm{Fukuda}, \binits{N.}},
\bauthor{\bsnm{Hachiuma}, \binits{I.}},
\bauthor{\bsnm{Inabe}, \binits{N.}},
\bauthor{\bsnm{Ishibashi}, \binits{Y.}},
\bauthor{\bsnm{Ito}, \binits{Y.}},
\bauthor{\bsnm{Kameda}, \binits{D.}},
\bauthor{\bsnm{Kubo}, \binits{T.}},
\bauthor{\bsnm{Kusaka}, \binits{K.}},
\bauthor{\bsnm{Lantz}, \binits{M.}},
\bauthor{\bsnm{Ma}, \binits{Y.}},
\bauthor{\bsnm{Matsuta}, \binits{K.}},
\bauthor{\bsnm{Mihara}, \binits{M.}},
\bauthor{\bsnm{Miyashita}, \binits{Y.}},
\bauthor{\bsnm{Momota}, \binits{S.}},
\bauthor{\bsnm{Namihira}, \binits{K.}},
\bauthor{\bsnm{Nagashima}, \binits{M.}},
\bauthor{\bsnm{Ohkuma}, \binits{Y.}},
\bauthor{\bsnm{Ohnishi}, \binits{T.}},
\bauthor{\bsnm{Ohtake}, \binits{M.}},
\bauthor{\bsnm{Ogawa}, \binits{K.}},
\bauthor{\bsnm{Sakurai}, \binits{H.}},
\bauthor{\bsnm{Shimbara}, \binits{Y.}},
\bauthor{\bsnm{Suda}, \binits{T.}},
\bauthor{\bsnm{Takeda}, \binits{H.}},
\bauthor{\bsnm{Takeuchi}, \binits{S.}},
\bauthor{\bsnm{Tanaka}, \binits{K.}},
\bauthor{\bsnm{Watanabe}, \binits{R.}},
\bauthor{\bsnm{Winkler}, \binits{M.}},
\bauthor{\bsnm{Yanagisawa}, \binits{Y.}},
\bauthor{\bsnm{Yasuda}, \binits{Y.}},
\bauthor{\bsnm{Yoshinaga}, \binits{K.}},
\bauthor{\bsnm{Yoshida}, \binits{A.}},
\bauthor{\bsnm{Yoshida}, \binits{K.}}:
\batitle{Interaction cross sections for ${ \text{Ne} }$ isotopes towards the
  island of inversion and halo structures of ${ {}^{29}\text{Ne} }$ and ${
  {}^{31}\text{Ne} }$}.
\bjtitle{Phys. Lett. B}
\bvolume{707},
\bfpage{357}
(\byear{2012})
\end{barticle}
\endbibitem

\bibitem{BeckerLED22}
\begin{barticle}
\bauthor{\bsnm{Becker}, \binits{K.S.}},
\bauthor{\bsnm{Launey}, \binits{K.D.}},
\bauthor{\bsnm{Ekstr\"om}, \binits{A.}},
\bauthor{\bsnm{Dytrych}, \binits{T.}}:
\batitle{{Ab initio symmetry-adapted emulator for studying emergent
  collectivity and clustering in nuclei}}.
\bjtitle{Front. in Phys.}
\bvolume{11},
\bfpage{1064601}
(\byear{2023})
{\href{https://arxiv.org/abs/2303.00667}{{arXiv:2303.00667}}}
{[nucl-th]}.
\doiurl{10.3389/fphy.2023.1064601}
\end{barticle}
\endbibitem

\bibitem{Becker23}
\begin{botherref}
\oauthor{\bsnm{Becker}, \binits{K.S.}}, et al.:
In preparation
(2022)
\end{botherref}
\endbibitem

\bibitem{PhysRevLett.129.042503}
\begin{barticle}
\bauthor{\bsnm{Hebborn}, \binits{C.}},
\bauthor{\bsnm{Hupin}, \binits{G.}},
\bauthor{\bsnm{Kravvaris}, \binits{K.}},
\bauthor{\bsnm{Quaglioni}, \binits{S.}},
\bauthor{\bsnm{Navr\'atil}, \binits{P.}},
\bauthor{\bsnm{Gysbers}, \binits{P.}}:
\batitle{Ab initio prediction of the ${ ^{4}\text{He}(d,\gamma)^{6}\text{Li} }$
  big bang radiative capture}.
\bjtitle{Phys. Rev. Lett.}
\bvolume{129},
\bfpage{5}
(\byear{2022})
\end{barticle}
\endbibitem

\bibitem{PhysRevC.102.014001}
\begin{barticle}
\bauthor{\bsnm{Gnech}, \binits{A.}},
\bauthor{\bsnm{Viviani}, \binits{M.}},
\bauthor{\bsnm{Marcucci}, \binits{L.E.}}:
\batitle{Calculation of the ${ ^{6}\text{Li} }$ ground state within the
  hyperspherical harmonic basis}.
\bjtitle{Phys. Rev. C}
\bvolume{102},
\bfpage{19}
(\byear{2020})
\end{barticle}
\endbibitem

\bibitem{Frame:2017fah}
\begin{barticle}
\bauthor{\bsnm{Frame}, \binits{D.}},
\bauthor{\bsnm{He}, \binits{R.}},
\bauthor{\bsnm{Ipsen}, \binits{I.}},
\bauthor{\bsnm{Lee}, \binits{D.}},
\bauthor{\bsnm{Lee}, \binits{D.}},
\bauthor{\bsnm{Rrapaj}, \binits{E.}}:
\batitle{Eigenvector continuation with subspace learning}.
\bjtitle{Phys. Rev. Lett.}
\bvolume{121},
\bfpage{032501}
(\byear{2018})
\end{barticle}
\endbibitem

\bibitem{Konig:2019adq}
\begin{barticle}
\bauthor{\bsnm{K\"onig}, \binits{S.}},
\bauthor{\bsnm{Ekstr\"om}, \binits{A.}},
\bauthor{\bsnm{Hebeler}, \binits{K.}},
\bauthor{\bsnm{Lee}, \binits{D.}},
\bauthor{\bsnm{Schwenk}, \binits{A.}}:
\batitle{Eigenvector continuation as an efficient and accurate emulator for
  uncertainty quantification}.
\bjtitle{Phys. Lett. B}
\bvolume{810},
\bfpage{135814}
(\byear{2020})
\end{barticle}
\endbibitem

\bibitem{Ekstrom:2019lss}
\begin{barticle}
\bauthor{\bsnm{Ekstr\"om}, \binits{A.}},
\bauthor{\bsnm{Hagen}, \binits{G.}}:
\batitle{Global sensitivity analysis of bulk properties of an atomic nucleus}.
\bjtitle{Phys. Rev. Lett.}
\bvolume{123},
\bfpage{252501}
(\byear{2019})
\end{barticle}
\endbibitem

\bibitem{crawford14_1607}
\begin{barticle}
\bauthor{\bsnm{Crawford}, \binits{H.L.}},
\bauthor{\bsnm{Fallon}, \binits{P.}},
\bauthor{\bsnm{Macchiavelli}, \binits{A.O.}},
\bauthor{\bsnm{Clark}, \binits{R.M.}},
\bauthor{\bsnm{Brown}, \binits{B.A.}},
\bauthor{\bsnm{Tostevin}, \binits{J.A.}},
\bauthor{\bsnm{Bazin}, \binits{D.}},
\bauthor{\bsnm{Aoi}, \binits{N.}},
\bauthor{\bsnm{Doornenbal}, \binits{P.}},
\bauthor{\bsnm{Matsushita}, \binits{M.}},
\bauthor{\bsnm{Scheit}, \binits{H.}},
\bauthor{\bsnm{Steppenbeck}, \binits{D.}},
\bauthor{\bsnm{Takeuchi}, \binits{S.}},
\bauthor{\bsnm{Baba}, \binits{H.}},
\bauthor{\bsnm{Campbell}, \binits{C.M.}},
\bauthor{\bsnm{Cromaz}, \binits{M.}},
\bauthor{\bsnm{Ideguchi}, \binits{E.}},
\bauthor{\bsnm{Kobayashi}, \binits{N.}},
\bauthor{\bsnm{Kondo}, \binits{Y.}},
\bauthor{\bsnm{Lee}, \binits{G.}},
\bauthor{\bsnm{Lee}, \binits{I.Y.}},
\bauthor{\bsnm{Lee}, \binits{J.}},
\bauthor{\bsnm{Li}, \binits{K.}},
\bauthor{\bsnm{Michimasa}, \binits{S.}},
\bauthor{\bsnm{Motobayashi}, \binits{T.}},
\bauthor{\bsnm{Nakamura}, \binits{T.}},
\bauthor{\bsnm{Ota}, \binits{S.}},
\bauthor{\bsnm{Paschalis}, \binits{S.}},
\bauthor{\bsnm{Petri}, \binits{M.}},
\bauthor{\bsnm{Sako}, \binits{T.}},
\bauthor{\bsnm{Sakurai}, \binits{H.}},
\bauthor{\bsnm{Shimoura}, \binits{S.}},
\bauthor{\bsnm{Takechi}, \binits{M.}},
\bauthor{\bsnm{Togano}, \binits{Y.}},
\bauthor{\bsnm{Wang}, \binits{H.}},
\bauthor{},
\bauthor{\bsnm{Yoneda}, \binits{K.}}:
\batitle{Shell and shape evolution at ${ N = 28 }$: {T}he ${ {}^{40}\text{Mg}
  }$ ground state}.
\bjtitle{Phys. Rev. C}
\bvolume{89},
\bfpage{041303}
(\byear{2014})
\end{barticle}
\endbibitem

\bibitem{crawford19_2359}
\begin{barticle}
\bauthor{\bsnm{Crawford}, \binits{H.L.}},
\bauthor{\bsnm{Fallon}, \binits{P.}},
\bauthor{\bsnm{Macchiavelli}, \binits{A.O.}},
\bauthor{\bsnm{Doornenbal}, \binits{P.}},
\bauthor{\bsnm{Aoi}, \binits{N.}},
\bauthor{\bsnm{Browne}, \binits{F.}},
\bauthor{\bsnm{Campbell}, \binits{C.M.}},
\bauthor{\bsnm{Chen}, \binits{S.}},
\bauthor{\bsnm{Clark}, \binits{R.M.}},
\bauthor{\bsnm{Cort\'es}, \binits{M.L.}},
\bauthor{\bsnm{Cromaz}, \binits{M.}},
\bauthor{\bsnm{Ideguchi}, \binits{E.}},
\bauthor{\bsnm{Jones}, \binits{M.D.}},
\bauthor{\bsnm{Kanungo}, \binits{R.}},
\bauthor{\bsnm{{MacCormick}}, \binits{M.}},
\bauthor{\bsnm{Momiyama}, \binits{S.}},
\bauthor{\bsnm{Murray}, \binits{I.}},
\bauthor{\bsnm{Niikura}, \binits{M.}},
\bauthor{\bsnm{Paschalis}, \binits{S.}},
\bauthor{\bsnm{Petri}, \binits{M.}},
\bauthor{\bsnm{Sakurai}, \binits{H.}},
\bauthor{\bsnm{Salathe}, \binits{M.}},
\bauthor{\bsnm{Schrock}, \binits{P.}},
\bauthor{\bsnm{Steppenbeck}, \binits{D.}},
\bauthor{\bsnm{Takeuchi}, \binits{S.}},
\bauthor{\bsnm{Tanaka}, \binits{Y.K.}},
\bauthor{\bsnm{Taniuchi}, \binits{R.}},
\bauthor{\bsnm{Wang}, \binits{H.}},
\bauthor{\bsnm{Wimmer}, \binits{K.}}:
\batitle{First spectroscopy of the near drip-line nucleus ${ {}^{40}\text{Mg}
  }$}.
\bjtitle{Phys. Rev. Lett.}
\bvolume{122},
\bfpage{052501}
(\byear{2019})
\end{barticle}
\endbibitem

\bibitem{refId0}
\begin{barticle}
\bauthor{\bsnm{Macchiavelli}, \binits{A.O.}},
\bauthor{\bsnm{Crawford}, \binits{H.L.}},
\bauthor{\bsnm{Fallon}, \binits{P.}},
\bauthor{\bsnm{Clark}, \binits{R.M.}},
\bauthor{\bsnm{Poves}, \binits{A.}}:
\batitle{Weak binding effects on the structure of ${ {}^{40}\text{Mg} }$}.
\bjtitle{Eur. Phys. J. A}
\bvolume{58},
\bfpage{66}
(\byear{2022})
\end{barticle}
\endbibitem

\bibitem{navratil07_733}
\begin{barticle}
\bauthor{\bsnm{Navr\'atil}, \binits{P.}},
\bauthor{\bsnm{Gueorguiev}, \binits{V.G.}},
\bauthor{\bsnm{Vary}, \binits{J.P.}},
\bauthor{\bsnm{Ormand}, \binits{W.E.}},
\bauthor{\bsnm{Nogga}, \binits{A.}}:
\batitle{Structure of ${ A = 10 - 13 }$ nuclei with two- plus three-nucleon
  interaction from chiral effective field theory}.
\bjtitle{Phys. Rev. Lett.}
\bvolume{99},
\bfpage{042501}
(\byear{2007})
\end{barticle}
\endbibitem

\bibitem{mccutchan12_2321}
\begin{barticle}
\bauthor{\bsnm{{McCutchan}}, \binits{E.A.}},
\bauthor{\bsnm{Lister}, \binits{C.J.}},
\bauthor{\bsnm{Elvers}, \binits{M.}},
\bauthor{\bsnm{Savran}, \binits{D.}},
\bauthor{\bsnm{Greene}, \binits{J.P.}},
\bauthor{\bsnm{Ahmed}, \binits{T.}},
\bauthor{\bsnm{Ahn}, \binits{T.}},
\bauthor{\bsnm{Cooper}, \binits{N.}},
\bauthor{\bsnm{Heinz}, \binits{A.}},
\bauthor{\bsnm{Hughes}, \binits{R.O.}},
\bauthor{\bsnm{Ilie}, \binits{G.}},
\bauthor{\bsnm{Pauerstein}, \binits{B.}},
\bauthor{\bsnm{Radeck}, \binits{D.}},
\bauthor{\bsnm{Shenkov}, \binits{N.}},
\bauthor{\bsnm{Werner}, \binits{V.}}:
\batitle{Precise ${ \gamma }$-ray intensity measurements in ${ {}^{10}\text{B}
  }$}.
\bjtitle{Phys. Rev. C}
\bvolume{86},
\bfpage{057306}
(\byear{2012})
\end{barticle}
\endbibitem

\bibitem{carlson15_1610}
\begin{barticle}
\bauthor{\bsnm{Carlson}, \binits{J.}},
\bauthor{\bsnm{Gandolfi}, \binits{S.}},
\bauthor{\bsnm{Pederiva}, \binits{F.}},
\bauthor{\bsnm{Pieper}, \binits{S.C.}},
\bauthor{\bsnm{Schiavilla}, \binits{R.}},
\bauthor{\bsnm{Schmidt}, \binits{K.E.}},
\bauthor{\bsnm{Wiringa}, \binits{R.B.}}:
\batitle{Quantum {M}onte {C}arlo methods for nuclear physics}.
\bjtitle{Rev. Mod. Phys.}
\bvolume{87},
\bfpage{1067}
(\byear{2015})
\end{barticle}
\endbibitem

\bibitem{lunderberg12_556}
\begin{barticle}
\bauthor{\bsnm{Lunderberg}, \binits{E.}},
\bauthor{\bsnm{{DeYoung}}, \binits{P.A.}},
\bauthor{\bsnm{Kohley}, \binits{Z.}},
\bauthor{\bsnm{Attanayake}, \binits{H.}},
\bauthor{\bsnm{Baumann}, \binits{T.}},
\bauthor{\bsnm{Bazin}, \binits{D.}},
\bauthor{\bsnm{Christian}, \binits{G.}},
\bauthor{\bsnm{Divaratne}, \binits{D.}},
\bauthor{\bsnm{Grimes}, \binits{S.M.}},
\bauthor{\bsnm{Haagsma}, \binits{A.}},
\bauthor{\bsnm{Finck}, \binits{J.E.}},
\bauthor{\bsnm{Frank}, \binits{N.}},
\bauthor{\bsnm{Luther}, \binits{B.}},
\bauthor{\bsnm{Mosby}, \binits{S.}},
\bauthor{\bsnm{Nagi}, \binits{T.}},
\bauthor{\bsnm{Peaslee}, \binits{G.F.}},
\bauthor{\bsnm{Schiller}, \binits{A.}},
\bauthor{\bsnm{Snyder}, \binits{J.}},
\bauthor{\bsnm{Spyrou}, \binits{A.}},
\bauthor{\bsnm{Strongman}, \binits{M.J.}},
\bauthor{\bsnm{Thoennessen}, \binits{M.}}:
\batitle{Evidence for the ground-state resonance of ${ {}^{26}\text{O} }$}.
\bjtitle{Phys. Rev. Lett.}
\bvolume{108},
\bfpage{142503}
(\byear{2012})
\end{barticle}
\endbibitem

\bibitem{kohley13_1541}
\begin{barticle}
\bauthor{\bsnm{Kohley}, \binits{Z.}},
\bauthor{\bsnm{Baumann}, \binits{T.}},
\bauthor{\bsnm{Bazin}, \binits{D.}},
\bauthor{\bsnm{Christian}, \binits{G.}},
\bauthor{\bsnm{{DeYoung}}, \binits{P.A.}},
\bauthor{\bsnm{Finck}, \binits{J.E.}},
\bauthor{\bsnm{Frank}, \binits{N.}},
\bauthor{\bsnm{Jones}, \binits{M.}},
\bauthor{\bsnm{Lunderberg}, \binits{E.}},
\bauthor{\bsnm{Luther}, \binits{B.}},
\bauthor{\bsnm{Mosby}, \binits{S.}},
\bauthor{\bsnm{Nagi}, \binits{T.}},
\bauthor{\bsnm{Smith}, \binits{J.K.}},
\bauthor{\bsnm{Snyder}, \binits{J.}},
\bauthor{\bsnm{Spyrou}, \binits{A.}},
\bauthor{\bsnm{Thoennessen}, \binits{M.}}:
\batitle{Study of two-neutron radioactivity in the decay of ${ {}^{26}\text{O}
  }$}.
\bjtitle{Phys. Rev. Lett.}
\bvolume{110},
\bfpage{152501}
(\byear{2013})
\end{barticle}
\endbibitem

\bibitem{caesar13_1765}
\begin{barticle}
\bauthor{\bsnm{Caesar}, \binits{C.}},
\bauthor{\bsnm{Simonis}, \binits{J.}},
\bauthor{\bsnm{Adachi}, \binits{T.}},
\bauthor{\bsnm{Aksyutina}, \binits{Y.}},
\bauthor{\bsnm{Alcantara}, \binits{J.}},
\bauthor{\bsnm{Altstadt}, \binits{S.}},
\bauthor{\bsnm{{Alvarez-Pol}}, \binits{H.}},
\bauthor{\bsnm{Ashwood}, \binits{N.}},
\bauthor{\bsnm{Aumann}, \binits{T.}},
\bauthor{\bsnm{Avdeichikov}, \binits{V.}},
\bauthor{\bsnm{Barr}, \binits{M.}},
\bauthor{\bsnm{Beceiro}, \binits{S.}},
\bauthor{\bsnm{Bemmerer}, \binits{D.}},
\bauthor{\bsnm{Benlliure}, \binits{J.}},
\bauthor{\bsnm{Bertulani}, \binits{C.A.}},
\bauthor{\bsnm{Boretzky}, \binits{K.}},
\bauthor{\bsnm{Borge}, \binits{M.J.G.}},
\bauthor{\bsnm{Burgunder}, \binits{G.}},
\bauthor{\bsnm{Caamano}, \binits{M.}},
\bauthor{\bsnm{Casarejos}, \binits{E.}},
\bauthor{\bsnm{Catford}, \binits{W.}},
\bauthor{\bsnm{Cederk\"all}, \binits{J.}},
\bauthor{\bsnm{Chakraborty}, \binits{S.}},
\bauthor{\bsnm{Chartier}, \binits{M.}},
\bauthor{\bsnm{Chulkov}, \binits{L.}},
\bauthor{\bsnm{{Cortina-Gil}}, \binits{D.}},
\bauthor{\bsnm{{Datta Pramanik}}, \binits{U.}},
\bauthor{\bsnm{{Diaz Fernandez}}, \binits{P.}},
\bauthor{\bsnm{Dillmann}, \binits{I.}},
\bauthor{\bsnm{Elekes}, \binits{Z.}},
\bauthor{\bsnm{Enders}, \binits{J.}},
\bauthor{\bsnm{Ershova}, \binits{O.}},
\bauthor{\bsnm{Estrade}, \binits{A.}},
\bauthor{\bsnm{Farinon}, \binits{F.}},
\bauthor{\bsnm{Fraile}, \binits{L.M.}},
\bauthor{\bsnm{Freer}, \binits{M.}},
\bauthor{\bsnm{Freudenberger}, \binits{M.}},
\bauthor{\bsnm{Fynbo}, \binits{H.O.U.}},
\bauthor{\bsnm{Galaviz}, \binits{D.}},
\bauthor{\bsnm{Geissel}, \binits{H.}},
\bauthor{\bsnm{Gernh\"auser}, \binits{R.}},
\bauthor{\bsnm{Golubev}, \binits{P.}},
\bauthor{\bsnm{{Gonzalez Diaz}}, \binits{D.}},
\bauthor{\bsnm{Hagdahl}, \binits{J.}},
\bauthor{\bsnm{Heftrich}, \binits{T.}},
\bauthor{\bsnm{Heil}, \binits{M.}},
\bauthor{\bsnm{Heine}, \binits{M.}},
\bauthor{\bsnm{Heinz}, \binits{A.}},
\bauthor{\bsnm{Henriques}, \binits{A.}},
\bauthor{\bsnm{Holl}, \binits{M.}},
\bauthor{\bsnm{Holt}, \binits{J.D.}},
\bauthor{\bsnm{Ickert}, \binits{G.}},
\bauthor{\bsnm{Ignatov}, \binits{A.}},
\bauthor{\bsnm{Jakobsson}, \binits{B.}},
\bauthor{\bsnm{Johansson}, \binits{H.T.}},
\bauthor{\bsnm{Jonson}, \binits{B.}},
\bauthor{\bsnm{{Kalantar-Nayestanaki}}, \binits{N.}},
\bauthor{\bsnm{Kanungo}, \binits{R.}},
\bauthor{\bsnm{{Kelic-Heil}}, \binits{A.}},
\bauthor{\bsnm{Kn\"obel}, \binits{R.}},
\bauthor{\bsnm{Kr\"oll}, \binits{T.}},
\bauthor{\bsnm{Kr\"ucken}, \binits{R.}},
\bauthor{\bsnm{Kurcewicz}, \binits{J.}},
\bauthor{\bsnm{Labiche}, \binits{M.}},
\bauthor{\bsnm{Langer}, \binits{C.}},
\bauthor{\bsnm{{Le Bleis}}, \binits{T.}},
\bauthor{\bsnm{Lemmon}, \binits{R.}},
\bauthor{\bsnm{Lepyoshkina}, \binits{O.}},
\bauthor{\bsnm{Lindberg}, \binits{S.}},
\bauthor{\bsnm{Machado}, \binits{J.}},
\bauthor{\bsnm{Marganiec}, \binits{J.}},
\bauthor{\bsnm{Maroussov}, \binits{V.}},
\bauthor{\bsnm{Men\'endez}, \binits{J.}},
\bauthor{\bsnm{Mostazo}, \binits{M.}},
\bauthor{\bsnm{Movsesyan}, \binits{A.}},
\bauthor{\bsnm{Najafi}, \binits{A.}},
\bauthor{\bsnm{Nilsson}, \binits{T.}},
\bauthor{\bsnm{Nociforo}, \binits{C.}},
\bauthor{\bsnm{Panin}, \binits{V.}},
\bauthor{\bsnm{Perea}, \binits{A.}},
\bauthor{\bsnm{Pietri}, \binits{S.}},
\bauthor{\bsnm{Plag}, \binits{R.}},
\bauthor{\bsnm{Prochazka}, \binits{A.}},
\bauthor{\bsnm{Rahaman}, \binits{A.}},
\bauthor{\bsnm{Rastrepina}, \binits{G.}},
\bauthor{\bsnm{Reifarth}, \binits{R.}},
\bauthor{\bsnm{Ribeiro}, \binits{G.}},
\bauthor{\bsnm{Ricciardi}, \binits{M.V.}},
\bauthor{\bsnm{Rigollet}, \binits{C.}},
\bauthor{\bsnm{Riisager}, \binits{K.}},
\bauthor{\bsnm{R\"oder}, \binits{M.}},
\bauthor{\bsnm{Rossi}, \binits{D.}},
\bauthor{\bsnm{{Sanchez del Rio}}, \binits{J.}},
\bauthor{\bsnm{Savran}, \binits{D.}},
\bauthor{\bsnm{Scheit}, \binits{H.}},
\bauthor{\bsnm{Schwenk}, \binits{A.}},
\bauthor{\bsnm{Simon}, \binits{H.}},
\bauthor{\bsnm{Sorlin}, \binits{O.}},
\bauthor{\bsnm{Stoica}, \binits{V.}},
\bauthor{\bsnm{Streicher}, \binits{B.}},
\bauthor{\bsnm{Taylor}, \binits{J.}},
\bauthor{\bsnm{Tengblad}, \binits{O.}},
\bauthor{\bsnm{Terashima}, \binits{S.}},
\bauthor{\bsnm{Thies}, \binits{R.}},
\bauthor{\bsnm{Togano}, \binits{Y.}},
\bauthor{\bsnm{Uberseder}, \binits{E.}},
\bauthor{\bsnm{{Van de Walle}}, \binits{J.}},
\bauthor{\bsnm{Velho}, \binits{P.}},
\bauthor{\bsnm{Volkov}, \binits{V.}},
\bauthor{\bsnm{Wagner}, \binits{A.}},
\bauthor{\bsnm{Wamers}, \binits{F.}},
\bauthor{\bsnm{Weick}, \binits{H.}},
\bauthor{\bsnm{Weigand}, \binits{M.}},
\bauthor{\bsnm{Wheldon}, \binits{C.}},
\bauthor{\bsnm{Wilson}, \binits{G.}},
\bauthor{\bsnm{Wimmer}, \binits{C.}},
\bauthor{\bsnm{Winfield}, \binits{J.S.}},
\bauthor{\bsnm{Woods}, \binits{P.}},
\bauthor{\bsnm{Yakorev}, \binits{D.}},
\bauthor{\bsnm{Zhukov}, \binits{M.V.}},
\bauthor{\bsnm{Zilges}, \binits{A.}},
\bauthor{\bsnm{Zoric}, \binits{M.}},
\bauthor{\bsnm{Zuber}, \binits{K.}},
\bauthor{\bsnm{{(R3B collaboration)}}}:
\batitle{Beyond the neutron drip line: {T}he unbound oxygen isotopes ${
  {}^{25}\text{O} }$ and ${ {}^{26}\text{O} }$}.
\bjtitle{Phys. Rev. C}
\bvolume{88},
\bfpage{034313}
(\byear{2013})
\end{barticle}
\endbibitem

\bibitem{kondo16_1439}
\begin{barticle}
\bauthor{\bsnm{Kondo}, \binits{Y.}},
\bauthor{\bsnm{Nakamura}, \binits{T.}},
\bauthor{\bsnm{Tanaka}, \binits{R.}},
\bauthor{\bsnm{Minakata}, \binits{R.}},
\bauthor{\bsnm{Ogoshi}, \binits{S.}},
\bauthor{\bsnm{Orr}, \binits{N.A.}},
\bauthor{\bsnm{Achouri}, \binits{N.L.}},
\bauthor{\bsnm{Aumann}, \binits{T.}},
\bauthor{\bsnm{Baba}, \binits{H.}},
\bauthor{\bsnm{Delaunay}, \binits{F.}},
\bauthor{\bsnm{Doomenbal}, \binits{P.}},
\bauthor{\bsnm{Fukuda}, \binits{N.}},
\bauthor{\bsnm{Gibelin}, \binits{J.}},
\bauthor{\bsnm{Hwang}, \binits{J.W.}},
\bauthor{\bsnm{Inabe}, \binits{N.}},
\bauthor{\bsnm{Isobe}, \binits{T.}},
\bauthor{\bsnm{Kameda}, \binits{D.}},
\bauthor{\bsnm{Kanno}, \binits{D.}},
\bauthor{\bsnm{Kim}, \binits{S.}},
\bauthor{\bsnm{Kobayashi}, \binits{N.}},
\bauthor{\bsnm{Kobayashi}, \binits{T.}},
\bauthor{\bsnm{Kubo}, \binits{T.}},
\bauthor{\bsnm{Leblond}, \binits{S.}},
\bauthor{\bsnm{Lee}, \binits{J.}},
\bauthor{\bsnm{Marqu\'es}, \binits{F.M.}},
\bauthor{\bsnm{Motobayashi}, \binits{T.}},
\bauthor{\bsnm{Murai}, \binits{D.}},
\bauthor{\bsnm{Murakami}, \binits{T.}},
\bauthor{\bsnm{Muto}, \binits{K.}},
\bauthor{\bsnm{Nakashima}, \binits{T.}},
\bauthor{\bsnm{Nakatsuka}, \binits{N.}},
\bauthor{\bsnm{Navin}, \binits{A.}},
\bauthor{\bsnm{Nishi}, \binits{S.}},
\bauthor{\bsnm{Otsu}, \binits{H.}},
\bauthor{\bsnm{Sato}, \binits{H.}},
\bauthor{\bsnm{Satou}, \binits{Y.}},
\bauthor{\bsnm{Shimizu}, \binits{Y.}},
\bauthor{\bsnm{Suzuki}, \binits{H.}},
\bauthor{\bsnm{Takahashi}, \binits{K.}},
\bauthor{\bsnm{Takeda}, \binits{H.}},
\bauthor{\bsnm{Takeuchi}, \binits{S.}},
\bauthor{\bsnm{Togano}, \binits{Y.}},
\bauthor{\bsnm{Tuff}, \binits{A.G.}},
\bauthor{\bsnm{Vandebrouck}, \binits{M.}},
\bauthor{\bsnm{Yoneda}, \binits{K.}}:
\batitle{Nucleus ${ {}^{26}\text{O} }$: A barely unbound system beyond the drip
  line}.
\bjtitle{Phys. Rev. Lett.}
\bvolume{116},
\bfpage{102503}
(\byear{2016})
\end{barticle}
\endbibitem

\bibitem{kalanee13_1909}
\begin{barticle}
\bauthor{\bsnm{{Al Kalanee}}, \binits{T.}},
\bauthor{\bsnm{Gibelin}, \binits{J.}},
\bauthor{\bsnm{{Roussel-Chomaz}}, \binits{P.}},
\bauthor{\bsnm{Keeley}, \binits{N.}},
\bauthor{\bsnm{Beaumel}, \binits{D.}},
\bauthor{\bsnm{Blumenfeld}, \binits{Y.}},
\bauthor{\bsnm{{Fern\'andez-Dom\'inguez}}, \binits{B.}},
\bauthor{\bsnm{Force}, \binits{C.}},
\bauthor{\bsnm{Gaudefroy}, \binits{L.}},
\bauthor{\bsnm{Gillibert}, \binits{A.}},
\bauthor{\bsnm{Guillot}, \binits{J.}},
\bauthor{\bsnm{Iwasaki}, \binits{H.}},
\bauthor{\bsnm{Krupko}, \binits{S.}},
\bauthor{\bsnm{Lapoux}, \binits{V.}},
\bauthor{\bsnm{Mittig}, \binits{W.}},
\bauthor{\bsnm{Mougeot}, \binits{X.}},
\bauthor{\bsnm{Nalpas}, \binits{L.}},
\bauthor{\bsnm{Pollacco}, \binits{E.}},
\bauthor{\bsnm{Rusek}, \binits{K.}},
\bauthor{\bsnm{Roger}, \binits{T.}},
\bauthor{\bsnm{Savajols}, \binits{H.}},
\bauthor{\bsnm{{de S\'er\'eville}}, \binits{N.}},
\bauthor{\bsnm{Sidorchuk}, \binits{S.}},
\bauthor{\bsnm{Suzuki}, \binits{D.}},
\bauthor{\bsnm{Strojek}, \binits{I.}},
\bauthor{\bsnm{Orr}, \binits{N.A.}}:
\batitle{Structure of unbound neutron-rich ${ {}^{9}\text{He} }$ studied using
  single-neutron transfer}.
\bjtitle{Phys. Rev. C}
\bvolume{88},
\bfpage{034301}
(\byear{2013})
\end{barticle}
\endbibitem

\bibitem{vorabbi18_1977}
\begin{barticle}
\bauthor{\bsnm{Vorabbi}, \binits{M.}},
\bauthor{\bsnm{Calci}, \binits{A.}},
\bauthor{\bsnm{Navr\'atil}, \binits{P.}},
\bauthor{\bsnm{Kruse}, \binits{M.K.G.}},
\bauthor{\bsnm{Quaglioni}, \binits{S.}},
\bauthor{\bsnm{Hupin}, \binits{G.}}:
\batitle{Structure of the exotic ${ {}^{9}\text{He} }$ nucleus from the no-core
  shell model with continuum}.
\bjtitle{Phys. Rev. C}
\bvolume{97},
\bfpage{034314}
(\byear{2018})
\end{barticle}
\endbibitem

\bibitem{votaw20_2353}
\begin{barticle}
\bauthor{\bsnm{Votaw}, \binits{D.}},
\bauthor{\bsnm{{DeYoung}}, \binits{P.A.}},
\bauthor{\bsnm{Baumann}, \binits{T.}},
\bauthor{\bsnm{Blake}, \binits{A.}},
\bauthor{\bsnm{Boone}, \binits{J.}},
\bauthor{\bsnm{Brown}, \binits{J.}},
\bauthor{\bsnm{Chrisman}, \binits{D.}},
\bauthor{\bsnm{Finck}, \binits{J.E.}},
\bauthor{\bsnm{Frank}, \binits{N.}},
\bauthor{\bsnm{Gombas}, \binits{J.}},
\bauthor{\bsnm{Gu\`eye}, \binits{P.}},
\bauthor{\bsnm{Hinnefeld}, \binits{J.}},
\bauthor{\bsnm{Karrick}, \binits{H.}},
\bauthor{\bsnm{Kuchera}, \binits{A.N.}},
\bauthor{\bsnm{Liu}, \binits{H.}},
\bauthor{\bsnm{Luther}, \binits{B.}},
\bauthor{\bsnm{Ndayisabye}, \binits{F.}},
\bauthor{\bsnm{Neal}, \binits{M.}},
\bauthor{\bsnm{{Owens-Fryar}}, \binits{J.}},
\bauthor{\bsnm{Pereira}, \binits{J.}},
\bauthor{\bsnm{Persch}, \binits{C.}},
\bauthor{\bsnm{Phan}, \binits{T.}},
\bauthor{\bsnm{Redpath}, \binits{T.}},
\bauthor{\bsnm{Rogers}, \binits{W.F.}},
\bauthor{\bsnm{Stephenson}, \binits{S.}},
\bauthor{\bsnm{Stiefel}, \binits{K.}},
\bauthor{\bsnm{Sword}, \binits{C.}},
\bauthor{\bsnm{Wantz}, \binits{A.}},
\bauthor{\bsnm{Thoennessen}, \binits{M.}}:
\batitle{Low-lying level structure of the neutron-unbound ${ N = 7 }$
  isotones}.
\bjtitle{Phys. Rev. C}
\bvolume{102},
\bfpage{014325}
(\byear{2020})
\end{barticle}
\endbibitem

\bibitem{BarrancoPRC2020}
\begin{barticle}
\bauthor{\bsnm{Barranco}, \binits{F.}},
\bauthor{\bsnm{Potel}, \binits{G.}},
\bauthor{\bsnm{Vigezzi}, \binits{E.}},
\bauthor{\bsnm{Broglia}, \binits{R.A.}}:
\batitle{${ ^{9}\text{Li}(d,p) }$ reaction as a specific probe of ${
  ^{10}\text{Li} }$, the paradigm of parity-inverted nuclei around the ${ N = 6
  }$ closed shell}.
\bjtitle{Phys. Rev. C}
\bvolume{101},
\bfpage{8}
(\byear{2020})
\end{barticle}
\endbibitem

\bibitem{Efimov:1990rzd}
\begin{barticle}
\bauthor{\bsnm{Efimov}, \binits{V.}}:
\batitle{Is a qualitative approach to the three-body problem useful?}
\bjtitle{Comments Nucl. Part. Phys.}
\bvolume{19},
\bfpage{271}
(\byear{1990})
\end{barticle}
\endbibitem

\bibitem{AmorimPRC1997}
\begin{barticle}
\bauthor{\bsnm{Amorim}, \binits{A.E.A.}},
\bauthor{\bsnm{Frederico}, \binits{T.}},
\bauthor{\bsnm{Tomio}, \binits{L.}}:
\batitle{Universal aspects of {E}fimov states and light halo nuclei}.
\bjtitle{Phys. Rev. C}
\bvolume{56},
\bfpage{0}
(\byear{1997})
\end{barticle}
\endbibitem

\bibitem{Hammer:2022lhx}
\begin{botherref}
\oauthor{\bsnm{Hammer}, \binits{H.-W.}}:
Theory of halo nuclei.
arXiv:2203.13074
(2022)
\end{botherref}
\endbibitem

\bibitem{Bringas:2004zz}
\begin{barticle}
\bauthor{\bsnm{Bringas}, \binits{F.}},
\bauthor{\bsnm{Yamashita}, \binits{M.T.}},
\bauthor{\bsnm{Frederico}, \binits{T.}}:
\batitle{Triatomic continuum resonances for large negative scattering lengths}.
\bjtitle{Phys. Rev. A}
\bvolume{69},
\bfpage{040702}
(\byear{2004})
\end{barticle}
\endbibitem

\bibitem{Deltuva:2020sdd}
\begin{barticle}
\bauthor{\bsnm{Deltuva}, \binits{A.}}:
\batitle{Energies and widths of {E}fimov states in the three-boson continuum}.
\bjtitle{Phys. Rev. C}
\bvolume{102},
\bfpage{034003}
(\byear{2020})
\end{barticle}
\endbibitem

\bibitem{Dietz:2021haj}
\begin{barticle}
\bauthor{\bsnm{Dietz}, \binits{S.}},
\bauthor{\bsnm{Hammer}, \binits{H.-W.}},
\bauthor{\bsnm{K\"onig}, \binits{S.}},
\bauthor{\bsnm{Schwenk}, \binits{A.}}:
\batitle{Three-body resonances in pionless effective field theory}.
\bjtitle{Phys. Rev. C}
\bvolume{105},
\bfpage{064002}
(\byear{2022})
\end{barticle}
\endbibitem

\bibitem{HiyamaPRC2019}
\begin{barticle}
\bauthor{\bsnm{Hiyama}, \binits{E.}},
\bauthor{\bsnm{Lazauskas}, \binits{R.}},
\bauthor{\bsnm{Marqu\'es}, \binits{F.M.}},
\bauthor{\bsnm{Carbonell}, \binits{J.}}:
\batitle{Modeling ${ ^{19}\text{B} }$ as a ${ ^{17}\text{B}-n-n }$ three-body
  system in the unitary limit}.
\bjtitle{Phys. Rev. C}
\bvolume{100},
\bfpage{5}
(\byear{2019})
\end{barticle}
\endbibitem

\bibitem{DuerNat2022}
\begin{barticle}
\bauthor{\bsnm{Duer}, \binits{M.}},
\bauthor{\bparticle{et} \bsnm{al.}}:
\batitle{Observation of a correlated free four-neutron system}.
\bjtitle{Nature}
\bvolume{606},
\bfpage{682}
(\byear{2022})
\end{barticle}
\endbibitem

\bibitem{Moschini:2018lwh}
\begin{barticle}
\bauthor{\bsnm{Moschini}, \binits{L.}},
\bauthor{\bsnm{Capel}, \binits{P.}}:
\batitle{Reliable extraction of the ${ dB(\text{E1})/dE }$ for ${
  ^{11}\text{Be} }$ from its breakup at 520 {MeV}/nucleon}.
\bjtitle{Phys. Lett. B}
\bvolume{790},
\bfpage{371}
(\byear{2019})
\end{barticle}
\endbibitem

\bibitem{Moschini:2019tyx}
\begin{barticle}
\bauthor{\bsnm{Moschini}, \binits{L.}},
\bauthor{\bsnm{Yang}, \binits{J.}},
\bauthor{\bsnm{Capel}, \binits{P.}}:
\batitle{${ ^{15}\text{C} }$ : From halo effective field theory structure to
  the study of transfer, breakup, and radiative-capture reactions}.
\bjtitle{Phys. Rev. C}
\bvolume{100},
\bfpage{044615}
(\byear{2019})
\end{barticle}
\endbibitem

\bibitem{Capel:2020obz}
\begin{barticle}
\bauthor{\bsnm{Capel}, \binits{P.}},
\bauthor{\bsnm{Phillips}, \binits{D.R.}},
\bauthor{\bsnm{Hammer}, \binits{H.-W.}}:
\batitle{Simulating core excitation in breakup reactions of halo nuclei using
  an effective three-body force}.
\bjtitle{Phys. Lett. B}
\bvolume{825},
\bfpage{136847}
(\byear{2022})
\end{barticle}
\endbibitem

\bibitem{Hebborn:2021mzi}
\begin{barticle}
\bauthor{\bsnm{Hebborn}, \binits{C.}},
\bauthor{\bsnm{Capel}, \binits{P.}}:
\batitle{Halo effective field theory analysis of one-neutron knockout reactions
  of be11 and c15}.
\bjtitle{Phys. Rev. C}
\bvolume{104},
\bfpage{024616}
(\byear{2021})
\end{barticle}
\endbibitem

\bibitem{BakerPRC1999}
\begin{barticle}
\bauthor{\bsnm{Baker}, \binits{G.A.}}:
\batitle{Neutron matter model}.
\bjtitle{Phys. Rev. C}
\bvolume{60},
\bfpage{6}
(\byear{1999})
\end{barticle}
\endbibitem

\bibitem{PhysRevA.65.010705}
\begin{barticle}
\bauthor{\bsnm{Esry}, \binits{B.D.}},
\bauthor{\bsnm{Greene}, \binits{C.H.}},
\bauthor{\bsnm{Suno}, \binits{H.}}:
\batitle{Threshold laws for three-body recombination}.
\bjtitle{Phys. Rev. A}
\bvolume{65},
\bfpage{4}
(\byear{2001})
\end{barticle}
\endbibitem

\bibitem{StecherNature}
\begin{barticle}
\bauthor{\bparticle{von} \bsnm{Stecher}, \binits{D.J.P.} \bsuffix{J.}},
\bauthor{\bsnm{Greene}, \binits{C.H.}}:
\batitle{Signatures of universal four-body phenomena and their relation to the
  {E}fimov effect}.
\bjtitle{Nat. Phys.}
\bvolume{5},
\bfpage{233201}
(\byear{2009})
\end{barticle}
\endbibitem

\bibitem{NatureKramer2006}
\begin{barticle}
\bauthor{\bsnm{Kraemer}, \binits{T.}},
\bauthor{\bsnm{Mark}, \binits{M.}},
\bauthor{\bsnm{Waldburger}, \binits{P.}},
\bauthor{\bsnm{Danzl}, \binits{J.G.}},
\bauthor{\bsnm{Chin}, \binits{C.}},
\bauthor{\bsnm{Engeser}, \binits{B.}},
\bauthor{\bsnm{Lange}, \binits{A.D.}},
\bauthor{\bsnm{Pilch}, \binits{K.}},
\bauthor{\bsnm{Jaakkola}, \binits{A.}},
\bauthor{\bsnm{Nägerl}, \binits{H.-C.}},
\bauthor{\bsnm{Grimm}, \binits{R.}}:
\batitle{Evidence for {E}fimov quantum states in an ultracold gas of caesium
  atoms}.
\bjtitle{Nature}
\bvolume{440},
\bfpage{4}
(\byear{2006})
\end{barticle}
\endbibitem

\bibitem{Zenesini_2013}
\begin{barticle}
\bauthor{\bsnm{Zenesini}, \binits{A.}},
\bauthor{\bsnm{Huang}, \binits{B.}},
\bauthor{\bsnm{Berninger}, \binits{M.}},
\bauthor{\bsnm{Besler}, \binits{S.}},
\bauthor{\bsnm{Nägerl}, \binits{H.-C.}},
\bauthor{\bsnm{Ferlaino}, \binits{F.}},
\bauthor{\bsnm{Grimm}, \binits{R.}},
\bauthor{\bsnm{Greene}, \binits{C.H.}},
\bauthor{\bparticle{von} \bsnm{Stecher}, \binits{J.}}:
\batitle{Resonant five-body recombination in an ultracold gas of bosonic
  atoms}.
\bjtitle{New J. Phys.}
\bvolume{15},
\bfpage{043040}
(\byear{2013})
\end{barticle}
\endbibitem

\bibitem{PhysRevA.79.030501}
\begin{barticle}
\bauthor{\bsnm{D'Incao}, \binits{J.P.}},
\bauthor{\bsnm{Rittenhouse}, \binits{S.T.}},
\bauthor{\bsnm{Mehta}, \binits{N.P.}},
\bauthor{\bsnm{Greene}, \binits{C.H.}}:
\batitle{Dimer-dimer collisions at finite energies in two-component fermi
  gases}.
\bjtitle{Phys. Rev. A}
\bvolume{79},
\bfpage{4}
(\byear{2009})
\end{barticle}
\endbibitem

\bibitem{PhysRevA.77.043619}
\begin{barticle}
\bauthor{\bparticle{von} \bsnm{Stecher}, \binits{J.}},
\bauthor{\bsnm{Greene}, \binits{C.H.}},
\bauthor{\bsnm{Blume}, \binits{D.}}:
\batitle{Energetics and structural properties of trapped two-component fermi
  gases}.
\bjtitle{Phys. Rev. A}
\bvolume{77},
\bfpage{20}
(\byear{2008})
\end{barticle}
\endbibitem

\bibitem{PhysRevLett.99.090402}
\begin{barticle}
\bauthor{\bparticle{von} \bsnm{Stecher}, \binits{J.}},
\bauthor{\bsnm{Greene}, \binits{C.H.}}:
\batitle{Spectrum and dynamics of the bcs-bec crossover from a few-body
  perspective}.
\bjtitle{Phys. Rev. Lett.}
\bvolume{99},
\bfpage{4}
(\byear{2007})
\end{barticle}
\endbibitem

\bibitem{KievskyARNP2021}
\begin{botherref}
\oauthor{\bsnm{Kievsky}, \binits{A.}},
\oauthor{\bsnm{Gattobigio}, \binits{M.}},
\oauthor{\bsnm{Girlanda}, \binits{L.}},
\oauthor{\bsnm{Viviani}, \binits{M.}}:
Efimov physics and connections to nuclear physics.
Annu. Rev. Nucl. Part. Sci.
\textbf{71}
(2021)
\end{botherref}
\endbibitem

\bibitem{PhysRevA.86.052709}
\begin{barticle}
\bauthor{\bsnm{Garrido}, \binits{E.}},
\bauthor{\bsnm{Romero-Redondo}, \binits{C.}},
\bauthor{\bsnm{Kievsky}, \binits{A.}},
\bauthor{\bsnm{Viviani}, \binits{M.}}:
\batitle{Integral relations and the adiabatic expansion method for 1+2
  reactions above the breakup threshold: Helium trimers with soft-core
  potentials}.
\bjtitle{Phys. Rev. A}
\bvolume{86},
\bfpage{9}
(\byear{2012})
\end{barticle}
\endbibitem

\bibitem{PhysRevC.102.034007}
\begin{barticle}
\bauthor{\bsnm{Viviani}, \binits{M.}},
\bauthor{\bsnm{Girlanda}, \binits{L.}},
\bauthor{\bsnm{Kievsky}, \binits{A.}},
\bauthor{\bsnm{Marcucci}, \binits{L.E.}}:
\batitle{${ n+^{3}\text{H} }$, ${ p+^{3}\text{He} }$, ${ p+^{3}\text{H} }$, and
  ${ n+^{3}\text{He} }$ scattering with the hyperspherical harmonic method}.
\bjtitle{Phys. Rev. C}
\bvolume{102},
\bfpage{32}
(\byear{2020})
\end{barticle}
\endbibitem

\bibitem{PhysRevC.105.014001}
\begin{barticle}
\bauthor{\bsnm{Viviani}, \binits{M.}},
\bauthor{\bsnm{Filandri}, \binits{E.}},
\bauthor{\bsnm{Girlanda}, \binits{L.}},
\bauthor{\bsnm{Gustavino}, \binits{C.}},
\bauthor{\bsnm{Kievsky}, \binits{A.}},
\bauthor{\bsnm{Marcucci}, \binits{L.E.}},
\bauthor{\bsnm{Schiavilla}, \binits{R.}}:
\batitle{${ X17 }$ boson and the ${ ^{3}\text{H}(p,e^+ e^-)^{4}\text{He} }$ and
  ${ ^{3}\text{He}(n,e^+ e^-)^{4}\text{He} }$ processes: A theoretical
  analysis}.
\bjtitle{Phys. Rev. C}
\bvolume{105},
\bfpage{30}
(\byear{2022})
\end{barticle}
\endbibitem

\bibitem{PhysRevLett.125.052501}
\begin{barticle}
\bauthor{\bsnm{Higgins}, \binits{M.D.}},
\bauthor{\bsnm{Greene}, \binits{C.H.}},
\bauthor{\bsnm{Kievsky}, \binits{A.}},
\bauthor{\bsnm{Viviani}, \binits{M.}}:
\batitle{Nonresonant density of states enhancement at low energies for three or
  four neutrons}.
\bjtitle{Phys. Rev. Lett.}
\bvolume{125},
\bfpage{6}
(\byear{2020})
\end{barticle}
\endbibitem

\bibitem{PhysRevC.103.024004}
\begin{barticle}
\bauthor{\bsnm{Higgins}, \binits{M.D.}},
\bauthor{\bsnm{Greene}, \binits{C.H.}},
\bauthor{\bsnm{Kievsky}, \binits{A.}},
\bauthor{\bsnm{Viviani}, \binits{M.}}:
\batitle{Comprehensive study of the three- and four-neutron systems at low
  energies}.
\bjtitle{Phys. Rev. C}
\bvolume{103},
\bfpage{12}
(\byear{2021})
\end{barticle}
\endbibitem

\bibitem{PhysRevC.77.054313}
\begin{barticle}
\bauthor{\bsnm{Romero-Redondo}, \binits{C.}},
\bauthor{\bsnm{Garrido}, \binits{E.}},
\bauthor{\bsnm{Fedorov}, \binits{D.V.}},
\bauthor{\bsnm{Jensen}, \binits{A.S.}}:
\batitle{Three-body structure of low-lying ${ ^{12}\text{Be} }$ states}.
\bjtitle{Phys. Rev. C}
\bvolume{77},
\bfpage{15}
(\byear{2008})
\end{barticle}
\endbibitem

\bibitem{PhysRevC.86.024310}
\begin{barticle}
\bauthor{\bsnm{Garrido}, \binits{E.}},
\bauthor{\bsnm{Jensen}, \binits{A.S.}},
\bauthor{\bsnm{Fedorov}, \binits{D.V.}},
\bauthor{\bsnm{Johansen}, \binits{J.G.}}:
\batitle{Three-body properties of low-lying ${ {}^{12}\text{Be} }$ resonances}.
\bjtitle{Phys. Rev. C}
\bvolume{86},
\bfpage{12}
(\byear{2012})
\end{barticle}
\endbibitem

\bibitem{Rittenhouse_2011}
\begin{barticle}
\bauthor{\bsnm{Rittenhouse}, \binits{S.T.}},
\bauthor{\bparticle{von} \bsnm{Stecher}, \binits{J.}},
\bauthor{\bsnm{D'Incao}, \binits{J.P.}},
\bauthor{\bsnm{Mehta}, \binits{N.P.}},
\bauthor{\bsnm{Greene}, \binits{C.H.}}:
\batitle{The hyperspherical four-fermion problem}.
\bjtitle{J. Phys. B: Atom. Mol. Opt. Phys.}
\bvolume{44},
\bfpage{172001}
(\byear{2011})
\end{barticle}
\endbibitem

\bibitem{BlumeGreene2000}
\begin{barticle}
\bauthor{\bsnm{Blume}, \binits{D.}},
\bauthor{\bsnm{Greene}, \binits{C.H.}}:
\batitle{Monte carlo hyperspherical description of helium cluster excited
  states}.
\bjtitle{J. Chem. Phys.}
\bvolume{112},
\bfpage{067}
(\byear{2000})
\end{barticle}
\endbibitem

\bibitem{PhysRevA.105.022824}
\begin{barticle}
\bauthor{\bsnm{Yates}, \binits{A.J.}},
\bauthor{\bsnm{Blume}, \binits{D.}}:
\batitle{Structural properties of ${ ^{4}\text{He}_{N} }$ ${ (N = 2-10) }$
  clusters for different potential models at the physical point and at
  unitarity}.
\bjtitle{Phys. Rev. A}
\bvolume{105},
\bfpage{14}
(\byear{2022})
\end{barticle}
\endbibitem

\bibitem{PhysRevLett.116.052501}
\begin{barticle}
\bauthor{\bsnm{Kisamori}, \binits{K.}},
\bauthor{\bsnm{Shimoura}, \binits{S.}},
\bauthor{\bsnm{Miya}, \binits{H.}},
\bauthor{\bsnm{Michimasa}, \binits{S.}},
\bauthor{\bsnm{Ota}, \binits{S.}},
\bauthor{\bsnm{Assie}, \binits{M.}},
\bauthor{\bsnm{Baba}, \binits{H.}},
\bauthor{\bsnm{Baba}, \binits{T.}},
\bauthor{\bsnm{Beaumel}, \binits{D.}},
\bauthor{\bsnm{Dozono}, \binits{M.}},
\bauthor{\bsnm{Fujii}, \binits{T.}},
\bauthor{\bsnm{Fukuda}, \binits{N.}},
\bauthor{\bsnm{Go}, \binits{S.}},
\bauthor{\bsnm{Hammache}, \binits{F.}},
\bauthor{\bsnm{Ideguchi}, \binits{E.}},
\bauthor{\bsnm{Inabe}, \binits{N.}},
\bauthor{\bsnm{Itoh}, \binits{M.}},
\bauthor{\bsnm{Kameda}, \binits{D.}},
\bauthor{\bsnm{Kawase}, \binits{S.}},
\bauthor{\bsnm{Kawabata}, \binits{T.}},
\bauthor{\bsnm{Kobayashi}, \binits{M.}},
\bauthor{\bsnm{Kondo}, \binits{Y.}},
\bauthor{\bsnm{Kubo}, \binits{T.}},
\bauthor{\bsnm{Kubota}, \binits{Y.}},
\bauthor{\bsnm{Kurata-Nishimura}, \binits{M.}},
\bauthor{\bsnm{Lee}, \binits{C.S.}},
\bauthor{\bsnm{Maeda}, \binits{Y.}},
\bauthor{\bsnm{Matsubara}, \binits{H.}},
\bauthor{\bsnm{Miki}, \binits{K.}},
\bauthor{\bsnm{Nishi}, \binits{T.}},
\bauthor{\bsnm{Noji}, \binits{S.}},
\bauthor{\bsnm{Sakaguchi}, \binits{S.}},
\bauthor{\bsnm{Sakai}, \binits{H.}},
\bauthor{\bsnm{Sasamoto}, \binits{Y.}},
\bauthor{\bsnm{Sasano}, \binits{M.}},
\bauthor{\bsnm{Sato}, \binits{H.}},
\bauthor{\bsnm{Shimizu}, \binits{Y.}},
\bauthor{\bsnm{Stolz}, \binits{A.}},
\bauthor{\bsnm{Suzuki}, \binits{H.}},
\bauthor{\bsnm{Takaki}, \binits{M.}},
\bauthor{\bsnm{Takeda}, \binits{H.}},
\bauthor{\bsnm{Takeuchi}, \binits{S.}},
\bauthor{\bsnm{Tamii}, \binits{A.}},
\bauthor{\bsnm{Tang}, \binits{L.}},
\bauthor{\bsnm{Tokieda}, \binits{H.}},
\bauthor{\bsnm{Tsumura}, \binits{M.}},
\bauthor{\bsnm{Uesaka}, \binits{T.}},
\bauthor{\bsnm{Yako}, \binits{K.}},
\bauthor{\bsnm{Yanagisawa}, \binits{Y.}},
\bauthor{\bsnm{Yokoyama}, \binits{R.}},
\bauthor{\bsnm{Yoshida}, \binits{K.}}:
\batitle{Candidate resonant tetraneutron state populated by the ${
  ^{4}\text{He}(^{8}\text{He},^{8}\text{Be}) }$ reaction}.
\bjtitle{Phys. Rev. Lett.}
\bvolume{116},
\bfpage{5}
(\byear{2016})
\end{barticle}
\endbibitem

\bibitem{faddeev1960mathematical}
\begin{barticle}
\bauthor{\bsnm{Faddeev}, \binits{L.D.}}:
\batitle{Mathematical aspects of the three-body problem in quantum scattering
  theory}.
\bjtitle{Zh. Eksp. Teor. Fiz}
\bvolume{39},
\bfpage{1459}
(\byear{196})
\end{barticle}
\endbibitem

\bibitem{alt1967reduction}
\begin{barticle}
\bauthor{\bsnm{Alt}, \binits{E.}},
\bauthor{\bsnm{Grassberger}, \binits{P.}},
\bauthor{\bsnm{Sandhas}, \binits{W.}}:
\batitle{Reduction of the three-particle collision problem to multi-channel
  two-particle lippmann-schwinger equations}.
\bjtitle{Nucl. Phys. B}
\bvolume{2},
\bfpage{180}
(\byear{1967})
\end{barticle}
\endbibitem

\bibitem{deltuva2005momentum}
\begin{barticle}
\bauthor{\bsnm{Deltuva}, \binits{A.}},
\bauthor{\bsnm{Fonseca}, \binits{A.}},
\bauthor{\bsnm{Sauer}, \binits{P.}}:
\batitle{Momentum-space treatment of the {C}oulomb interaction in three-nucleon
  reactions with two protons}.
\bjtitle{Phys. Rev. C}
\bvolume{71},
\bfpage{054005}
(\byear{2005})
\end{barticle}
\endbibitem

\bibitem{deltuva2005benchmark}
\begin{barticle}
\bauthor{\bsnm{Deltuva}, \binits{A.}},
\bauthor{\bsnm{Fonseca}, \binits{A.}},
\bauthor{\bsnm{Kievsky}, \binits{A.}},
\bauthor{\bsnm{Rosati}, \binits{S.}},
\bauthor{\bsnm{Sauer}, \binits{P.}},
\bauthor{\bsnm{Viviani}, \binits{M.}}:
\batitle{Benchmark calculation for proton-deuteron elastic scattering
  observables including the {C}oulomb interaction}.
\bjtitle{Phys. Rev. C}
\bvolume{71},
\bfpage{064003}
(\byear{2005})
\end{barticle}
\endbibitem

\bibitem{deltuva2006coulomb}
\begin{barticle}
\bauthor{\bsnm{Deltuva}, \binits{A.}}:
\batitle{Coulomb force effects in low-energy ${ \alpha }$-deuteron scattering}.
\bjtitle{Phys. Rev. C}
\bvolume{74},
\bfpage{064001}
(\byear{2006})
\end{barticle}
\endbibitem

\bibitem{deltuva2007three}
\begin{barticle}
\bauthor{\bsnm{Deltuva}, \binits{A.}},
\bauthor{\bsnm{Moro}, \binits{A.}},
\bauthor{\bsnm{Cravo}, \binits{E.}},
\bauthor{\bsnm{Nunes}, \binits{F.M.}},
\bauthor{\bsnm{Fonseca}, \binits{A.}}:
\batitle{Three-body description of direct nuclear reactions: Comparison with
  the continuum discretized coupled channels method}.
\bjtitle{Phys. Rev. C}
\bvolume{76},
\bfpage{064602}
(\byear{2007})
\end{barticle}
\endbibitem

\bibitem{MERKURIEV1980395:1980}
\begin{barticle}
\bauthor{\bsnm{Merkuriev}, \binits{S.P.}}:
\batitle{On the three-body {C}oulomb scattering problem}.
\bjtitle{Ann. Phys.}
\bvolume{130},
\bfpage{26}
(\byear{1980})
\end{barticle}
\endbibitem

\bibitem{lazauskas2011application}
\begin{barticle}
\bauthor{\bsnm{Lazauskas}, \binits{R.}},
\bauthor{\bsnm{Carbonell}, \binits{J.}}:
\batitle{Application of the complex-scaling method to few-body scattering}.
\bjtitle{Phys. Rev. C}
\bvolume{84},
\bfpage{034002}
(\byear{2011})
\end{barticle}
\endbibitem

\bibitem{Deltuva2014}
\begin{bbook}
\bauthor{\bsnm{Deltuva}, \binits{A.}},
\bauthor{\bsnm{Fonseca}, \binits{A.C.}},
\bauthor{\bsnm{Lazauskas}, \binits{R.}}:
In: \beditor{\bsnm{Beck}, \binits{C.}} (ed.)
\bbtitle{Faddeev Equation Approach for Three-Cluster Nuclear Reactions},
p. \bfpage{1}.
\bpublisher{Springer},
\blocation{Cham}
(\byear{2014})
\end{bbook}
\endbibitem

\bibitem{Nuttall:1969nap}
\begin{barticle}
\bauthor{\bsnm{Nuttall}, \binits{J.}},
\bauthor{\bsnm{Cohen}, \binits{H.L.}}:
\batitle{Method of complex coordinates for three-body calculations above the
  breakup threshold}.
\bjtitle{Phys. Rev.}
\bvolume{188},
\bfpage{1542}
(\byear{1966})
\end{barticle}
\endbibitem

\bibitem{Balslev:1971vb}
\begin{barticle}
\bauthor{\bsnm{Balslev}, \binits{E.}},
\bauthor{\bsnm{Combes}, \binits{J.M.}}:
\batitle{Spectral properties of many-body schroedinger operators with
  dilatation-analytic interactions}.
\bjtitle{Commun. Math. Phys.}
\bvolume{22},
\bfpage{280}
(\byear{1971})
\end{barticle}
\endbibitem

\bibitem{Mukhamedzhanov:2000qg}
\begin{barticle}
\bauthor{\bsnm{Mukhamedzhanov}, \binits{A.M.}},
\bauthor{\bsnm{Alt}, \binits{E.O.}},
\bauthor{\bsnm{Avakov}, \binits{G.V.}}:
\batitle{Momentum space integral equations for three charged particles:
  Nondiagonal kernels}.
\bjtitle{Phys. Rev. C}
\bvolume{61},
\bfpage{064006}
(\byear{2000})
\end{barticle}
\endbibitem

\bibitem{Mukhamedzhanov:2000nt}
\begin{barticle}
\bauthor{\bsnm{Mukhamedzhanov}, \binits{A.M.}},
\bauthor{\bsnm{Alt}, \binits{E.O.}},
\bauthor{\bsnm{Avakov}, \binits{G.V.}}:
\batitle{Momentum space integral equations for three charged particles:
  Diagonal kernels}.
\bjtitle{Phys. Rev. C}
\bvolume{63},
\bfpage{044005}
(\byear{2001})
\end{barticle}
\endbibitem

\bibitem{Mukhamedzhanov:2012qv}
\begin{barticle}
\bauthor{\bsnm{Mukhamedzhanov}, \binits{A.M.}},
\bauthor{\bsnm{Eremenko}, \binits{V.}},
\bauthor{\bsnm{Sattarov}, \binits{A.I.}}:
\batitle{Generalized {F}addeev equations in the ags form for deuteron stripping
  with explicit inclusion of target excitations and {C}oulomb interaction}.
\bjtitle{Phys. Rev. C}
\bvolume{86},
\bfpage{034001}
(\byear{2012})
\end{barticle}
\endbibitem

\bibitem{Alt:1970xe}
\begin{barticle}
\bauthor{\bsnm{Alt}, \binits{E.O.}},
\bauthor{\bsnm{Grassberger}, \binits{P.}},
\bauthor{\bsnm{Sandhas}, \binits{W.}}:
\batitle{Treatment of the three- and four-nucleon systems by a generalized
  separable-potential model}.
\bjtitle{Phys. Rev. C}
\bvolume{1},
\bfpage{85}
(\byear{1970})
\end{barticle}
\endbibitem

\bibitem{Deltuva:2007xv}
\begin{barticle}
\bauthor{\bsnm{Deltuva}, \binits{A.}},
\bauthor{\bsnm{Fonseca}, \binits{A.C.}}:
\batitle{{Four-body calculation of proton-$^3$He scattering}}.
\bjtitle{Phys. Rev. Lett.}
\bvolume{98},
\bfpage{162502}
(\byear{2007})
\end{barticle}
\endbibitem

\bibitem{Deltuva:2006sz}
\begin{barticle}
\bauthor{\bsnm{Deltuva}, \binits{A.}},
\bauthor{\bsnm{Fonseca}, \binits{A.C.}}:
\batitle{Four-nucleon scattering: \textit{Ab initio} calculations in momentum
  space}.
\bjtitle{Phys. Rev. C}
\bvolume{75},
\bfpage{014005}
(\byear{2007})
\end{barticle}
\endbibitem

\bibitem{Lazauskas:2012jc}
\begin{barticle}
\bauthor{\bsnm{Lazauskas}, \binits{R.}}:
\batitle{Application of the complex-scaling method to four-nucleon scattering
  above break-up threshold}.
\bjtitle{Phys. Rev. C}
\bvolume{86},
\bfpage{044002}
(\byear{2012})
\end{barticle}
\endbibitem

\bibitem{Lazauskas:2017paz}
\begin{barticle}
\bauthor{\bsnm{Lazauskas}, \binits{R.}}:
\batitle{Solution of the ${ n-^{4}\text{He} }$ elastic scattering problem using
  the {F}addeev-{Y}akubovsky equations}.
\bjtitle{Phys. Rev. C}
\bvolume{97},
\bfpage{044002}
(\byear{2018})
\end{barticle}
\endbibitem

\bibitem{IAV85}
\begin{barticle}
\bauthor{\bsnm{Ichimura}, \binits{M.}},
\bauthor{\bsnm{Austern}, \binits{N.}},
\bauthor{\bsnm{Vincent}, \binits{C.M.}}:
\batitle{Equivalence of post and prior sum rules for inclusive breakup
  reactions}.
\bjtitle{Phys. Rev. C}
\bvolume{32},
\bfpage{0}
(\byear{1985})
\end{barticle}
\endbibitem

\bibitem{LMo15}
\begin{barticle}
\bauthor{\bsnm{Lei}, \binits{J.}},
\bauthor{\bsnm{Moro}, \binits{A.M.}}:
\batitle{Reexamining closed-form formulae for inclusive breakup: Application to
  deuteron- and ${ ^{6}\text{Li} }$-induced reactions}.
\bjtitle{Phys. Rev. C}
\bvolume{92},
\bfpage{14}
(\byear{2015})
\end{barticle}
\endbibitem

\bibitem{LMo17}
\begin{barticle}
\bauthor{\bsnm{Lei}, \binits{J.}},
\bauthor{\bsnm{Moro}, \binits{A.M.}}:
\batitle{Comprehensive analysis of large ${ \alpha }$ yields observed in ${
  ^{6}\text{Li} }$-induced reactions}.
\bjtitle{Phys. Rev. C}
\bvolume{95},
\bfpage{11}
(\byear{2017})
\end{barticle}
\endbibitem

\bibitem{LMo19}
\begin{barticle}
\bauthor{\bsnm{Lei}, \binits{J.}},
\bauthor{\bsnm{Moro}, \binits{A.M.}}:
\batitle{Unraveling the reaction mechanisms leading to partial fusion of weakly
  bound nuclei}.
\bjtitle{Phys. Rev. Lett.}
\bvolume{123},
\bfpage{6}
(\byear{2019})
\end{barticle}
\endbibitem

\bibitem{SCC21}
\begin{barticle}
\bauthor{\bsnm{Souza}, \binits{L.A.}},
\bauthor{\bsnm{Chimanski}, \binits{E.V.}},
\bauthor{\bsnm{Carlson}, \binits{B.V.}}:
\batitle{Inclusive breakup cross sections in reactions induced by the nuclides
  ${ ^{6}\text{He} }$ and ${ ^{6,7}\text{Li} }$ in the two-body cluster model}.
\bjtitle{Phys. Rev. C}
\bvolume{104},
\bfpage{11}
(\byear{2021})
\end{barticle}
\endbibitem

\bibitem{YLY21}
\begin{barticle}
\bauthor{\bsnm{Yang}, \binits{L.}},
\bauthor{\bsnm{Lin}, \binits{C.J.}},
\bauthor{\bsnm{Yamaguchi}, \binits{H.}},
\bauthor{\bsnm{Lei}, \binits{J.}},
\bauthor{\bsnm{Wen}, \binits{P.W.}},
\bauthor{\bsnm{Mazzocco}, \binits{M.}},
\bauthor{\bsnm{Ma}, \binits{N.R.}},
\bauthor{\bsnm{Sun}, \binits{L.J.}},
\bauthor{\bsnm{Wang}, \binits{D.X.}},
\bauthor{\bsnm{Zhang}, \binits{G.X.}},
\bauthor{\bsnm{Abe}, \binits{K.}},
\bauthor{\bsnm{Cha}, \binits{S.M.}},
\bauthor{\bsnm{Chae}, \binits{K.Y.}},
\bauthor{\bsnm{Diaz-Torres}, \binits{A.}},
\bauthor{\bsnm{Ferreira}, \binits{J.L.}},
\bauthor{\bsnm{Hayakawa}, \binits{S.}},
\bauthor{\bsnm{Jia}, \binits{H.M.}},
\bauthor{\bsnm{Kahl}, \binits{D.}},
\bauthor{\bsnm{Kim}, \binits{A.}},
\bauthor{\bsnm{Kwag}, \binits{M.S.}},
\bauthor{\bsnm{{La Commara}}, \binits{M.}},
\bauthor{\bsnm{{Navarro Pérez}}, \binits{R.}},
\bauthor{\bsnm{Parascandolo}, \binits{C.}},
\bauthor{\bsnm{Pierroutsakou}, \binits{D.}},
\bauthor{\bsnm{Rangel}, \binits{J.}},
\bauthor{\bsnm{Sakaguchi}, \binits{Y.}},
\bauthor{\bsnm{Signorini}, \binits{C.}},
\bauthor{\bsnm{Strano}, \binits{E.}},
\bauthor{\bsnm{Xu}, \binits{X.X.}},
\bauthor{\bsnm{Yang}, \binits{F.}},
\bauthor{\bsnm{Yang}, \binits{Y.Y.}},
\bauthor{\bsnm{Zhang}, \binits{G.L.}},
\bauthor{\bsnm{Zhong}, \binits{F.P.}},
\bauthor{\bsnm{Lubian}, \binits{J.}}:
\batitle{Insight into the reaction dynamics of proton drip-line nuclear system
  ${ ^{17}\text{F} + ^{58}\text{Ni} }$ at near-barrier energies}.
\bjtitle{Phys. Lett. B}
\bvolume{813},
\bfpage{136045}
(\byear{2021})
\end{barticle}
\endbibitem

\bibitem{RCL20}
\begin{barticle}
\bauthor{\bsnm{Rangel}, \binits{J.}},
\bauthor{\bsnm{Cortes}, \binits{M.R.}},
\bauthor{\bsnm{Lubian}, \binits{J.}},
\bauthor{\bsnm{Canto}, \binits{L.F.}}:
\batitle{Theory of complete and incomplete fusion of weakly bound systems}.
\bjtitle{Phys. Lett. B}
\bvolume{803},
\bfpage{135337}
(\byear{2020})
\end{barticle}
\endbibitem

\bibitem{CRF20}
\begin{barticle}
\bauthor{\bsnm{Cortes}, \binits{M.R.}},
\bauthor{\bsnm{Rangel}, \binits{J.}},
\bauthor{\bsnm{Ferreira}, \binits{J.L.}},
\bauthor{\bsnm{Lubian}, \binits{J.}},
\bauthor{\bsnm{Canto}, \binits{L.F.}}:
\batitle{Complete and incomplete fusion of ${ ^{7}\text{Li} }$ projectiles on
  heavy targets}.
\bjtitle{Phys. Rev. C}
\bvolume{102},
\bfpage{16}
(\byear{2020})
\end{barticle}
\endbibitem

\bibitem{LFR22}
\begin{barticle}
\bauthor{\bsnm{Lubian}, \binits{J.}},
\bauthor{\bsnm{Ferreira}, \binits{J.L.}},
\bauthor{\bsnm{Rangel}, \binits{J.}},
\bauthor{\bsnm{Cortes}, \binits{M.R.}},
\bauthor{\bsnm{Canto}, \binits{L.F.}}:
\batitle{Fusion processes in collisions of ${ ^{6}\text{Li} }$ beams on heavy
  targets}.
\bjtitle{Phys. Rev. C}
\bvolume{105},
\bfpage{054601}
(\byear{2022})
\end{barticle}
\endbibitem

\bibitem{CCG02}
\begin{barticle}
\bauthor{\bsnm{Chamon}, \binits{L.C.}},
\bauthor{\bsnm{Carlson}, \binits{B.V.}},
\bauthor{\bsnm{Gasques}, \binits{L.R.}},
\bauthor{\bsnm{Pereira}, \binits{D.}},
\bauthor{\bsnm{De~Conti}, \binits{C.}},
\bauthor{\bsnm{Alvarez}, \binits{M.A.G.}},
\bauthor{\bsnm{Hussein}, \binits{M.S.}},
\bauthor{\bsnm{C\^andido~Ribeiro}, \binits{M.A.}},
\bauthor{\bsnm{Rossi}, \binits{E.S.}},
\bauthor{\bsnm{Silva}, \binits{C.P.}}:
\batitle{Toward a global description of the nucleus-nucleus interaction}.
\bjtitle{Phys. Rev. C}
\bvolume{66},
\bfpage{014610}
(\byear{2002})
\end{barticle}
\endbibitem

\bibitem{CCG21}
\begin{barticle}
\bauthor{\bsnm{Chamon}, \binits{L.C.}},
\bauthor{\bsnm{Carlson}, \binits{B.V.}},
\bauthor{\bsnm{Gasques}, \binits{L.R.}}:
\batitle{São paulo potential version 2 (spp2) and brazilian nuclear potential
  (bnp)}.
\bjtitle{Comput. Phys. Commun.}
\bvolume{267},
\bfpage{108061}
(\byear{2021})
\end{barticle}
\endbibitem

\bibitem{RNT01}
\begin{barticle}
\bauthor{\bsnm{Raman}, \binits{S.}},
\bauthor{\bsnm{Nestor}, \binits{C.W.}},
\bauthor{\bsnm{Tikkanen}, \binits{P.}}:
\batitle{Transition probability from the ground to the first-excited $2^+$
  state of even–even nuclides}.
\bjtitle{At. Data Nucl. Data Tables}
\bvolume{78},
\bfpage{28}
(\byear{2001})
\end{barticle}
\endbibitem

\bibitem{KiS02}
\begin{barticle}
\bauthor{\bsnm{Kibédi}, \binits{T.}},
\bauthor{\bsnm{Spear}, \binits{R.H.}}:
\batitle{Reduced electric-octupole transition probabilities, ${ B(E3;0_1^+ \to
  3_1^-) }$—an update}.
\bjtitle{At. Data Nucl. Data Tables}
\bvolume{80},
\bfpage{2}
(\byear{2002})
\end{barticle}
\endbibitem

\bibitem{Potel:10}
\begin{barticle}
\bauthor{\bsnm{Potel}, \binits{G.}},
\bauthor{\bsnm{Barranco}, \binits{F.}},
\bauthor{\bsnm{Vigezzi}, \binits{E.}},
\bauthor{\bsnm{Broglia}, \binits{R.A.}}:
\batitle{Evidence for phonon mediated pairing interaction in the halo of the
  nucleus ${ ^{11}\text{Li} }$}.
\bjtitle{Phys. Rev. Lett.}
\bvolume{105},
\bfpage{172502}
(\byear{2010})
\end{barticle}
\endbibitem

\bibitem{Descouvemont:21}
\begin{barticle}
\bauthor{\bsnm{Descouvemont}, \binits{P.}}:
\batitle{Halo effects in the ${ ^{11}\text{Li}(p,t)^{9}\text{Li} }$ reaction}.
\bjtitle{Phys. Rev. C}
\bvolume{104},
\bfpage{9}
(\byear{2021})
\end{barticle}
\endbibitem

\bibitem{Cappuzzello:15}
\begin{barticle}
\bauthor{\bsnm{Cappuzzello}, \binits{F.}},
\bauthor{\bsnm{Carbone}, \binits{D.}},
\bauthor{\bsnm{Cavallaro}, \binits{M.}},
\bauthor{\bsnm{{Bond\'i}}, \binits{M.}},
\bauthor{\bsnm{Agodi}, \binits{C.}},
\bauthor{\bsnm{Azaiez}, \binits{F.}},
\bauthor{\bsnm{Bonaccorso}, \binits{A.}},
\bauthor{\bsnm{Cunsolo}, \binits{A.}},
\bauthor{\bsnm{Fortunato}, \binits{L.}},
\bauthor{\bsnm{Foti}, \binits{A.}},
\bauthor{\bsnm{Franchoo}, \binits{S.}},
\bauthor{\bsnm{Khan}, \binits{E.}},
\bauthor{\bsnm{Linares}, \binits{R.}},
\bauthor{\bsnm{Lubian}, \binits{J.}},
\bauthor{\bsnm{Scarpaci}, \binits{J.A.}},
\bauthor{\bsnm{Vitturi}, \binits{A.}}:
\batitle{Signatures of the giant pairing vibration in the ${ ^{14}\text{C} }$
  and ${ ^{15}\text{C} }$ atomic nuclei.}
\bjtitle{Nature Communications}
\bvolume{6},
\bfpage{6743}
(\byear{2015})
\end{barticle}
\endbibitem

\bibitem{Barranco:17b}
\begin{barticle}
\bauthor{\bsnm{Barranco}, \binits{F.}},
\bauthor{\bsnm{Potel}, \binits{G.}},
\bauthor{\bsnm{Broglia}, \binits{R.A.}},
\bauthor{\bsnm{Vigezzi}, \binits{E.}}:
\batitle{Structure and reactions of ${ ^{11}\text{Be} }$: many-body basis for
  single-neutron halo}.
\bjtitle{Phys. Rev. Lett.}
\bvolume{119},
\bfpage{082501}
(\byear{2017})
\end{barticle}
\endbibitem

\bibitem{Cavallaro:19}
\begin{barticle}
\bauthor{\bsnm{Cavallaro}, \binits{M.}},
\bauthor{\bsnm{Cappuzzello}, \binits{F.}},
\bauthor{\bsnm{Carbone}, \binits{D.}},
\bauthor{\bsnm{Agodi}, \binits{C.}}:
\batitle{Giant pairing vibrations in light nuclei}.
\bjtitle{Eur. Phys. J. A}
\bvolume{55},
\bfpage{244}
(\byear{2019})
\end{barticle}
\endbibitem

\bibitem{Assie:19}
\begin{barticle}
\bauthor{\bsnm{{Assi\'e, M.}}},
\bauthor{\bsnm{{Dasso, C. H.}}},
\bauthor{\bsnm{{Liotta, R. J.}}},
\bauthor{\bsnm{{Macchiavelli, A. O.}}},
\bauthor{\bsnm{{Vitturi, A.}}}:
\batitle{The giant pairing vibration in heavy nuclei - present status and
  future studies}.
\bjtitle{Eur. Phys. J. A}
\bvolume{55}(\bissue{12}),
\bfpage{245}
(\byear{2019}).
\doiurl{10.1140/epja/i2019-12829-8}
\end{barticle}
\endbibitem

\bibitem{AUMANN2021103847}
\begin{barticle}
\bauthor{\bsnm{Aumann}, \binits{T.}},
\bauthor{\bsnm{Barbieri}, \binits{C.}},
\bauthor{\bsnm{Bazin}, \binits{D.}},
\bauthor{\bsnm{Bertulani}, \binits{C.A.}},
\bauthor{\bsnm{Bonaccorso}, \binits{A.}},
\bauthor{\bsnm{Dickhoff}, \binits{W.H.}},
\bauthor{\bsnm{Gade}, \binits{A.}},
\bauthor{\bsnm{Gómez-Ramos}, \binits{M.}},
\bauthor{\bsnm{Kay}, \binits{B.P.}},
\bauthor{\bsnm{Moro}, \binits{A.M.}},
\bauthor{\bsnm{Nakamura}, \binits{T.}},
\bauthor{\bsnm{Obertelli}, \binits{A.}},
\bauthor{\bsnm{Ogata}, \binits{K.}},
\bauthor{\bsnm{Paschalis}, \binits{S.}},
\bauthor{\bsnm{Uesaka}, \binits{T.}}:
\batitle{Quenching of single-particle strength from direct reactions with
  stable and rare-isotope beams}.
\bjtitle{Prog. Part. Nucl. Phys.}
\bvolume{118},
\bfpage{103847}
(\byear{2021})
\end{barticle}
\endbibitem

\bibitem{GM18}
\begin{barticle}
\bauthor{\bsnm{Gómez-Ramos}, \binits{M.}},
\bauthor{\bsnm{Moro}, \binits{A.M.}}:
\batitle{Binding-energy independence of reduced spectroscopic strengths derived
  from ${ (p,2p) }$ and ${ (p,pn) }$ reactions with nitrogen and oxygen
  isotopes}.
\bjtitle{Phys. Lett. B}
\bvolume{785},
\bfpage{511}
(\byear{2018})
\end{barticle}
\endbibitem

\bibitem{LeePRC06}
\begin{barticle}
\bauthor{\bsnm{Lee}, \binits{J.}},
\bauthor{\bsnm{Tostevin}, \binits{J.A.}},
\bauthor{\bsnm{Brown}, \binits{B.A.}},
\bauthor{\bsnm{Delaunay}, \binits{F.}},
\bauthor{\bsnm{Lynch}, \binits{W.G.}},
\bauthor{\bsnm{Saelim}, \binits{M.J.}},
\bauthor{\bsnm{Tsang}, \binits{M.B.}}:
\batitle{Reduced neutron spectroscopic factors when using potential geometries
  constrained by hartree-fock calculations}.
\bjtitle{Phys. Rev. C}
\bvolume{73},
\bfpage{044608}
(\byear{2006})
\end{barticle}
\endbibitem

\bibitem{Tetal09}
\begin{barticle}
\bauthor{\bsnm{Tsang}, \binits{M.B.}},
\bauthor{\bsnm{Lee}, \binits{J.}},
\bauthor{\bsnm{Su}, \binits{S.C.}},
\bauthor{\bsnm{Dai}, \binits{J.Y.}},
\bauthor{\bsnm{Horoi}, \binits{M.}},
\bauthor{\bsnm{Liu}, \binits{H.}},
\bauthor{\bsnm{Lynch}, \binits{W.G.}},
\bauthor{\bsnm{Warren}, \binits{S.}}:
\batitle{Survey of excited state neutron spectroscopic factors for ${ Z = 8-28
  }$ nuclei}.
\bjtitle{Phys. Rev. Lett.}
\bvolume{102},
\bfpage{062501}
(\byear{2009})
\end{barticle}
\endbibitem

\bibitem{PhysRevC.103.054610}
\begin{barticle}
\bauthor{\bsnm{Tostevin}, \binits{J.A.}},
\bauthor{\bsnm{Gade}, \binits{A.}}:
\batitle{Updated systematics of intermediate-energy single-nucleon removal
  cross sections}.
\bjtitle{Phys. Rev. C}
\bvolume{103},
\bfpage{7}
(\byear{2021})
\end{barticle}
\endbibitem

\bibitem{KRAMER2001267}
\begin{barticle}
\bauthor{\bsnm{Kramer}, \binits{G.J.}},
\bauthor{\bsnm{Blok}, \binits{H.P.}},
\bauthor{\bsnm{Lapikás}, \binits{L.}}:
\batitle{A consistent analysis of (e,e'p) and (d,3he) experiments}.
\bjtitle{Nucl. Phys. A}
\bvolume{679},
\bfpage{86}
(\byear{2001})
\end{barticle}
\endbibitem

\bibitem{PhysRevLett.129.152501}
\begin{barticle}
\bauthor{\bsnm{Kay}, \binits{B.P.}},
\bauthor{\bsnm{Tang}, \binits{T.L.}},
\bauthor{\bsnm{Tolstukhin}, \binits{I.A.}},
\bauthor{\bsnm{Roderick}, \binits{G.B.}},
\bauthor{\bsnm{Mitchell}, \binits{A.J.}},
\bauthor{\bsnm{Ayyad}, \binits{Y.}},
\bauthor{\bsnm{Bennett}, \binits{S.A.}},
\bauthor{\bsnm{Chen}, \binits{J.}},
\bauthor{\bsnm{Chipps}, \binits{K.A.}},
\bauthor{\bsnm{Crawford}, \binits{H.L.}},
\bauthor{\bsnm{Freeman}, \binits{S.J.}},
\bauthor{\bsnm{Garrett}, \binits{K.}},
\bauthor{\bsnm{Gott}, \binits{M.D.}},
\bauthor{\bsnm{Hall}, \binits{M.R.}},
\bauthor{\bsnm{Hoffman}, \binits{C.R.}},
\bauthor{\bsnm{Jayatissa}, \binits{H.}},
\bauthor{\bsnm{Macchiavelli}, \binits{A.O.}},
\bauthor{\bsnm{MacGregor}, \binits{P.T.}},
\bauthor{\bsnm{Sharp}, \binits{D.K.}},
\bauthor{\bsnm{Wilson}, \binits{G.L.}}:
\batitle{Quenching of single-particle strength in ${ A=15 }$ nuclei}.
\bjtitle{Phys. Rev. Lett.}
\bvolume{129},
\bfpage{152501}
(\byear{2022})
\end{barticle}
\endbibitem

\bibitem{B88}
\begin{barticle}
\bauthor{\bsnm{Brown}, \binits{B.A.}},
\bauthor{\bsnm{Wildenthal}, \binits{B.H.}}:
\batitle{Status of the nuclear shell model}.
\bjtitle{Ann. Rev. Nucl. Part. Sci.}
\bvolume{38},
\bfpage{29}--\blpage{66}
(\byear{1988})
\end{barticle}
\endbibitem

\bibitem{PhysRevLett.107.032501}
\begin{barticle}
\bauthor{\bsnm{Jensen}, \binits{O.}},
\bauthor{\bsnm{Hagen}, \binits{G.}},
\bauthor{\bsnm{Hjorth-Jensen}, \binits{M.}},
\bauthor{\bsnm{Brown}, \binits{B.A.}},
\bauthor{\bsnm{Gade}, \binits{A.}}:
\batitle{Quenching of spectroscopic factors for proton removal in oxygen
  isotopes}.
\bjtitle{Phys. Rev. Lett.}
\bvolume{107},
\bfpage{4}
(\byear{2011})
\end{barticle}
\endbibitem

\bibitem{PhysRevLett.103.202502}
\begin{barticle}
\bauthor{\bsnm{Barbieri}, \binits{C.}}:
\batitle{Role of long-range correlations in the quenching of spectroscopic
  factors}.
\bjtitle{Phys. Rev. Lett.}
\bvolume{103},
\bfpage{4}
(\byear{2009})
\end{barticle}
\endbibitem

\bibitem{PhysRevC.92.014306}
\begin{barticle}
\bauthor{\bsnm{Cipollone}, \binits{A.}},
\bauthor{\bsnm{Barbieri}, \binits{C.}},
\bauthor{\bsnm{Navr\'atil}, \binits{P.}}:
\batitle{Chiral three-nucleon forces and the evolution of correlations along
  the oxygen isotopic chain}.
\bjtitle{Phys. Rev. C}
\bvolume{92},
\bfpage{12}
(\byear{2015})
\end{barticle}
\endbibitem

\bibitem{PhysRevC.104.L061301}
\begin{barticle}
\bauthor{\bsnm{Wylie}, \binits{J.}},
\bauthor{\bsnm{Oko\l{}owicz}, \binits{J.}},
\bauthor{\bsnm{Nazarewicz}, \binits{W.}},
\bauthor{\bsnm{P\l{}oszajczak}, \binits{M.}},
\bauthor{\bsnm{Wang}, \binits{S.M.}},
\bauthor{\bsnm{Mao}, \binits{X.}},
\bauthor{\bsnm{Michel}, \binits{N.}}:
\batitle{Spectroscopic factors in dripline nuclei}.
\bjtitle{Phys. Rev. C}
\bvolume{104},
\bfpage{7}
(\byear{2021})
\end{barticle}
\endbibitem

\bibitem{PhysRevC.92.034313}
\begin{barticle}
\bauthor{\bsnm{Duguet}, \binits{T.}},
\bauthor{\bsnm{Hergert}, \binits{H.}},
\bauthor{\bsnm{Holt}, \binits{J.D.}},
\bauthor{\bsnm{Som\`a}, \binits{V.}}:
\batitle{Nonobservable nature of the nuclear shell structure: Meaning,
  illustrations, and consequences}.
\bjtitle{Phys. Rev. C}
\bvolume{92},
\bfpage{15}
(\byear{2015})
\end{barticle}
\endbibitem

\bibitem{PhysRevC.104.034311}
\begin{barticle}
\bauthor{\bsnm{Tropiano}, \binits{A.J.}},
\bauthor{\bsnm{Bogner}, \binits{S.K.}},
\bauthor{\bsnm{Furnstahl}, \binits{R.J.}}:
\batitle{Short-range correlation physics at low renormalization group
  resolution}.
\bjtitle{Phys. Rev. C}
\bvolume{104},
\bfpage{16}
(\byear{2021})
\end{barticle}
\endbibitem

\bibitem{PhysRevC.106.024324}
\begin{barticle}
\bauthor{\bsnm{Tropiano}, \binits{A.J.}},
\bauthor{\bsnm{Bogner}, \binits{S.K.}},
\bauthor{\bsnm{Furnstahl}, \binits{R.J.}},
\bauthor{\bsnm{Hisham}, \binits{M.A.}}:
\batitle{Quasi-deuteron model at low renormalization group resolution}.
\bjtitle{Phys. Rev. C}
\bvolume{106},
\bfpage{8}
(\byear{2022})
\end{barticle}
\endbibitem

\bibitem{PhysRevC.106.024616}
\begin{barticle}
\bauthor{\bsnm{Hisham}, \binits{M.A.}},
\bauthor{\bsnm{Furnstahl}, \binits{R.J.}},
\bauthor{\bsnm{Tropiano}, \binits{A.J.}}:
\batitle{Renormalization group evolution of optical potentials: Explorations
  using a ``toy'' model}.
\bjtitle{Phys. Rev. C}
\bvolume{106},
\bfpage{13}
(\byear{2022})
\end{barticle}
\endbibitem

\bibitem{PhysRevC.100.054607}
\begin{barticle}
\bauthor{\bsnm{Hebborn}, \binits{C.}},
\bauthor{\bsnm{Capel}, \binits{P.}}:
\batitle{Sensitivity of one-neutron knockout to the nuclear structure of halo
  nuclei}.
\bjtitle{Phys. Rev. C}
\bvolume{100},
\bfpage{10}
(\byear{2019})
\end{barticle}
\endbibitem

\bibitem{HCinprep}
\begin{botherref}
\oauthor{\bsnm{Hebborn}, \binits{C.}},
\oauthor{\bsnm{Capel}, \binits{P.}}
in preparation
(2022)
\end{botherref}
\endbibitem

\bibitem{PhysRevC.77.044306}
\begin{barticle}
\bauthor{\bsnm{Gade}, \binits{A.}},
\bauthor{\bsnm{Adrich}, \binits{P.}},
\bauthor{\bsnm{Bazin}, \binits{D.}},
\bauthor{\bsnm{Bowen}, \binits{M.D.}},
\bauthor{\bsnm{Brown}, \binits{B.A.}},
\bauthor{\bsnm{Campbell}, \binits{C.M.}},
\bauthor{\bsnm{Cook}, \binits{J.M.}},
\bauthor{\bsnm{Glasmacher}, \binits{T.}},
\bauthor{\bsnm{Hansen}, \binits{P.G.}},
\bauthor{\bsnm{Hosier}, \binits{K.}},
\bauthor{\bsnm{McDaniel}, \binits{S.}},
\bauthor{\bsnm{McGlinchery}, \binits{D.}},
\bauthor{\bsnm{Obertelli}, \binits{A.}},
\bauthor{\bsnm{Siwek}, \binits{K.}},
\bauthor{\bsnm{Riley}, \binits{L.A.}},
\bauthor{\bsnm{Tostevin}, \binits{J.A.}},
\bauthor{\bsnm{Weisshaar}, \binits{D.}}:
\batitle{Reduction of spectroscopic strength: Weakly-bound and strongly-bound
  single-particle states studied using one-nucleon knockout reactions}.
\bjtitle{Phys. Rev. C}
\bvolume{77},
\bfpage{10}
(\byear{2008})
\end{barticle}
\endbibitem

\bibitem{PhysRevLett.119.262501}
\begin{barticle}
\bauthor{\bsnm{Aumann}, \binits{T.}},
\bauthor{\bsnm{Bertulani}, \binits{C.A.}},
\bauthor{\bsnm{Schindler}, \binits{F.}},
\bauthor{\bsnm{Typel}, \binits{S.}}:
\batitle{Peeling off neutron skins from neutron-rich nuclei: Constraints on the
  symmetry energy from neutron-removal cross sections}.
\bjtitle{Phys. Rev. Lett.}
\bvolume{119},
\bfpage{5}
(\byear{2017})
\end{barticle}
\endbibitem

\bibitem{ATKINSON2019135027}
\begin{barticle}
\bauthor{\bsnm{Atkinson}, \binits{M.C.}},
\bauthor{\bsnm{Dickhoff}, \binits{W.H.}}:
\batitle{Investigating the link between proton reaction cross sections and the
  quenching of proton spectroscopic factors in ${}^{48}$ca}.
\bjtitle{Phys. Lett. B}
\bvolume{798},
\bfpage{135027}
(\byear{2019})
\end{barticle}
\endbibitem

\bibitem{PASCHALIS2020135110}
\begin{barticle}
\bauthor{\bsnm{Paschalis}, \binits{S.}},
\bauthor{\bsnm{Petri}, \binits{M.}},
\bauthor{\bsnm{Macchiavelli}, \binits{A.O.}},
\bauthor{\bsnm{Hen}, \binits{O.}},
\bauthor{\bsnm{Piasetzky}, \binits{E.}}:
\batitle{Nucleon-nucleon correlations and the single-particle strength in
  atomic nuclei}.
\bjtitle{Phys. Lett. B}
\bvolume{800},
\bfpage{135110}
(\byear{2020})
\end{barticle}
\endbibitem

\bibitem{PhysRevLett.108.252501}
\begin{barticle}
\bauthor{\bsnm{Flavigny}, \binits{F.}},
\bauthor{\bsnm{Obertelli}, \binits{A.}},
\bauthor{\bsnm{Bonaccorso}, \binits{A.}},
\bauthor{\bsnm{Grinyer}, \binits{G.F.}},
\bauthor{\bsnm{Louchart}, \binits{C.}},
\bauthor{\bsnm{Nalpas}, \binits{L.}},
\bauthor{\bsnm{Signoracci}, \binits{A.}}:
\batitle{Nonsudden limits of heavy-ion induced knockout reactions}.
\bjtitle{Phys. Rev. Lett.}
\bvolume{108},
\bfpage{5}
(\byear{2012})
\end{barticle}
\endbibitem

\bibitem{PhysRevC.83.011601}
\begin{barticle}
\bauthor{\bsnm{Louchart}, \binits{C.}},
\bauthor{\bsnm{Obertelli}, \binits{A.}},
\bauthor{\bsnm{Boudard}, \binits{A.}},
\bauthor{\bsnm{Flavigny}, \binits{F.}}:
\batitle{Nucleon removal from unstable nuclei investigated via intranuclear
  cascade}.
\bjtitle{Phys. Rev. C}
\bvolume{83},
\bfpage{5}
(\byear{2011})
\end{barticle}
\endbibitem

\bibitem{GOMEZRAMOS2022137252}
\begin{barticle}
\bauthor{\bsnm{Gómez-Ramos}, \binits{M.}},
\bauthor{\bsnm{Gómez-Camacho}, \binits{J.}},
\bauthor{\bsnm{Moro}, \binits{A.M.}}:
\batitle{Binding-energy asymmetry in absorption explored through cdcc extended
  for complex potentials}.
\bjtitle{Phys. Lett. B}
\bvolume{832},
\bfpage{137252}
(\byear{2022})
\end{barticle}
\endbibitem

\bibitem{HP2022}
\begin{barticle}
\bauthor{\bsnm{Hebborn}, \binits{C.}},
\bauthor{\bsnm{Potel}, \binits{G.}}:
\batitle{{Green's function knockout formalism}}.
\bjtitle{Phys. Rev. C}
\bvolume{107}(\bissue{1}),
\bfpage{014607}
(\byear{2023})
{\href{https://arxiv.org/abs/2206.09948}{{arXiv:2206.09948}}}
{[nucl-th]}.
\doiurl{10.1103/PhysRevC.107.014607}
\end{barticle}
\endbibitem

\bibitem{GRI11}
\begin{barticle}
\bauthor{\bsnm{Grinyer}, \binits{G.F.}},
\bauthor{\bsnm{Bazin}, \binits{D.}},
\bauthor{\bsnm{Gade}, \binits{A.}},
\bauthor{\bsnm{Tostevin}, \binits{J.A.}},
\bauthor{\bsnm{Adrich}, \binits{P.}},
\bauthor{\bsnm{Bowen}, \binits{M.D.}},
\bauthor{\bsnm{Brown}, \binits{B.A.}},
\bauthor{\bsnm{Campbell}, \binits{C.M.}},
\bauthor{\bsnm{Cook}, \binits{J.M.}},
\bauthor{\bsnm{Glasmacher}, \binits{T.}},
\bauthor{\bsnm{McDaniel}, \binits{S.}},
\bauthor{\bsnm{Navr\'atil}, \binits{P.}},
\bauthor{\bsnm{Obertelli}, \binits{A.}},
\bauthor{\bsnm{Quaglioni}, \binits{S.}},
\bauthor{\bsnm{Siwek}, \binits{K.}},
\bauthor{\bsnm{Terry}, \binits{J.R.}},
\bauthor{\bsnm{Weisshaar}, \binits{D.}},
\bauthor{\bsnm{Wiringa}, \binits{R.B.}}:
\batitle{Knockout reactions from ${ p }$-shell nuclei: Tests of ab initio
  structure models}.
\bjtitle{Phys. Rev. Lett.}
\bvolume{106},
\bfpage{4}
(\byear{2011})
\end{barticle}
\endbibitem

\bibitem{KUC22}
\begin{barticle}
\bauthor{\bsnm{Kuchera}, \binits{A.N.}},
\bauthor{\bsnm{Bazin}, \binits{D.}},
\bauthor{\bsnm{Phan}, \binits{T.}},
\bauthor{\bsnm{Tostevin}, \binits{J.A.}},
\bauthor{\bsnm{Babo}, \binits{M.}},
\bauthor{\bsnm{Baumann}, \binits{T.}},
\bauthor{\bsnm{Bender}, \binits{P.C.}},
\bauthor{\bsnm{Bowry}, \binits{M.}},
\bauthor{\bsnm{Bradt}, \binits{J.}},
\bauthor{\bsnm{Brown}, \binits{J.}},
\bauthor{\bsnm{DeYoung}, \binits{P.A.}},
\bauthor{\bsnm{Elman}, \binits{B.}},
\bauthor{\bsnm{Finck}, \binits{J.E.}},
\bauthor{\bsnm{Gade}, \binits{A.}},
\bauthor{\bsnm{Grinyer}, \binits{G.F.}},
\bauthor{\bsnm{Jones}, \binits{M.D.}},
\bauthor{\bsnm{Longfellow}, \binits{B.}},
\bauthor{\bsnm{Lunderberg}, \binits{E.}},
\bauthor{\bsnm{Redpath}, \binits{T.H.}},
\bauthor{\bsnm{Rogers}, \binits{W.F.}},
\bauthor{\bsnm{Stiefel}, \binits{K.}},
\bauthor{\bsnm{Thoennessen}, \binits{M.}},
\bauthor{\bsnm{Votaw}, \binits{D.}},
\bauthor{\bsnm{Weisshaar}, \binits{D.}},
\bauthor{\bsnm{Whitmore}, \binits{K.}},
\bauthor{\bsnm{Wiringa}, \binits{R.B.}}:
\batitle{Mirror nucleon removal reactions in ${ p }$-shell nuclei}.
\bjtitle{Phys. Rev. C}
\bvolume{105},
\bfpage{8}
(\byear{2022})
\end{barticle}
\endbibitem

\bibitem{Lovell_2020}
\begin{barticle}
\bauthor{\bsnm{Lovell}, \binits{A.E.}},
\bauthor{\bsnm{Nunes}, \binits{F.M.}},
\bauthor{\bsnm{Catacora-Rios}, \binits{M.}},
\bauthor{\bsnm{King}, \binits{G.B.}}:
\batitle{Recent advances in the quantification of uncertainties in reaction
  theory}.
\bjtitle{J. Phys. G: Nucl. Part. Phys}
\bvolume{48},
\bfpage{014001}
(\byear{2020})
\end{barticle}
\endbibitem

\bibitem{Ioannides19781331}
\begin{barticle}
\bauthor{\bsnm{Ioannides}, \binits{A.A.}},
\bauthor{\bsnm{Johnson}, \binits{R.C.}}:
\batitle{Propagation of a deuteron in nuclear matter and the spin dependence of
  the deuteron optical potential}.
\bjtitle{Phys. Rev. C}
\bvolume{17},
\bfpage{1331}
(\byear{1978})
\end{barticle}
\endbibitem

\bibitem{Johnson1981750}
\begin{botherref}
\oauthor{\bsnm{Johnson}, \binits{R.C.}},
\oauthor{\bsnm{Tostevin}, \binits{J.A.}},
\oauthor{\bsnm{Lopes}, \binits{M.H.}}:
Antisymmetrization effects in deuteron-nucleus elastic scattering.
Nucl. Phys. 1980,
750
(1981)
\end{botherref}
\endbibitem

\bibitem{Johnson1982348}
\begin{barticle}
\bauthor{\bsnm{Johnson}, \binits{R.C.}},
\bauthor{\bsnm{Austern}, \binits{N.}},
\bauthor{\bsnm{Lopes}, \binits{M.H.}}:
\batitle{Antisymmetrized deuteron stripping}.
\bjtitle{Phys. Rev. C}
\bvolume{26},
\bfpage{348}--\blpage{356}
(\byear{1982})
\end{barticle}
\endbibitem

\bibitem{TOSTEVIN198783}
\begin{barticle}
\bauthor{\bsnm{Tostevin}, \binits{J.A.}},
\bauthor{\bsnm{Lopes}, \binits{M.H.}},
\bauthor{\bsnm{Johnson}, \binits{R.C.}}:
\batitle{Antisymmetrization corrections in deuteron elastic scattering and
  deuteron-induced transfer reactions}.
\bjtitle{Nucl. Phys. A}
\bvolume{465},
\bfpage{22}
(\byear{1987})
\end{barticle}
\endbibitem

\bibitem{Dinmore:2019ouu}
\begin{barticle}
\bauthor{\bsnm{Dinmore}, \binits{M.J.}},
\bauthor{\bsnm{Timofeyuk}, \binits{N.K.}},
\bauthor{\bsnm{Al-Khalili}, \binits{J.S.}},
\bauthor{\bsnm{Johnson}, \binits{R.C.}}:
\batitle{Effects of an induced three-body force in the incident channel of
  $(d,p)$ reactions}.
\bjtitle{Phys. Rev. C}
\bvolume{99},
\bfpage{064612}
(\byear{2019})
\end{barticle}
\endbibitem

\bibitem{Theeten:2007}
\begin{barticle}
\bauthor{\bsnm{Theeten}, \binits{M.}},
\bauthor{\bsnm{Matsumura}, \binits{H.}},
\bauthor{\bsnm{Orabi}, \binits{M.}},
\bauthor{\bsnm{Baye}, \binits{D.}},
\bauthor{\bsnm{Descouvemont}, \binits{P.}},
\bauthor{\bsnm{Fujiwara}, \binits{Y.}},
\bauthor{\bsnm{Suzuki}, \binits{Y.}}:
\batitle{Three-body model of light nuclei with microscopic nonlocal
  interactions}.
\bjtitle{Phys. Rev. C}
\bvolume{76},
\bfpage{11}
(\byear{2007})
\end{barticle}
\endbibitem

\bibitem{Avila17}
\begin{barticle}
\bauthor{\bsnm{Avila}, \binits{M.L.}},
\bauthor{\bsnm{Rehm}, \binits{K.E.}},
\bauthor{\bsnm{Almaraz-Calderon}, \binits{S.}},
\bauthor{\bsnm{Ayangeakaa}, \binits{A.D.}},
\bauthor{\bsnm{Dickerson}, \binits{C.}},
\bauthor{\bsnm{Hoffman}, \binits{C.R.}},
\bauthor{\bsnm{Jiang}, \binits{C.L.}},
\bauthor{\bsnm{Kay}, \binits{B.P.}},
\bauthor{\bsnm{Lai}, \binits{J.}},
\bauthor{\bsnm{Nusair}, \binits{O.}},
\bauthor{\bsnm{Pardo}, \binits{R.C.}},
\bauthor{\bsnm{Santiago-Gonzalez}, \binits{D.}},
\bauthor{\bsnm{Talwar}, \binits{R.}},
\bauthor{\bsnm{Ugalde}, \binits{C.}}:
\batitle{Study of ${ (\alpha,p) }$ and ${ (\alpha,n) }$ reactions with a
  multi-sampling ionization chamber}.
\bjtitle{Nucl. Instrum. Methods. Phys. Res. A}
\bvolume{859},
\bfpage{8}
(\byear{2017})
\end{barticle}
\endbibitem

\bibitem{Schiffer12}
\begin{barticle}
\bauthor{\bsnm{Schiffer}, \binits{J.P.}},
\bauthor{\bsnm{Hoffman}, \binits{C.R.}},
\bauthor{\bsnm{Kay}, \binits{B.P.}},
\bauthor{\bsnm{Clark}, \binits{J.A.}},
\bauthor{\bsnm{Deibel}, \binits{C.M.}},
\bauthor{\bsnm{Freeman}, \binits{S.J.}},
\bauthor{\bsnm{Howard}, \binits{A.M.}},
\bauthor{\bsnm{Mitchell}, \binits{A.J.}},
\bauthor{\bsnm{Parker}, \binits{P.D.}},
\bauthor{\bsnm{Sharp}, \binits{D.K.}},
\bauthor{\bsnm{Thomas}, \binits{J.S.}}:
\batitle{Test of sum rules in nucleon transfer reactions}.
\bjtitle{Phys. Rev. Lett.}
\bvolume{108},
\bfpage{5}
(\byear{2012})
\end{barticle}
\endbibitem

\bibitem{Schiffer13}
\begin{barticle}
\bauthor{\bsnm{Schiffer}, \binits{J.P.}},
\bauthor{\bsnm{Hoffman}, \binits{C.R.}},
\bauthor{\bsnm{Kay}, \binits{B.P.}},
\bauthor{\bsnm{Clark}, \binits{J.A.}},
\bauthor{\bsnm{Deibel}, \binits{C.M.}},
\bauthor{\bsnm{Freeman}, \binits{S.J.}},
\bauthor{\bsnm{Honma}, \binits{M.}},
\bauthor{\bsnm{Howard}, \binits{A.M.}},
\bauthor{\bsnm{Mitchell}, \binits{A.J.}},
\bauthor{\bsnm{Otsuka}, \binits{T.}},
\bauthor{\bsnm{Parker}, \binits{P.D.}},
\bauthor{\bsnm{Sharp}, \binits{D.K.}},
\bauthor{\bsnm{Thomas}, \binits{J.S.}}:
\batitle{{Valence nucleon populations in the Ni isotopes}}.
\bjtitle{Phys. Rev. C}
\bvolume{87},
\bfpage{15}
(\byear{2013})
\end{barticle}
\endbibitem

\bibitem{Wimmer18}
\begin{barticle}
\bauthor{\bsnm{Wimmer}, \binits{K.}}:
\batitle{Nucleon transfer reactions with radioactive beams}.
\bjtitle{J. Phys. G: Nucl. Part. Phys.}
\bvolume{45},
\bfpage{033002}
(\byear{2018})
\end{barticle}
\endbibitem

\bibitem{capel2007}
\begin{barticle}
\bauthor{\bsnm{Capel}, \binits{P.}},
\bauthor{\bsnm{Nunes}, \binits{F.M.}}:
\batitle{Peripherality of breakup reactions}.
\bjtitle{Phys. Rev. C}
\bvolume{75},
\bfpage{6}
(\byear{2007})
\end{barticle}
\endbibitem

\bibitem{Phillips:2020dmw}
\begin{barticle}
\bauthor{\bsnm{Phillips}, \binits{D.R.}}, \betal:
\batitle{{Get on the BAND wagon: a Bayesian framework for quantifying model
  uncertainties in nuclear dynamics}}.
\bjtitle{J. Phys. G: Nucl. Part. Phys.}
\bvolume{48},
\bfpage{072001}
(\byear{2021})
\end{barticle}
\endbibitem

\bibitem{Zhang:2015ajn}
\begin{barticle}
\bauthor{\bsnm{Zhang}, \binits{X.}},
\bauthor{\bsnm{Nollett}, \binits{K.M.}},
\bauthor{\bsnm{Phillips}, \binits{D.R.}}:
\batitle{Halo effective field theory constrains the solar ${ ^7\text{Be} + p
  \to ^8\text{B} + \gamma }$ rate}.
\bjtitle{Phys. Lett. B}
\bvolume{751},
\bfpage{535}
(\byear{2015})
\end{barticle}
\endbibitem

\bibitem{Iliadis:2016vkw}
\begin{barticle}
\bauthor{\bsnm{Iliadis}, \binits{C.}},
\bauthor{\bsnm{Anderson}, \binits{K.}},
\bauthor{\bsnm{Coc}, \binits{A.}},
\bauthor{\bsnm{Timmes}, \binits{F.}},
\bauthor{\bsnm{Starrfield}, \binits{S.}}:
\batitle{Bayesian estimation of thermonuclear reaction rates}.
\bjtitle{Astrophys. J.}
\bvolume{831},
\bfpage{107}
(\byear{2016})
\end{barticle}
\endbibitem

\bibitem{Acharya:2016kfl}
\begin{barticle}
\bauthor{\bsnm{Acharya}, \binits{B.}},
\bauthor{\bsnm{Carlsson}, \binits{B.D.}},
\bauthor{\bsnm{Ekstr\"om}, \binits{A.}},
\bauthor{\bsnm{Forss\'en}, \binits{C.}},
\bauthor{\bsnm{Platter}, \binits{L.}}:
\batitle{Uncertainty quantification for proton\textendash{}proton fusion in
  chiral effective field theory}.
\bjtitle{Phys. Lett. B}
\bvolume{760},
\bfpage{584}
(\byear{2016})
\end{barticle}
\endbibitem

\bibitem{Wesolowski:2018lzj}
\begin{barticle}
\bauthor{\bsnm{Wesolowski}, \binits{S.}},
\bauthor{\bsnm{Furnstahl}, \binits{R.J.}},
\bauthor{\bsnm{Melendez}, \binits{J.A.}},
\bauthor{\bsnm{Phillips}, \binits{D.R.}}:
\batitle{Exploring bayesian parameter estimation for chiral effective field
  theory using nucleon-nucleon phase shifts}.
\bjtitle{J. Phys. G: Nucl. Part. Phys.}
\bvolume{46},
\bfpage{045102}
(\byear{2019})
\end{barticle}
\endbibitem

\bibitem{Zhang:2019odg}
\begin{barticle}
\bauthor{\bsnm{Zhang}, \binits{X.}},
\bauthor{\bsnm{Nollett}, \binits{K.M.}},
\bauthor{\bsnm{Phillips}, \binits{D.R.}}:
\batitle{${ S }$-factor and scattering-parameter extractions from ${
  {}^{3}\text{He} +{}^{4}\text{He} \rightarrow {}^{7}\text{Be} + \gamma }$}.
\bjtitle{J. Phys. G: Nucl. Part. Phys.}
\bvolume{47},
\bfpage{054002}
(\byear{2020})
\end{barticle}
\endbibitem

\bibitem{Premarathna:2019tup}
\begin{barticle}
\bauthor{\bsnm{Premarathna}, \binits{P.}},
\bauthor{\bsnm{Rupak}, \binits{G.}}:
\batitle{Bayesian analysis of capture reactions ${
  ^3\text{He}(\alpha,\gamma)^7\text{Be} }$ and ${
  ^3\text{H}(\alpha,\gamma)^7\text{Li} }$}.
\bjtitle{Eur. Phys. J. A}
\bvolume{56},
\bfpage{166}
(\byear{2022})
\end{barticle}
\endbibitem

\bibitem{Acharya:2019fij}
\begin{barticle}
\bauthor{\bsnm{Acharya}, \binits{B.}},
\bauthor{\bsnm{Bacca}, \binits{S.}}:
\batitle{Neutrino-deuteron scattering: Uncertainty quantification and new ${
  L_{1,A} }$ constraints}.
\bjtitle{Phys. Rev. C}
\bvolume{101},
\bfpage{015505}
(\byear{2020})
\end{barticle}
\endbibitem

\bibitem{Melendez:2019izc}
\begin{barticle}
\bauthor{\bsnm{Melendez}, \binits{J.A.}},
\bauthor{\bsnm{Furnstahl}, \binits{R.J.}},
\bauthor{\bsnm{Phillips}, \binits{D.R.}},
\bauthor{\bsnm{Pratola}, \binits{M.T.}},
\bauthor{\bsnm{Wesolowski}, \binits{S.}}:
\batitle{Quantifying correlated truncation errors in effective field theory}.
\bjtitle{Phys. Rev. C}
\bvolume{100},
\bfpage{044001}
(\byear{2019})
\end{barticle}
\endbibitem

\bibitem{Maris:2020qne}
\begin{barticle}
\bauthor{\bsnm{Maris}, \binits{P.}}, \betal:
\batitle{Light nuclei with semilocal momentum-space regularized chiral
  interactions up to third order}.
\bjtitle{Phys. Rev. C}
\bvolume{103},
\bfpage{054001}
(\byear{2021})
\end{barticle}
\endbibitem

\bibitem{Higa:2020pvj}
\begin{botherref}
\oauthor{\bsnm{Higa}, \binits{R.}},
\oauthor{\bsnm{Premarathna}, \binits{P.}},
\oauthor{\bsnm{Rupak}, \binits{G.}}:
Coupled-channel treatment of ${ ^7\text{Be}(p,\gamma)^8\text{B} }$ in effective
  field theory.
arXiv:2010.13003
(2020)
\end{botherref}
\endbibitem

\bibitem{Drischler:2020hwi}
\begin{barticle}
\bauthor{\bsnm{Drischler}, \binits{C.}},
\bauthor{\bsnm{Furnstahl}, \binits{R.J.}},
\bauthor{\bsnm{Melendez}, \binits{J.A.}},
\bauthor{\bsnm{Phillips}, \binits{D.R.}}:
\batitle{{How well do we know the neutron-matter equation of state at the
  densities inside neutron stars? A Bayesian approach with correlated
  uncertainties}}.
\bjtitle{Phys. Rev. Lett.}
\bvolume{125},
\bfpage{202702}
(\byear{2020})
\end{barticle}
\endbibitem

\bibitem{Drischler:2020yad}
\begin{barticle}
\bauthor{\bsnm{Drischler}, \binits{C.}},
\bauthor{\bsnm{Melendez}, \binits{J.A.}},
\bauthor{\bsnm{Furnstahl}, \binits{R.J.}},
\bauthor{\bsnm{Phillips}, \binits{D.R.}}:
\batitle{Quantifying uncertainties and correlations in the nuclear-matter
  equation of state}.
\bjtitle{Phys. Rev. C}
\bvolume{102},
\bfpage{054315}
(\byear{2020})
\end{barticle}
\endbibitem

\bibitem{Poudel:2021mii}
\begin{barticle}
\bauthor{\bsnm{Poudel}, \binits{M.}},
\bauthor{\bsnm{Phillips}, \binits{D.R.}}:
\batitle{Effective field theory analysis of ${ ^{3}\text{He}-\alpha }$
  scattering data}.
\bjtitle{J. Phys. G: Nucl. Part. Phys.}
\bvolume{49},
\bfpage{045102}
(\byear{2022})
\end{barticle}
\endbibitem

\bibitem{Acharya:2021lrv}
\begin{barticle}
\bauthor{\bsnm{Acharya}, \binits{B.}},
\bauthor{\bsnm{Bacca}, \binits{S.}}:
\batitle{Gaussian process error modeling for chiral effective-field-theory
  calculations of $np{\leftrightarrow}d{\gamma}$ at low energies}.
\bjtitle{Phys. Lett. B}
\bvolume{827},
\bfpage{137011}
(\byear{2022})
\end{barticle}
\endbibitem

\bibitem{Odell:2021nmp}
\begin{barticle}
\bauthor{\bsnm{Odell}, \binits{D.}},
\bauthor{\bsnm{Brune}, \binits{C.R.}},
\bauthor{\bsnm{Phillips}, \binits{D.R.}},
\bauthor{\bsnm{deBoer}, \binits{R.J.}},
\bauthor{\bsnm{Paneru}, \binits{S.N.}}:
\batitle{{Performing Bayesian analyses with AZURE2 using BRICK: an application
  to the $^7$Be system}}.
\bjtitle{Front. in Phys.}
\bvolume{10},
\bfpage{888476}
(\byear{2022})
\end{barticle}
\endbibitem

\bibitem{Wesolowski:2021cni}
\begin{barticle}
\bauthor{\bsnm{Wesolowski}, \binits{S.}},
\bauthor{\bsnm{Svensson}, \binits{I.}},
\bauthor{\bsnm{Ekstr\"om}, \binits{A.}},
\bauthor{\bsnm{Forss\'en}, \binits{C.}},
\bauthor{\bsnm{Furnstahl}, \binits{R.J.}},
\bauthor{\bsnm{Melendez}, \binits{J.A.}},
\bauthor{\bsnm{Phillips}, \binits{D.R.}}:
\batitle{Rigorous constraints on three-nucleon forces in chiral effective field
  theory from fast and accurate calculations of few-body observables}.
\bjtitle{Phys. Rev. C}
\bvolume{104},
\bfpage{064001}
(\byear{2021})
\end{barticle}
\endbibitem

\bibitem{Acharya:2022drl}
\begin{botherref}
\oauthor{\bsnm{Acharya}, \binits{B.}},
\oauthor{\bsnm{Bacca}, \binits{S.}},
\oauthor{\bsnm{Bonaiti}, \binits{F.}},
\oauthor{\bsnm{Muli}, \binits{S.S.L.}},
\oauthor{\bsnm{Sobczyk}, \binits{J.E.}}:
Uncertainty quantification in electromagnetic observables of nuclei.
arXiv:2210.04632
(2022)
\end{botherref}
\endbibitem

\bibitem{lovell2018}
\begin{barticle}
\bauthor{\bsnm{Lovell}, \binits{A.E.}},
\bauthor{\bsnm{Nunes}, \binits{F.M.}}:
\batitle{Constraining transfer cross sections using bayes' theorem}.
\bjtitle{Phys. Rev. C}
\bvolume{97},
\bfpage{16}
(\byear{2018})
\end{barticle}
\endbibitem

\bibitem{king2018}
\begin{barticle}
\bauthor{\bsnm{King}, \binits{G.B.}},
\bauthor{\bsnm{Lovell}, \binits{A.E.}},
\bauthor{\bsnm{Nunes}, \binits{F.M.}}:
\batitle{Uncertainty quantification due to optical potentials in models for ${
  (d,p) }$ reactions}.
\bjtitle{Phys. Rev. C}
\bvolume{98},
\bfpage{9}
(\byear{2018})
\end{barticle}
\endbibitem

\bibitem{catacora2019}
\begin{barticle}
\bauthor{\bsnm{Catacora-Rios}, \binits{M.}},
\bauthor{\bsnm{King}, \binits{G.B.}},
\bauthor{\bsnm{Lovell}, \binits{A.E.}},
\bauthor{\bsnm{Nunes}, \binits{F.M.}}:
\batitle{Exploring experimental conditions to reduce uncertainties in the
  optical potential}.
\bjtitle{Phys. Rev. C}
\bvolume{100},
\bfpage{10}
(\byear{2019})
\end{barticle}
\endbibitem

\bibitem{catacora2021}
\begin{barticle}
\bauthor{\bsnm{Catacora-Rios}, \binits{M.}},
\bauthor{\bsnm{King}, \binits{G.B.}},
\bauthor{\bsnm{Lovell}, \binits{A.E.}},
\bauthor{\bsnm{Nunes}, \binits{F.M.}}:
\batitle{Statistical tools for a better optical model}.
\bjtitle{Phys. Rev. C}
\bvolume{104},
\bfpage{9}
(\byear{2021})
\end{barticle}
\endbibitem

\bibitem{king2019}
\begin{barticle}
\bauthor{\bsnm{King}, \binits{G.B.}},
\bauthor{\bsnm{Lovell}, \binits{A.E.}},
\bauthor{\bsnm{Neufcourt}, \binits{L.}},
\bauthor{\bsnm{Nunes}, \binits{F.M.}}:
\batitle{Direct comparison between bayesian and frequentist uncertainty
  quantification for nuclear reactions}.
\bjtitle{Phys. Rev. Lett.}
\bvolume{122},
\bfpage{5}
(\byear{2019})
\end{barticle}
\endbibitem

\bibitem{Melendez:2022kid}
\begin{botherref}
\oauthor{\bsnm{Melendez}, \binits{J.A.}},
\oauthor{\bsnm{Drischler}, \binits{C.}},
\oauthor{\bsnm{Furnstahl}, \binits{R.J.}},
\oauthor{\bsnm{Garcia}, \binits{A.J.}},
\oauthor{\bsnm{Zhang}, \binits{X.}}:
Model reduction methods for nuclear emulators.
arXiv:2203.05528
(2022)
\end{botherref}
\endbibitem

\bibitem{Bonilla:2022rph}
\begin{barticle}
\bauthor{\bsnm{Bonilla}, \binits{E.}},
\bauthor{\bsnm{Giuliani}, \binits{P.}},
\bauthor{\bsnm{Godbey}, \binits{K.}},
\bauthor{\bsnm{Lee}, \binits{D.}}:
\batitle{{Training and projecting: A reduced basis method emulator for
  many-body physics}}.
\bjtitle{Phys. Rev. C}
\bvolume{106}(\bissue{5}),
\bfpage{054322}
(\byear{2022})
{\href{https://arxiv.org/abs/2203.05284}{{arXiv:2203.05284}}}
{[nucl-th]}.
\doiurl{10.1103/PhysRevC.106.054322}
\end{barticle}
\endbibitem

\bibitem{surer2022}
\begin{barticle}
\bauthor{\bsnm{S\"urer}, \binits{O.}},
\bauthor{\bsnm{Nunes}, \binits{F.M.}},
\bauthor{\bsnm{Plumlee}, \binits{M.}},
\bauthor{\bsnm{Wild}, \binits{S.M.}}:
\batitle{Uncertainty quantification in breakup reactions}.
\bjtitle{Phys. Rev. C}
\bvolume{106},
\bfpage{12}
(\byear{2022})
\end{barticle}
\endbibitem

\bibitem{Furnstahl:2020abp}
\begin{barticle}
\bauthor{\bsnm{Furnstahl}, \binits{R.J.}},
\bauthor{\bsnm{Garcia}, \binits{A.J.}},
\bauthor{\bsnm{Millican}, \binits{P.J.}},
\bauthor{\bsnm{Zhang}, \binits{X.}}:
\batitle{Efficient emulators for scattering using eigenvector continuation}.
\bjtitle{Phys. Lett. B}
\bvolume{809},
\bfpage{135719}
(\byear{2020})
\end{barticle}
\endbibitem

\bibitem{Drischler:2021qoy}
\begin{barticle}
\bauthor{\bsnm{Drischler}, \binits{C.}},
\bauthor{\bsnm{Quinonez}, \binits{M.}},
\bauthor{\bsnm{Giuliani}, \binits{P.G.}},
\bauthor{\bsnm{Lovell}, \binits{A.E.}},
\bauthor{\bsnm{Nunes}, \binits{F.M.}}:
\batitle{Toward emulating nuclear reactions using eigenvector continuation}.
\bjtitle{Phys. Lett. B}
\bvolume{823},
\bfpage{136777}
(\byear{2021})
\end{barticle}
\endbibitem

\bibitem{Melendez:2021lyq}
\begin{barticle}
\bauthor{\bsnm{Melendez}, \binits{J.A.}},
\bauthor{\bsnm{Drischler}, \binits{C.}},
\bauthor{\bsnm{Garcia}, \binits{A.J.}},
\bauthor{\bsnm{Furnstahl}, \binits{R.J.}},
\bauthor{\bsnm{Zhang}, \binits{X.}}:
\batitle{Fast \& accurate emulation of two-body scattering observables without
  wave functions}.
\bjtitle{Phys. Lett. B}
\bvolume{821},
\bfpage{136608}
(\byear{2021})
\end{barticle}
\endbibitem

\bibitem{Bai:2021xok}
\begin{barticle}
\bauthor{\bsnm{Bai}, \binits{D.}},
\bauthor{\bsnm{Ren}, \binits{Z.}}:
\batitle{Generalizing the calculable ${ R }$-matrix theory and eigenvector
  continuation to the incoming wave boundary condition}.
\bjtitle{Phys. Rev. C}
\bvolume{103},
\bfpage{014612}
(\byear{2021})
\end{barticle}
\endbibitem

\bibitem{Zhang:2021jmi}
\begin{barticle}
\bauthor{\bsnm{Zhang}, \binits{X.}},
\bauthor{\bsnm{Furnstahl}, \binits{R.J.}}:
\batitle{Fast emulation of quantum three-body scattering}.
\bjtitle{Phys. Rev. C}
\bvolume{105},
\bfpage{064004}
(\byear{2021})
\end{barticle}
\endbibitem

\bibitem{Macchiavelli:2015xml}
\begin{barticle}
\bauthor{\bsnm{Macchiavelli}, \binits{A.O.}}:
\batitle{{How to Study Efimov States in Exotic Nuclei?}}
\bjtitle{Few Body Syst.}
\bvolume{56}(\bissue{11-12}),
\bfpage{773}--\blpage{778}
(\byear{2015})
\end{barticle}
\endbibitem

\bibitem{Efimov:1970zz}
\begin{barticle}
\bauthor{\bsnm{Efimov}, \binits{V.}}:
\batitle{Energy levels arising form the resonant two-body forces in a
  three-body system}.
\bjtitle{Phys. Lett. B}
\bvolume{33},
\bfpage{563}
(\byear{1970})
\end{barticle}
\endbibitem

\bibitem{Efimov:1971zz}
\begin{barticle}
\bauthor{\bsnm{Efimov}, \binits{V.N.}}:
\batitle{Weakly-bound states of 3 resonantly-interacting particles}.
\bjtitle{Sov. J. Nucl. Phys.}
\bvolume{12},
\bfpage{589}
(\byear{1971})
\end{barticle}
\endbibitem

\bibitem{Landau:1991wop}
\begin{bbook}
\bauthor{\bsnm{Landau}, \binits{L.D.}},
\bauthor{\bsnm{Lifshits}, \binits{E.M.}}:
\bbtitle{{Q}uantum {M}echanics: {N}on-{R}elativistic {T}heory}
vol. \bseriesno{3}.
\bpublisher{Butterworth-Heinemann},
\blocation{Oxford}
(\byear{1991})
\end{bbook}
\endbibitem

\bibitem{Adhikari:1982zz}
\begin{barticle}
\bauthor{\bsnm{Adhikari}, \binits{S.K.}},
\bauthor{\bsnm{Tomio}, \binits{L.}}:
\batitle{Efimov effect in the three-nucleon system}.
\bjtitle{Phys. Rev. C}
\bvolume{26},
\bfpage{83}
(\byear{1982})
\end{barticle}
\endbibitem

\bibitem{Rupak:2018gnc}
\begin{barticle}
\bauthor{\bsnm{Rupak}, \binits{G.}},
\bauthor{\bsnm{Vaghani}, \binits{A.}},
\bauthor{\bsnm{Higa}, \binits{R.}},
\bauthor{\bparticle{van} \bsnm{Kolck}, \binits{U.}}:
\batitle{Fate of the neutron\textendash{}deuteron virtual state as an {E}fimov
  level}.
\bjtitle{Phys. Lett. B}
\bvolume{791},
\bfpage{414}
(\byear{2019})
\end{barticle}
\endbibitem

\bibitem{Kuhn:2021dvu}
\begin{barticle}
\bauthor{\bsnm{Kuhn}, \binits{K.}}, \betal:
\batitle{{Experimental study of the nature of the $1^{-}$ and $2^{-}$ excited
  states in $^{10}$Be using the $^{11}$Be$(p,d)$ reaction in inverse
  kinematics}}.
\bjtitle{Phys. Rev. C}
\bvolume{104},
\bfpage{044601}
(\byear{2021})
\end{barticle}
\endbibitem

\end{thebibliography}


\end{document}